\documentclass[twocolumn]{aastex61}
\usepackage{graphicx}	
\usepackage{amsmath}	
\usepackage{amssymb}	
\usepackage{enumitem}
\usepackage{natbib}
\usepackage{cleveref}
\usepackage{booktabs}
\defcitealias{Lehner2018}{CCC~I}
\defcitealias{Wotta2019}{CCC~II}
\defcitealias{Lehner2019}{CCC~III}
\defcitealias{Savage2014}{S14}
\defcitealias{Tchernyshyov2022}{T22}
\DeclareOldFontCommand{\rm}{\normalfont\rmfamily}{\mathrm}
\DeclareOldFontCommand{\sf}{\normalfont\sffamily}{\mathsf}
\DeclareOldFontCommand{\tt}{\normalfont\ttfamily}{\mathtt}
\DeclareOldFontCommand{\bf}{\normalfont\bfseries}{\mathbf}
\DeclareOldFontCommand{\it}{\normalfont\itshape}{\mathit}
\DeclareOldFontCommand{\sl}{\normalfont\slshape}{\@nomath\sl}
\DeclareOldFontCommand{\sc}{\normalfont\scshape}{\@nomath\sc}

\newcommand{\neviii}{\mbox{Ne\,{\sc viii}}}

\newcommand{\ciii}{\mbox{C\,{\sc iii}}}
\newcommand{\cii}{\mbox{C\,{\sc ii}}}

\newcommand{\siiii}{\mbox{Si\,{\sc iii}}}
\newcommand{\siii}{\mbox{Si\,{\sc ii}}}
\newcommand{\siv}{\mbox{S\,{\sc iv}}}
\newcommand{\sv}{\mbox{S\,{\sc v}}}
\newcommand{\svi}{\mbox{S\,{\sc vi}}}

\newcommand{\oii}{\mbox{O\,{\sc ii}}}
\newcommand{\oiii}{\mbox{O\,{\sc iii}}}
\newcommand{\oiv}{\mbox{O\,{\sc iv}}}

\newcommand{\ovi}{\mbox{O\,{\sc vi}}}

\newcommand{\mgii}{\ifmmode {\rm Mg}{\textsc{ii}} \else Mg\,{\sc ii}\fi}
\newcommand{\heii}{\ifmmode {\rm He}{\textsc{ii}} \else He\,{\sc ii}\fi}

\newcommand{\hi}{\mbox{H\,{\sc i}}}
\newcommand{\mgi}{\ifmmode {\rm Mg}{\textsc{i}} \else Mg\,{\sc i}\fi}

\newcommand{\sn}{\ensuremath{\rm S/N}}

\newcommand{\hst}{\it HST}
\newcommand{\lya}{Ly$\alpha$}

\newcommand{\colden}{$\log N(\hi)$}

\newcommand{\rvir}{\ensuremath{R_{\rm vir}}}

\renewcommand{\arraystretch}{1.225}

\def\kms{\hbox{km\,s$^{-1}$}}
\def\cmsq{\hbox{cm$^{-2}$}}

\def\HI{\hbox{{\rm H~}\kern 0.1em{\sc i}}}
\newcommand{\mean}{$\mu$}
\newcommand{\median}{$\mu_{1/2}$}
\newcommand{\std}{$\sigma$} 
\newcommand{\sdmu}{$\sigma_{\mu}$}
\newcommand{\sdmedian}{$\sigma_{\mu_{1/2}}$}

\newcommand{\nhi}{$N_{\rm H\,I}$}

\newcommand{\coldenovi}{$\log N(${\ovi}$)$}

\newcommand{\xh}{\ensuremath{\rm [X/H]}}
\newcommand{\vmin}{\ensuremath{v_{\rm min}}}
\newcommand{\vmax}{\ensuremath{v_{\rm max}}}
\newcommand{\nav}{\ensuremath{N_a(v)}}
\newcommand{\vhi}{\ensuremath{v_{\rm H\,I}}}
\newcommand{\vovi}{\ensuremath{v_{\rm O\,VI}}}
\newcommand{\vcii}{\ensuremath{v_{\rm C\,II}}}
\newcommand{\vciii}{\ensuremath{v_{\rm C\,III}}}
\newcommand{\meancoldenovi}{\ensuremath{\log\,\langle\,N(${\ovi}$)\,\rangle}}
\newcommand{\geomeancoldenovi}{\ensuremath{\langle\,\log\,N(${\ovi}$)\,\rangle}}
\shortauthors{Sameer et al.}
\shorttitle{CCC. V: the {\ovi} dichotomy in lower-ionization gas}
\begin{document}

\title{The COS CGM Compendium V: The Dichotomy of {\ovi} Associated with Low- and High-Metallicity Cool Gas at $\lowercase{z}<1$}

\author[0000-0001-9966-6790]{Sameer}
\affiliation{Department of Physics and Astronomy, University of Notre Dame, Notre Dame, IN 46556, USA}

\author[0000-0001-9158-0829]{Nicolas Lehner}
\affiliation{Department of Physics and Astronomy, University of Notre Dame, Notre Dame, IN 46556, USA}

\author[0000-0002-2591-3792]{J. Christopher Howk}
\affiliation{Department of Physics and Astronomy, University of Notre Dame, Notre Dame, IN 46556, USA}

\author[0000-0003-0724-4115]{Andrew J. Fox}
\affiliation{Space Telescope Science Institute, 3700 San Martin Drive, Baltimore, MD 21218, USA}
\affiliation{Department of Physics \& Astronomy, Johns Hopkins University, 3400 N. Charles St., Baltimore, MD 21218, USA}

\author[0000-0002-7893-1054]{John M. O'Meara}
\affiliation{W.M. Keck Observatory 65-1120 Mamalahoa Highway, Kamuela, HI 96743, USA}

\author[0000-0002-3391-2116]{Benjamin D. Oppenheimer}
\affiliation{CASA, Department of Astrophysical and Planetary Sciences, University of Colorado, Boulder, CO 80309, USA}

\begin{abstract}

We analyze the \ovi\ content and kinematics for 126 {\hi}-selected absorbers at $0.14 \lesssim z \lesssim 0.73$  for which the metallicities of their cool photoionized phase have been determined. We separate the absorbers into 100 strong {\lya} forest systems (SLFSs with $15 \la$\,{\colden}\,$< 16.2$) and 26 partial Lyman Limit systems (pLLSs with $16.2\le$\,{\colden}\,$\le 17.2$). The sample is drawn from the COS CGM Compendium (CCC) and has \ovi\ coverage in $\sn \geq 8$ {\it HST}/COS G130M/G160M QSO spectra, yielding a $2\sigma$ completeness level of {\coldenovi}$\,\geq 13.6$. The \ovi\ detection rates differ substantially between low-metallicity (LM; {\xh} $\leq -1.4$) and high-metallicity (HM; {\xh} $> -1.4$) SLFSs, with  20\% and 60\% detection rates, respectively. The  \ovi\ detection frequency for the HM and LM pLLSs is, however, similar at $\sim$60\%. The SLFSs and pLLSs without detected \ovi\ are consistent with the absorbing gas being in a single phase, while those with \ovi\ trace multiphase gas. We show that the \ovi\ velocity widths and column densities have different distributions in LM and HM gas. We find a strong correlation between \ovi\ column density and metallicity. The strongest (\coldenovi$\,\ga 14$) and broadest {\ovi} absorbers are nearly always associated with HM absorbers, while weaker \ovi\ absorbers are found in both LM and HM absorbers. From comparisons with galaxy-selected and blind \ovi\ surveys, we conclude absorbers with \coldenovi$\,\ga 14$ most likely arise in the circumgalactic medium (CGM) of star-forming galaxies. Absorbers with weak \ovi\ likely trace the extended CGM or intergalactic medium (IGM), while those without \ovi\ likely originate in the IGM.
\end{abstract}


\section{Introduction}\label{s-intro}

The ionized gaseous halos surrounding galaxies play a crucial role in their evolution. These halos, a.k.a. the circumgalactic media (CGM), are important reservoirs of baryons and metals, harboring more than 50\% of the baryonic and metal budget of the universe~\citep[e.g.,][]{Werk2014,tumlinson2017circumgalactic,PH2020}. They regulate the processes of accretion and feedback in galaxies, which are necessary for galaxies to sustain star formation over cosmological timescales~\citep[e.g.,][]{Maller2004,Dekel2006}, but also to avoid forming more stars than observed~\citep[e.g.,][]{Kacprzak2008,Oppenheimer2010,Oppenheimer2012,Faucher2011,Fumagalli2011b}. The growth and evolution of galaxies are driven by the exchange of matter between them, their CGM, and the diffuse intergalactic medium (IGM)~(e.g., \citealt{dave2012}). 

This exchange of matter can be probed by QSO absorption line systems, whose \hi\ column density can serve as a metric of the overdensity ($\delta$) of the gas \citep{Schaye2001}. Absorbers with {\colden}\,$= 15$--19 fall between the diffuse IGM probed by the \lya\ forest ($\log\delta < 1$, {\colden}\,$\la 14.5$) and the galaxies that can be probed by damped \lya\ absorbers ({\colden}\,$\ga 20.3$). These absorbers at the interface between the IGM and galaxies comprise the strong {\lya} forest systems (SLFSs; {\colden}\,$= 15$--16.2), the partial Lyman limit systems (pLLSs; {\colden}\,$= 16.2$--17.2), and the Lyman limit systems (LLSs; {\colden}\,$= 17.2$--19) following the nomenclature defined in \citet{Lehner2018} (hereafter \citetalias{Lehner2018}). Studies of galaxies in QSO fields where these absorbers have been observed show that indeed they often probe the CGM of galaxies~\citep[e.g.,][]{Steidel1995,Chen2000,Cooper2021,Wilde2021,Berg2023}. The  BASIC survey \citep{Berg2023} shows that while the metal-enriched pLLSs at $z\la 1$ are indeed always associated with (relatively massive) galaxies, the very metal-poor pLLSs, on the other hand, can track a more diverse set of physical structures, including the CGM of galaxies, and also dense regions of the IGM devoid of any galaxies down to very low masses. The MAGG survey similarly shows that LLSs at $z\sim 3$--4 are either located in regions close to galaxies or likely in filaments of the IGM \citep{Fumagalli2016,Lofthouse2020,Lofthouse2023}. While future and ongoing surveys will refine these findings, it is evident already that strong \hi\ absorbers with $15 \la$\,{\colden}\,$\la 19$ probe a wide range of environments, especially for the low-metallicity population; the metallicity is a key diagnostic for deciphering the origins of the strong \hi\ absorbers. In low-$z$ simulations, absorbers with {\colden}\,$\ga 15$ generally track the CGM of galaxies within the impact parameters of 300\,kpc~(e.g., \citealt{Hafen2017,Weng2024}). However, very high-resolution cosmological simulations of the IGM at $z\sim 3$ also show that LLSs can be found in small-scale, dense-cloud structures of the IGM that were not resolved in lower-resolution simulations \citep{Mandelker2019,Mandelker2021}.

\citet{Fumagalli2016} (and see also \citealt{Wotta2016}; \citealt{Wotta2019}---hereafter \citetalias{Wotta2019}; \citealt{Lehner2019}---hereafter \citetalias{Lehner2019}; \citealt{Gibson2022}) showed that the metallicity estimates in the (predominantly) ionized regions of the universe are less dependent on the ionization assumptions, such as the slope of the extreme ultraviolet background (EUVB), than other gas properties such as hydrogen density, size scale, or the total hydrogen column density, making metallicity a robust probe of the origin and chemical enrichment history of the gas associated with strong \hi\ absorbers. For example, \citet{Gibson2022} find that the average uncertainty of inferred metallicities of lower-ionization gas only increased from 0.08 dex to 0.14 dex when switching from models with a fixed EUVB slope to those with slope varying over $-2.0$ to $-1.4$ for the KS19 ionizing background models \citep{Khaire2019}.

The strong \hi\ absorbers (SLFSs, pLLSs, and LLSs) exhibit absorption in low ions (e.g., {\cii}, {\mgii}, {\siii}), intermediate ions (e.g., \oii, \oiii, \siiii, \ciii), and sometimes \oiv\  that have velocity profiles that are similar to the associated \hi\ absorption profiles \citepalias{Lehner2018}.\footnote{The separation between the low and intermediate ions is based on their ionization potentials, with low ions ({\cii}, {\mgii}, {\siii}) having ionization potentials between 15.04--24.38 eV, and intermediate ions (\oii, \oiii, \siiii, \ciii) having ionization potentials between 33.49--54.94 eV. \oiv\, with an ionization potential of 77.41 eV, represents a higher ionization state.} All these ions can also be in many cases modeled with a single-phase, photoionization model where the ionization is produced by the EUVB (e.g., \citealt{Cooksey2008,Lehner2013}; \citetalias{Wotta2019, Lehner2019}; \citealt{zahedy2019}), although in some cases (especially in the pLLS and LLS regimes), they may require more than a single ionization phase (e.g., \citealt{Lehner2019,haislmaier2021,Qu2022,Sameer2024CMBM}). Some of these \hi\ absorbers also have \ovi\ absorption found at similar redshifts (especially, for absorbers {\colden}\,$\ga 16$), and these photoionization models generally underestimate the amount of \ovi\ at $z\la 1$, implying that,  when \ovi\ is detected, the gas probed by strong \hi\ absorbers must have multiple gas-phases. The reason is that the EUVB is typically not hard enough at $z\la 1$ and the densities are not low enough to produce the \ovi\ with the other ions in a single-phase photoionized gas for these absorbers (e.g., \citealt{Cooksey2008,Lehner2009,Kacprzak2012,Crighton2013,Lehner2013}; \citealt{Savage2014}---hereafter \citetalias{Savage2014}; \citealt{Werk2016,Rosenwasser2018,Cooper2021,haislmaier2021,sameer2021,Sameer2024CMBM,zahedy2019}).  As these studies show, this does not necessarily mean that \ovi\ always traces collisionally ionized gas. \ovi\ can also be present in photoionized gas, but with much lower densities than in the gas probed by the bulk of the cool \hi\ observed in these absorbers (e.g., \citetalias{Savage2014}; \citealt{Muzahid2015,Werk2016,Rosenwasser2018,haislmaier2021}). This means the \ovi\ must be produced in a different gas phase than the lower ions. The \ovi\ velocity profiles in some instances also show a lack of kinematic correspondence with the {\hi}; indicating that the {\ovi}-bearing gas resides in a separate phase (e.g., \citealt{Lehner2014}; \citetalias{Savage2014}; \citealt{Werk2016}). 

Owing to its high frequency of occurrence and high ionization state, \ovi\ is therefore an important and complementary diagnostic of very low density cool photoionized gas and warm-hot, collisionally ionized gas \citepalias[e.g.,][]{Savage2014}. Strong \ovi\ absorbers with large column densities (\coldenovi$\, \ga 14.5$) and velocity widths  ($\Delta v($\ovi$)\ga 100$\,\kms) have been directly identified in some instances with large-scale outflows from starburst galaxies \citep[e.g.,][]{Grimes2009,Tripp2011,Muzahid2015,Rosenwasser2018}. The CGM$^2$ survey (a galaxy-centric survey at $0.1\la z\la 0.6$ of 52 galaxies with stellar masses of $\sim 3 \times 10^{10}$\,M$_\sun$) recently shows that \ovi\ absorbers with \coldenovi$\,\ga 14.3$ are nearly always found around star-forming galaxies (covering fraction $\ga 90\%$), but rarely around passive galaxies (covering fraction $\la 20\%$) with similar stellar or halo masses (\citealt{Tchernyshyov2023}; and see also \citealt{Tumlinson2011}, \citealt{Tchernyshyov2022}; hereafter \citetalias{Tchernyshyov2022}), implying a CGM transformation when galaxies quench and a relationship between the star formation level in galaxy and the amount of \ovi\ in its CGM. As for the low metallicity pLLSs and LLSs, weaker \ovi\ absorbers have a wider range of origins, including the IGM, extended CGM or intragroup gas, or CGM of dwarf galaxies (e.g., \citealt{Stocke2014,Stocke2017}; \citetalias{Savage2014,Tchernyshyov2022}; \citealt{Tchernyshyov2023}).  

The investigation of the relationship between the metallicity of low ionization CGM gas with the properties of the \ovi\ gas can lead to new and additional insights on the origins of the \ovi\ absorbers and multiple gas phases of these absorbers. At $z<1$, an earlier study of \citet{Fox2013} using a sample 23 {\hi}-selected pLLSs shows a high-frequency rate of detection of \ovi\ ($\sim 70\%$) and strongly hints that the strength of \ovi\ absorption is correlated with the metallicity of the photoionized gas; with metal-poor pLLSs showing an absence of strong \ovi\ with \coldenovi$\, \ga 14.5$. At $z>2$, the Keck Observatory Database of Ionized Absorption toward Quasars (KODIAQ; \citealt{Lehner2014}) survey led to a similar conclusion where the strongest \ovi\ absorbers associated with  {\colden}\,$\ga 17$ absorbers were found in the highest metallicity systems \citep{Lehner2014,Lehner2017}.\footnote{Owing to the cosmic evolution, low-metallicity gas is defined as $\xh \la -2.4$ and $\la -1.4$ at  $2.2 \la z\la 3.5$ and $z\la 1$, respectively (\citealt{Lehner2022}; \citetalias{Lehner2019}). Below these thresholds, it is very rare to find DLAs, but about 50\% of the SLFSs, pLLSs, and LLSs are found at all redshifts surveyed so far.} These two surveys have a relatively small sample size, with only about 5--10 metal-poor absorbers. Neither explores the \ovi\ associated with lower column density SLFSs. 

Here we revisit and expand the \citet{Fox2013} study with a much larger sample of absorbers drawn from the COS CGM Compendium survey \citepalias[CCC,][]{Lehner2018, Wotta2019, Lehner2019} that includes both pLLSs and SLFSs, spanning the redshift range of $0.14 \lesssim z \lesssim 0.73$. We also aim to understand the origins of the \ovi\ properties in this \hi-selected sample by comparing with several recent galaxy-centric and blind \ovi\ surveys that were not available at the time of the \citeauthor{Fox2013} survey \citepalias[e.g.,][]{Savage2014, Tchernyshyov2022}. The CCC is a high-resolution survey of {\hi}-selected absorbers at $z<1$ that has focused so far on the metallicity of cool photoionized gas. The \ovi-bearing gas in these absorbers was not detailed beyond basic measurements, except noting that the {\ovi} could not be reproduced by the same photoionization models that produce the lower ions in the pLLSs and LLSs (see \citealt{Lehner2013}). The \hi-selection of the CCC survey eliminates any metallicity bias, allowing one to detect both high- and low-metallicity gas. It is also blind in terms of the presence of the \ovi\ absorption that might be associated (or not) with the \hi\ and lower ion absorption. This experiment with its selection criteria can be easily replicated in cosmological simulations for a comparison between theory and observations (e.g., \citetalias{Lehner2019}, \citealt{Lehner2022}). This \hi-selection approach leads to a widely different distribution of the \hi\ column densities (and as we will show of the \ovi\ column density distribution as well) compared to blind \ovi\ surveys where the absorbers have typically \hi\ column densities of {\colden}\,$\la 15$ \citep[e.g.,][]{Danforth2005,Tripp2008,Thom2008b,Thom2008a,Danforth2016}. 

Our paper is structured as follows. In Section~\ref{s-data}, we describe the sample. In Section~\ref{s-measurements}, we present our empirical approach to investigating the relationship between \hi\ and \ovi. Our main results are presented in Section~\ref{s-results} and discussed in Section~\ref{s-discussion}. In Section~\ref{s-summary}, we summarize our main findings. 

\begin{deluxetable*}{lccccc}
\tabcolsep=5pt
\tablecolumns{4}
\tablewidth{0pc}
\tablecaption{Number of spectra with {\ovi} coverage\label{ovi-sample}}
\tabletypesize{\normalem}
\tablehead{\colhead{Sample} & \colhead{SLFSs} & \colhead{pLLSs} & \colhead{Total} & $\langle z \rangle$ & [X/H]$^a$}
\startdata
Entire     & 133 & 37 & 170 & 0.44 & $-1.22_{-0.11}^{+0.23}$\\
Robust: {\sn} $\geq$ 8    & 100  &  26 & 126 & 0.41 & $-1.22_{-0.11}^{+0.23}$\\
HM: {\sn} $\geq$ 8 \& {\xh} $> -1.4$   & 64 & 18 & 82 & 0.40 & $-0.76_{-0.14}^{+0.15}$\\
LM: {\sn} $\geq$ 8 \& {\xh} $\leq -1.4$  & 36 & 8 & 44 & 0.42 & $-2.01_{-0.25}^{+0.05}$\\
\enddata
\tablecomments{$a$: median metallicity and upper/lower bounds of the 95\% confidence interval of the sample, which is determined using the KM estimator, accounting for the upper limits.}
\end{deluxetable*}

We adopt the notations \mean, \median, \std, \sdmu, and \sdmedian, to denote the mean, median, intrinsic scatter of the distribution, and measurement uncertainty on the mean and median, respectively. Throughout this work, we consider a $p$-value $\leq$ 0.05 (corresponding to a significance level of $\approx2\sigma$) as the threshold for rejecting the null hypothesis and claiming statistical significance. In this study, we employed survival analysis techniques to handle the presence of upper limits in our data. Survival analysis methods independently do not account for measurement errors. To address this limitation, we implemented a Monte Carlo sampling approach in conjunction with survival analysis~\citep[e.g.,][]{Shy2022}. We generated 1000 replicated datasets, where each measurement in the dataset was randomly drawn from an underlying distribution modeled as a Fechner distribution~\citep{wallis2014} or a Gaussian distribution, depending on the nature of uncertainties. The Fechner distribution is chosen to treat the asymmetric uncertainty associated with the measurements appropriately. We utilized the \textsc{fanplot} package~\citep{fanplot} in R to facilitate the modeling of the asymmetric uncertainty. We then performed the relevant statistical test (e.g., Kaplan-Meier estimator (KM)~\citep{kaplan}, Akritas-Theil-Sen (ATS)~\citep{Akritas95} test, log-rank test~\citep{mantel1966evaluation}, Anderson-Darling (AD) test~\citep{adtest}, Mann-Whitney U (MWU) test~\citep{mann1947}, or Kolmogorov-Smirnov~(KS) test~\citep{Hodges1958}) on each of the 1000 datasets. The resulting distribution of the statistic of interest (e.g., mean or p-value) is summarized using its mean and standard deviation.

\section{Sample}\label{s-data}

\citetalias{Lehner2018} assembled a sample of 222 {\hi}-selected absorbers with $15.1 \la$\,{\colden}\,$\la 19$ at $z\la1$. The {\hst}/COS G130M and/or G160M spectral data were retrieved from the HST Spectroscopic Legacy Archive (HSLA, \citealt{Peeples2017}) available at the Barbara A. Mikulski Archive for Space Telescopes (MAST). These data were uniformly analyzed to derive the column densities of {\hi} and metal ions and atoms. As pointed out in \citetalias{Lehner2018}, an additional 39 absorbers are available from earlier surveys \citep{Lehner2013,Wotta2016}. These additional absorbers were observed with {\hst}/STIS E140M, {\it FUSE}, and {\hst}/COS G140L; their column densities were not reassessed in \citetalias{Lehner2018}. 

We adopt the column densities for {\hi}, {\cii}, and {\ciii} from \citetalias{Lehner2018} and the metallicities of the cool photoionized gas probed by the SLFSs and pLLSs from  \citetalias{Lehner2019}. As explained in \citetalias{Wotta2019} and \citetalias{Lehner2019}, the photoionization modeling of these absorbers was motivated by the empirical findings of \citetalias{Lehner2018} that showed that the properties of the low (e.g., \cii, \mgii) and intermediate (e.g., \oii, \ciii) ions and \hi\ in the SLFSs and pLLSs are characteristics of the gas being photoionized (and even in some cases \oiv)\footnote{\citetalias{Lehner2019}, as part of their analysis of the SLFSs, revisited the inclusion of \oiv\ in the single-phase photoionization models (which was not included for the pLLSs in \citetalias{Wotta2019}) and found that 12/17 SLFSs and 3/8 pLLSs could be modeled with the inclusion of \oiv. The other absorbers must probe multiphase gas. As demonstrated in \citetalias{Lehner2019} (see their Fig.~2) the metallicities derived with or without \oiv\ in the photoionization models do not drastically change (on average, a 0.06 dex difference in the mean metallicity of the sample with and without \oiv). In this work, we adopt the metallicities from  \citetalias{Lehner2019} since the same methodology was applied for all the SLFSs and pLLSs.}. The metallicity determination was therefore done with a single-phase photonization model using an MCMC technique that employs Bayesian statistics (see \citealt{Cooper2015,Crighton2015, Fumagalli2016,Wotta2019,Lehner2019}). The result is a posterior probability density function (PDF) for the metallicity of each absorber. This method also formally treats the lower and upper limits.

The signal-to-noise levels of the COS spectra are not high enough to pursue an unbiased survey of the metallicity at $z\la 1$ in the more diffuse gas probed by weak \hi\ absorbers with {\colden}\,$\ga 15$, and this sets the lower limit on {\colden} for CCC and therefore for our sample.  Two systems, the $z =$ 0.3905 towards J020930.74-043826.2 and $z = 0.47404$ towards J154553.63+093620.5, have {\colden} $= 19$ and 18.2, respectively. Since there are only two LLSs with \hi\ column with \ovi\ information, we set the upper limit of {\colden}\,$=17.2$ for our sample. Hence we consider only SLFSs and pLLSs in our analysis, totaling 171 absorbers with \ovi\ coverage of at least one of the doublet transitions. \footnote{For two absorbers---the $z=0.1385$ system toward PG1116$+$215 and the $z=0.1671$ absorber toward PKS0405$-$123---the {\ovi} information was not available in \citetalias{Lehner2019}, as at these redshifts STIS and $FUSE$ are required to determine $N($\hi$)$. We maximize the number of pLLSs in our sample by adopting the \hi\ column densities for these absorbers from \citet{Lehner2013} and use FUSE data for the estimation of the {\ovi} properties.} To minimize the influence of non-informative upper limits, we created a robust sample adopting signal-to-noise $\sn \geq 8$ per resolution element near the {\ovi} doublet, reducing the sample to 127 absorbers with \ovi\ coverage and a sensitivity level of {\coldenovi}$\,\geq$ 13.6 at the $2\sigma$ level.

We do not a priori exclude proximate absorbers, i.e., absorbers with small velocity separation from the background QSO ($\Delta\,v <3000$\,\kms), in our analysis, except for one case. They constitute a small percentage of our sample, with $\approx$ 8\% being proximate. \citetalias{Lehner2018} showed that for all but one, their properties are quite similar to intervening absorbers. The exception is the absorber at $z = 0.470800$ toward the quasar J161916.54+334238.4, where there is extremely strong absorption of intermediate to high ions, and the \ovi\ might even be saturated. This absorber also has strong absorption features of \svi, \sv, \siv, not typically observed except in the proximity of a very hard ionizing source, implying that this particular absorber is not only a proximate but also an associated system. We, therefore, exclude this absorber, reducing our total sample to 126 absorbers; this is the primary sample for this paper (see Section~\ref{sec:detectionrate}).  In Fig.~\ref{fig:example}, as an example, we present the apparent optical column density profiles for {\ovi}, {\cii}, {\ciii}, and {\hi} for two closely redshift-separated absorbers from our sample. The complete set of {\nav} and normalized flux profiles for the entire sample of {\ovi} absorbers is provided as online supplementary material. Each plot features a robustness flag (RF), where RF $=1$ indicates the absorption system is part of the primary (robust) sample, and RF $=0$ indicates it is not.

In Table~\ref{ovi-sample}, we summarize the numbers of \hi-selected absorbers grouped in \nhi, metallicity, S/N groupings. The mean redshift and median metallicity, $\xh$, of these samples is also noted. For the remainder of the paper, except otherwise noted, we only consider the robust sample for our analysis and interpretation of the \ovi\ associated with the {\hi}-selected absorbers. We select a threshold of the metallicity at $\xh=-1.4$ following \citetalias{Wotta2019}, which corresponds to the $2\sigma$ lower bound of the DLA metallicities at $z\la 1$. The absorbers below this metallicity were defined in \citetalias{Wotta2019} and \citetalias{Lehner2019}  as ``very metal-poor'' absorbers; here we simply refer to them as low-metal (LM) absorbers, and absorbers with $\xh>-1.4$ as high-metal (HM) absorbers. 

LM absorbers are very rare for DLAs at $z\la1$, but are very common for lower {\hi} column density absorbers with {\colden}\,$<19$ (see \citetalias{Lehner2019}). Low redshift DLAs typically trace the interstellar medium or extended gaseous disks of galaxies within approximately 20 kpc, as evidenced by the correlation between \hi\ column density and impact parameter \citep[e.g.,][]{Lehner2013,Berg2023,weng2023}. In contrast, lower {\hi} column density absorbers (pLLSs and LLSs) are found over a wider range of impact parameters, typically beyond 20\,kpc from galaxies.

\section{{Kinematics and Column density Measurements}}\label{s-measurements}

\begin{figure*}[tbp]
\centering
\includegraphics[scale=0.5]{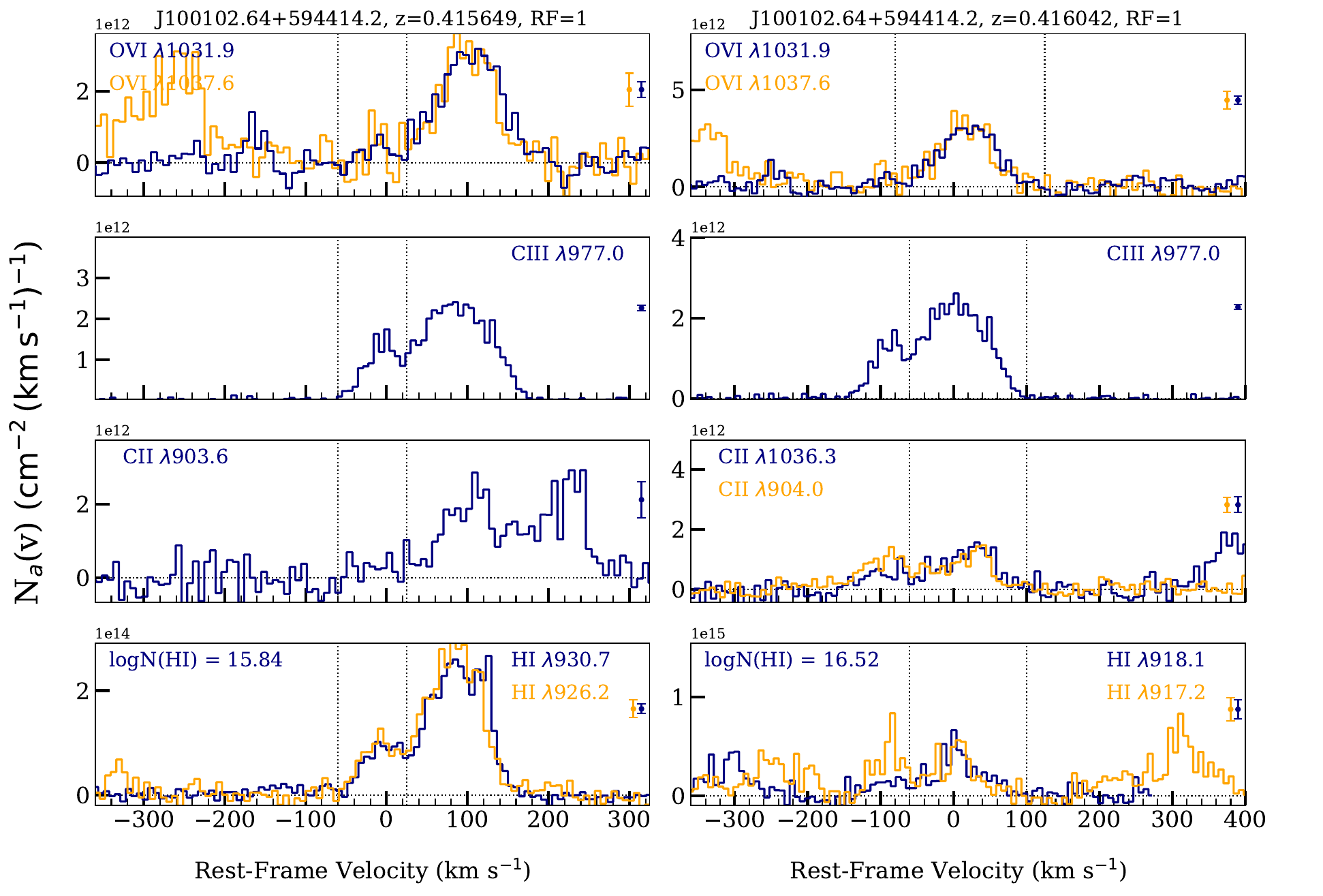}
\caption{Apparent optical column density profiles for {\ovi}, {\cii}, {\ciii}, and {\hi}, for two closely redshift-separated absorbers at  $z= 0.415649$ ({\it left}) and  0.416042 ({\it right}) toward the quasar J100102.64$+$594414.2. These two absorbers are separated by 100 {\kms}. The integrated velocity ranges are indicated by two vertical lines in each panel. {\ovi} is detected only at $z=0.416042$. In this case, these are HM absorbers with a similar metallicity within $1\sigma$. The average uncertainty on the data is indicated to the right in each of the panels. Each plot includes a robustness flag (RF) indicating whether the absorption system is part of the robust sample (RF $=1$) or not (RF $=0$).  The profiles are binned by 3 pixels for display purposes. \label{fig:example}}
\end{figure*}

\citetalias{Lehner2018} adopted the apparent optical depth (AOD) method \citep{Savage1991} for estimating the column densities of the metal lines associated with the \hi. We adopt the same method here to derive the properties of the \ovi\ absorbers. In short, with this method, the absorption profiles are converted into apparent optical depth per unit velocity, $\tau_a(v) = \ln[F_{\rm c}(v)/F_{\rm obs}(v)]$,  where $F_c(v)$ and $F_{\rm obs}(v)$ are the modeled continuum and observed fluxes as a function of velocity.  The AOD, $\tau_a(v)$, is related to the apparent column density per unit velocity, $N_a(v)$, through the relation $N_a(v)=3.768 \times 10^{14} \tau_a(v)/(f \lambda)$)  ${\rm cm}^{-2}$ $({\rm km\,s^{-1}})^{-1}$, where $f$ is the oscillator strength of the transition and $\lambda$ is the wavelength in \AA. The total column density is obtained by integrating the profile over a pre-defined velocity interval, $N = \int_{v_{\rm min}}^{v_{\rm max}} N_a(v) dv $, where  $[v_{\rm min},v_{\rm max}]$ are the boundaries of the absorption. We computed the \nav-weigthed average line centroids through the first moment of the AOD, $v_a = \int v N_a(v) dv/\int N_a(v)dv$ \kms. With the AOD method, both contamination and saturation can be determined using the prescriptions described in \citetalias{Lehner2018} (see their Section~4.2).

Besides the velocity centroid and column density, we characterize the velocity width of absorption using the velocity interval that encloses the central 90\% of the integrated AOD in the absorption line, $\Delta v_{90}$ \citep{Prochaska1997}. It is measured by integrating the absorption line between {\vmin} and {\vmax}, and finding the velocity difference between pixels at which 5\% and 95\% of the total apparent optical depth is attained. For a saturated line, we derive an upper limit to $\Delta v_{90}$ for the true column density because saturation in the absorption line can lead to an underestimation of the true optical depth. We thus treat the measured  $\Delta v_{90}$ as an upper limit for saturated absorption.

\addtolength{\tabcolsep}{-2pt}
\startlongtable
\begin{longrotatetable}
\begin{deluxetable*}{lccccccccccccc}
\footnotesize
\tablecaption{Properties of the {\ovi} absorbers in the CCC survey\label{tab:raw-ovi-properties}}
\tablehead{
\colhead{Target} & \colhead{$z_{\rm abs}$} & \colhead{$v_{\rm min}$} & \colhead{$v_{\rm max}$} & \colhead{$\langle v \rangle$} & \colhead{$\langle v \rangle$} & \colhead{$\Delta v_{90}$}  & \colhead{$\Delta v_{90}$} & \colhead{$\log N$} & \colhead{Detection} & \colhead{$\log N$} & \colhead{Detection} & \colhead{$S/N$} & \colhead{$S/N$}\\
\colhead{} & \colhead{} & \colhead{({\kms})} & \colhead{({\kms})} & \colhead{({\kms})} & \colhead{({\kms})} & \colhead{({\kms})}  & \colhead{({\kms})} & \colhead{[\cmsq]} & \colhead{Flag} & \colhead{[\cmsq]} & \colhead{Flag} & \colhead{} & \colhead{}\\
\colhead{} & \colhead{} & \colhead{} & \colhead{} & \colhead{$\lambda$1031} & \colhead{$\lambda$1037} & \colhead{$\lambda$1031}  & \colhead{$\lambda$1037} & \colhead{$\lambda$1031} & \colhead{} & \colhead{$\lambda$1037} & \colhead{} & \colhead{$\lambda$1031} & \colhead{$\lambda$1037}\\
\colhead{(1)} & \colhead{(2)} & \colhead{(3)} & \colhead{(4)} & \colhead{(5)} & \colhead{(6)} & \colhead{(7)}  & \colhead{(8)}  & \colhead{(9)} & \colhead{(10)} & \colhead{(11)} & \colhead{(12)} & \colhead{(13)} & \colhead{(14)}
}
\startdata
\hline
\midrule
J000559.23+160948.9 & $0.3058$ & $-100.0$ & $40.0$ & $-52.0 \pm 8.8$ & {\nodata} & $46.9 \pm 12.5$ & {\nodata} & $13.38_{-0.10}^{+0.08}$ & $0$ & $<13.26$ & $-1$ & $21.6$ & $23.3$\\[2pt]
J000559.23+160948.9 & $0.3479$ & $-50.0$ & $20.0$ & $-18.5 \pm 1.0$ & $-22.0 \pm 1.9$ & $58.1 \pm 1.4$ & $51.4 \pm 2.6$ & $13.99_{-0.02}^{+0.02}$ & $0$ & $13.97_{-0.04}^{+0.04}$ & $0$ & $21.0$ & $17.4$\\[2pt]
J000559.23+160948.9 & $0.3662$ & $-50.0$ & $50.0$ & $-3.3 \pm 1.1$ & $-0.3 \pm 1.9$ & $74.2 \pm 1.6$ & $82.5 \pm 2.6$ & $14.07_{-0.02}^{+0.02}$ & $-3$ & $14.02_{-0.03}^{+0.03}$ & $0$ & $21.5$ & $27.8$\\[2pt]
J004222.29-103743.8 & $0.3161$ & $-60.0$ & $70.0$ & $-4.4 \pm 7.8$ & {\nodata} & $72.7 \pm 11.1$ & {\nodata} & $13.97_{-0.12}^{+0.09}$ & $0$ & $<13.84$ & $-1$ & $8.5$ & $8.9$\\[2pt]
J004705.89+031954.9 & $0.3139$ & $-50.0$ & $50.0$ & {\nodata} & {\nodata} & {\nodata} & {\nodata} & $<13.39$ & $-1$ & $<13.69$ & $-1$ & $10.0$ & $9.6$\\[2pt]
J004705.89+031954.9 & $0.3143$ & $-50.0$ & $100.0$ & {\nodata} & {\nodata} & {\nodata} & {\nodata} & $<13.48$ & $-1$ & $<13.77$ & $-1$ & $9.9$ & $9.8$\\[2pt]
J011013.14-021952.8 & $0.2272$ & $-110.0$ & $100.0$ & $-24.6 \pm 2.5$ & $-13.6 \pm 5.2$ & $120.4 \pm 3.5$ & $134.1 \pm 7.4$ & $14.48_{-0.02}^{+0.02}$ & $0$ & $14.48_{-0.05}^{+0.04}$ & $0$ & $12.6$ & $12.9$\\[2pt]
J011013.14-021952.8 & $0.5365$ & $-20.0$ & $40.0$ & $4.1 \pm 4.2$ & {\nodata} & $34.8 \pm 6.0$ & {\nodata} & $13.59_{-0.15}^{+0.11}$ & $0$ & $<13.39$ & $-1$ & $9.9$ & $11.9$\\[2pt]
J011013.14-021952.8 & $0.7190$ & $-50.0$ & $70.0$ & {\nodata} & {\nodata} & {\nodata} & {\nodata} & $<13.53$ & $-1$ & $<13.87$ & $-1$ & $7.1$ & $6.6$\\[2pt]
J011016.25-021851.0 & $0.3991$ & $-100.0$ & $80.0$ & $-8.6 \pm 4.8$ & $-4.6 \pm 4.5$ & $105.7 \pm 6.7$ & $125.3 \pm 6.3$ & $14.34_{-0.05}^{+0.05}$ & $0$ & $14.37_{-0.05}^{+0.04}$ & $0$ & $8.9$ & $14.7$\\[2pt]
J011016.25-021851.0 & $0.5354$ & $-50.0$ & $40.0$ & {\nodata} & {\nodata} & {\nodata} & {\nodata} & $<13.44$ & $-1$ & $<13.60$ & $-1$ & $8.8$ & $9.8$\\[2pt]
J011016.25-021851.0 & $0.7180$ & $-70.0$ & $70.0$ & $-2.2 \pm 6.5$ & $-14.2 \pm 9.8$ & $80.9 \pm 9.1$ & $92.6 \pm 13.9$ & $14.31_{-0.11}^{+0.08}$ & $0$ & $14.54_{-0.21}^{+0.14}$ & $-3$ & $5.7$ & $5.4$\\[2pt]
J011623.04+142940.5 & $0.3338$ & $-100.0$ & $100.0$ & $-9.3 \pm 6.2$ & $16.2 \pm 12.5$ & $137.2 \pm 8.8$ & $129.8 \pm 17.7$ & $14.33_{-0.06}^{+0.05}$ & $0$ & $14.36_{-0.11}^{+0.09}$ & $0$ & $7.8$ & $7.6$\\[2pt]
J011935.69-282131.4 & $0.3483$ & $-60.0$ & $10.0$ & {\nodata} & {\nodata} & {\nodata} & {\nodata} & {\nodata} & $-4$ & $<13.09$ & $-1$ & {\nodata} & $24.2$\\[2pt]
J011935.69-282131.4 & $0.3487$ & $-100.0$ & $40.0$ & {\nodata} & $-34.4 \pm 3.3$ & {\nodata} & $89.8 \pm 4.7$ & {\nodata} & $-4$ & $14.07_{-0.04}^{+0.04}$ & $0$ & {\nodata} & $23.7$\\[2pt]
J012236.76-284321.3 & $0.3650$ & $-45.0$ & $80.0$ & $18.2 \pm 2.4$ & $19.7 \pm 5.3$ & $76.5 \pm 3.4$ & $76.1 \pm 7.5$ & $13.98_{-0.04}^{+0.03}$ & $0$ & $13.93_{-0.07}^{+0.06}$ & $0$ & $15.9$ & $15.0$\\[2pt]
J015513.20-450611.9 & $0.2260$ & $-100.0$ & $120.0$ & $7.3 \pm 2.1$ & $10.2 \pm 4.6$ & $107.0 \pm 3.0$ & $113.5 \pm 6.6$ & $14.18_{-0.02}^{+0.02}$ & $0$ & $14.16_{-0.04}^{+0.03}$ & $0$ & $25.9$ & $24.5$\\[2pt]
J020157.16-113233.1 & $0.3226$ & $-50.0$ & $50.0$ & {\nodata} & {\nodata} & {\nodata} & {\nodata} & $<12.89$ & $-1$ & $<13.19$ & $-1$ & $19.4$ & $20.1$\\[2pt]
J020157.16-113233.1 & $0.3231$ & $-40.0$ & $25.0$ & $-3.4 \pm 3.1$ & $-1.1 \pm 2.4$ & $52.6 \pm 4.4$ & $43.5 \pm 3.4$ & $13.27_{-0.09}^{+0.07}$ & $0$ & $13.70_{-0.07}^{+0.06}$ & $-3$ & $24.7$ & $23.1$\\[2pt]
J020157.16-113233.1 & $0.3234$ & $-55.0$ & $70.0$ & $22.1 \pm 1.6$ & {\nodata} & $87.5 \pm 2.2$ & {\nodata} & $14.13_{-0.02}^{+0.02}$ & $-3$ & $<13.29$ & $-1$ & $19.2$ & $20.3$\\[2pt]
J020157.16-113233.1 & $0.3245$ & $-45.0$ & $85.0$ & $16.2 \pm 1.4$ & $16.5 \pm 2.8$ & $98.6 \pm 2.0$ & $104.6 \pm 3.9$ & $14.11_{-0.02}^{+0.02}$ & $0$ & $14.12_{-0.04}^{+0.03}$ & $0$ & $25.1$ & $21.3$\\[2pt]
J023507.38-040205.6 & $0.3225$ & $-40.0$ & $40.0$ & $-1.6 \pm 2.9$ & {\nodata} & $66.1 \pm 4.1$ & {\nodata} & $13.58_{-0.06}^{+0.05}$ & $0$ & $<13.28$ & $-1$ & $18.7$ & $17.2$\\[2pt]
J023507.38-040205.6 & $0.7389$ & $-80.0$ & $100.0$ & $12.4 \pm 13.2$ & {\nodata} & $135.2 \pm 18.6$ & {\nodata} & $14.14_{-0.14}^{+0.11}$ & $0$ & {\nodata} & $-4$ & $5.1$ & {\nodata}\\[2pt]
J024337.66-303048.0 & $0.3037$ & $-70.0$ & $70.0$ & {\nodata} & {\nodata} & {\nodata} & {\nodata} & $<13.24$ & $-1$ & $<13.58$ & $-1$ & $13.4$ & $13.6$\\[2pt]
J024649.86-300741.3 & $0.3123$ & $-60.0$ & $100.0$ & $25.6 \pm 6.1$ & $-5.2 \pm 5.2$ & $106.2 \pm 8.6$ & $149.2 \pm 7.3$ & $14.13_{-0.07}^{+0.06}$ & $0$ & $14.68_{-0.06}^{+0.05}$ & $-3$ & $9.1$ & $9.1$\\[2pt]
J024649.86-300741.3 & $0.3359$ & $-50.0$ & $70.0$ & {\nodata} & {\nodata} & {\nodata} & {\nodata} & $<13.44$ & $-1$ & $<13.74$ & $-1$ & $9.4$ & $8.7$\\[2pt]
J035128.56-142908.0 & $0.3285$ & $-70.0$ & $100.0$ & $-0.9 \pm 11.6$ & {\nodata} & $92.3 \pm 16.3$ & {\nodata} & $13.57_{-0.12}^{+0.10}$ & $0$ & $<13.54$ & $-1$ & $14.8$ & $13.9$\\[2pt]
J035128.56-142908.0 & $0.3569$ & $-160.0$ & $120.0$ & $-13.2 \pm 2.0$ & $-23.2 \pm 4.0$ & $83.2 \pm 2.8$ & $159.5 \pm 5.7$ & $14.91_{-0.04}^{+0.04}$ & $-3$ & $14.69_{-0.03}^{+0.03}$ & $0$ & $13.5$ & $13.7$\\[2pt]
J035128.56-142908.0 & $0.3572$ & $0.0$ & $70.0$ & {\nodata} & {\nodata} & {\nodata} & {\nodata} & {\nodata} & $-4$ & $<13.34$ & $-1$ & {\nodata} & $13.7$\\[2pt]
J035128.56-142908.0 & $0.4398$ & $-100.0$ & $100.0$ & $-13.6 \pm 7.1$ & $9.6 \pm 11.4$ & $81.6 \pm 10.0$ & $96.0 \pm 16.1$ & $13.94_{-0.06}^{+0.06}$ & $0$ & $14.00_{-0.10}^{+0.08}$ & $0$ & $15.2$ & $14.8$\\[2pt]
J040148.98-054056.5 & $0.2195$ & $-50.0$ & $140.0$ & $42.3 \pm 3.2$ & $35.9 \pm 5.7$ & $136.0 \pm 4.6$ & $127.9 \pm 8.1$ & $14.54_{-0.03}^{+0.03}$ & $0$ & $14.48_{-0.05}^{+0.05}$ & $0$ & $10.4$ & $10.5$\\[2pt]
J040148.98-054056.5 & $0.3238$ & $-70.0$ & $70.0$ & {\nodata} & {\nodata} & {\nodata} & {\nodata} & $<13.45$ & $-1$ & $<13.75$ & $-1$ & $9.2$ & $9.4$\\[2pt]
J040748.42-121136.3 & $0.1672$ & $-125.0$ & $95.0$ & $-34.1 \pm 0.4$ & $-31.2 \pm 1.1$ & $149.2 \pm 0.5$ & $149.1 \pm 1.6$ & $14.63_{-0.00}^{+0.00}$ & $0$ & $14.63_{-0.01}^{+0.01}$ & $0$ & $72.2$ & $59.5$\\[2pt]
J040748.42-121136.3 & $0.3607$ & $-30.0$ & $30.0$ & $0.3 \pm 2.5$ & {\nodata} & $45.1 \pm 3.5$ & {\nodata} & $12.81_{-0.08}^{+0.07}$ & $0$ & {\nodata} & $-4$ & $56.1$ & {\nodata}\\[2pt]
J044011.90-524818.0 & $0.3279$ & $-80.0$ & $80.0$ & {\nodata} & {\nodata} & {\nodata} & {\nodata} & $<13.18$ & $-1$ & $<13.49$ & $-1$ & $16.3$ & $16.0$\\[2pt]
J044011.90-524818.0 & $0.6150$ & $-120.0$ & $135.0$ & $8.9 \pm 2.7$ & $1.7 \pm 4.3$ & $178.7 \pm 3.8$ & $171.5 \pm 6.1$ & $14.84_{-0.02}^{+0.02}$ & $0$ & $14.89_{-0.03}^{+0.03}$ & $0$ & $10.8$ & $10.8$\\[2pt]
J044011.90-524818.0 & $0.6157$ & $-20.0$ & $80.0$ & $16.0 \pm 5.3$ & {\nodata} & $72.8 \pm 7.5$ & {\nodata} & $13.85_{-0.09}^{+0.07}$ & $0$ & $<13.68$ & $-1$ & $9.3$ & $10.0$\\[2pt]
J055224.49-640210.7 & $0.3451$ & $-25.0$ & $25.0$ & {\nodata} & $-2.1 \pm 3.0$ & {\nodata} & $42.8 \pm 4.3$ & $<12.67$ & $-1$ & $13.40_{-0.11}^{+0.09}$ & $-3$ & $26.4$ & $23.4$\\[2pt]
J055224.49-640210.7 & $0.4461$ & $-65.0$ & $65.0$ & $-0.6 \pm 3.5$ & $-7.6 \pm 7.5$ & $81.9 \pm 5.0$ & $81.1 \pm 10.6$ & $13.66_{-0.05}^{+0.04}$ & $0$ & $13.64_{-0.10}^{+0.08}$ & $0$ & $24.6$ & $25.9$\\[2pt]
J063546.49-751616.8 & $0.4175$ & $-50.0$ & $95.0$ & $16.9 \pm 1.5$ & $24.1 \pm 3.1$ & $98.1 \pm 2.1$ & $112.7 \pm 4.3$ & $14.17_{-0.02}^{+0.02}$ & $0$ & $14.14_{-0.03}^{+0.03}$ & $0$ & $23.4$ & $22.7$\\[2pt]
J063546.49-751616.8 & $0.4528$ & $-20.0$ & $100.0$ & $41.0 \pm 4.1$ & {\nodata} & $80.9 \pm 5.8$ & {\nodata} & $13.63_{-0.06}^{+0.05}$ & $0$ & $<13.30$ & $-1$ & $20.1$ & $18.9$\\[2pt]
J063546.49-751616.8 & $0.4685$ & $-40.0$ & $80.0$ & $8.6 \pm 3.7$ & $15.5 \pm 7.1$ & $80.0 \pm 5.2$ & $86.9 \pm 10.0$ & $13.69_{-0.05}^{+0.05}$ & $0$ & $13.66_{-0.10}^{+0.08}$ & $0$ & $19.5$ & $19.0$\\[2pt]
J071950.89+742757.0 & $0.3777$ & $-40.0$ & $70.0$ & $9.1 \pm 3.4$ & {\nodata} & $69.4 \pm 4.8$ & {\nodata} & $13.77_{-0.05}^{+0.05}$ & $0$ & $<13.61$ & $-1$ & $15.7$ & $12.2$\\[2pt]
J075112.30+291938.3 & $0.4318$ & $-60.0$ & $80.0$ & $13.2 \pm 3.8$ & $6.0 \pm 5.3$ & $89.7 \pm 5.4$ & $88.9 \pm 7.4$ & $13.52_{-0.05}^{+0.04}$ & $0$ & $13.76_{-0.06}^{+0.05}$ & $-3$ & $26.1$ & $33.9$\\[2pt]
J075112.30+291938.3 & $0.4945$ & $-30.0$ & $50.0$ & $4.4 \pm 6.2$ & {\nodata} & $57.2 \pm 8.7$ & {\nodata} & $13.02_{-0.14}^{+0.10}$ & $0$ & {\nodata} & $-4$ & $29.6$ & {\nodata}\\[2pt]
J080908.13+461925.6 & $0.6191$ & $-90.0$ & $90.0$ & $10.0 \pm 4.5$ & $1.5 \pm 3.2$ & $204.2 \pm 6.4$ & $92.5 \pm 4.5$ & $15.08_{-0.07}^{+0.06}$ & $-3$ & $14.74_{-0.04}^{+0.04}$ & $0$ & $10.5$ & $10.8$\\[2pt]
J084349.47+411741.6 & $0.5335$ & $-40.0$ & $40.0$ & {\nodata} & {\nodata} & {\nodata} & {\nodata} & $<13.63$ & $-1$ & $<13.91$ & $-1$ & $5.6$ & $6.3$\\[2pt]
J084349.47+411741.6 & $0.5411$ & $-40.0$ & $40.0$ & $9.0 \pm 7.4$ & {\nodata} & $76.7 \pm 10.4$ & {\nodata} & $13.85_{-0.18}^{+0.12}$ & $0$ & $<13.85$ & $-1$ & $6.2$ & $5.9$\\[2pt]
J084349.47+411741.6 & $0.5437$ & $-30.0$ & $40.0$ & $2.5 \pm 4.7$ & {\nodata} & $48.5 \pm 6.7$ & {\nodata} & $13.91_{-0.15}^{+0.11}$ & $0$ & $<13.82$ & $-1$ & $6.2$ & $5.9$\\[2pt]
J091440.38+282330.6 & $0.2442$ & $-90.0$ & $70.0$ & $-18.6 \pm 4.8$ & $-17.4 \pm 7.4$ & $105.2 \pm 6.8$ & $111.6 \pm 10.4$ & $14.44_{-0.06}^{+0.05}$ & $0$ & $14.37_{-0.09}^{+0.07}$ & $0$ & $7.5$ & $8.5$\\[2pt]
J091440.38+282330.6 & $0.5995$ & $-70.0$ & $70.0$ & {\nodata} & {\nodata} & {\nodata} & {\nodata} & $<13.61$ & $-1$ & $<13.91$ & $-1$ & $7.4$ & $7.0$\\[2pt]
J092554.44+453544.5 & $0.3096$ & $-50.0$ & $50.0$ & {\nodata} & {\nodata} & {\nodata} & {\nodata} & $<13.12$ & $-1$ & $<13.39$ & $-1$ & $14.9$ & $15.6$\\[2pt]
J093518.19+020415.5 & $0.3547$ & $-50.0$ & $50.0$ & {\nodata} & {\nodata} & {\nodata} & {\nodata} & $<13.61$ & $-1$ & $<13.89$ & $-1$ & $6.4$ & $6.7$\\[2pt]
J093603.88+320709.3 & $0.3904$ & $-60.0$ & $60.0$ & {\nodata} & {\nodata} & {\nodata} & {\nodata} & $<13.97$ & $-1$ & {\nodata} & $-4$ & $4.0$ & {\nodata}\\[2pt]
J094331.61+053131.4 & $0.2290$ & $-50.0$ & $50.0$ & {\nodata} & {\nodata} & {\nodata} & {\nodata} & $<13.50$ & $-1$ & $<14.03$ & $-1$ & $8.3$ & $5.6$\\[2pt]
J094331.61+053131.4 & $0.3546$ & $-60.0$ & $40.0$ & {\nodata} & {\nodata} & {\nodata} & {\nodata} & $<13.61$ & $-1$ & $<13.90$ & $-1$ & $6.6$ & $6.8$\\[2pt]
J094331.61+053131.4 & $0.3549$ & $-40.0$ & $40.0$ & {\nodata} & {\nodata} & {\nodata} & {\nodata} & $<13.53$ & $-1$ & $<13.84$ & $-1$ & $6.7$ & $6.8$\\[2pt]
J100102.64+594414.2 & $0.4156$ & $-60.0$ & $25.0$ & {\nodata} & {\nodata} & {\nodata} & {\nodata} & $<13.05$ & $-1$ & $<13.37$ & $-1$ & $15.0$ & $15.5$\\[2pt]
J100102.64+594414.2 & $0.4160$ & $-80.0$ & $125.0$ & $18.8 \pm 2.3$ & $14.7 \pm 4.1$ & $121.1 \pm 3.2$ & $127.7 \pm 5.8$ & $14.46_{-0.02}^{+0.02}$ & $0$ & $14.45_{-0.04}^{+0.03}$ & $0$ & $15.5$ & $16.4$\\[2pt]
J100110.20+291137.5 & $0.5565$ & $-40.0$ & $90.0$ & $25.4 \pm 5.9$ & $-8.1 \pm 7.2$ & $75.6 \pm 8.3$ & $90.9 \pm 10.2$ & $14.37_{-0.14}^{+0.11}$ & $0$ & $14.85_{-0.12}^{+0.09}$ & $-3$ & $5.7$ & $6.0$\\[2pt]
J100535.25+013445.5 & $0.4185$ & $-60.0$ & $60.0$ & $-7.9 \pm 7.7$ & {\nodata} & $90.7 \pm 10.8$ & {\nodata} & $13.39_{-0.11}^{+0.09}$ & $0$ & $<13.38$ & $-1$ & $19.2$ & $18.6$\\[2pt]
J100535.25+013445.5 & $0.4197$ & $-40.0$ & $40.0$ & {\nodata} & {\nodata} & {\nodata} & {\nodata} & $<12.94$ & $-1$ & $<13.28$ & $-1$ & $18.7$ & $18.6$\\[2pt]
J100902.06+071343.8 & $0.3372$ & $-65.0$ & $40.0$ & {\nodata} & {\nodata} & {\nodata} & {\nodata} & $<13.52$ & $-1$ & $<13.84$ & $-1$ & $8.1$ & $7.6$\\[2pt]
J100902.06+071343.8 & $0.3554$ & $-50.0$ & $25.0$ & $-13.2 \pm 4.0$ & $-12.6 \pm 4.6$ & $59.8 \pm 5.6$ & $57.6 \pm 6.6$ & $14.57_{-0.24}^{+0.15}$ & $-3$ & $14.25_{-0.12}^{+0.10}$ & $0$ & $7.3$ & $6.1$\\[2pt]
J101622.60+470643.3 & $0.2638$ & $-60.0$ & $45.0$ & {\nodata} & {\nodata} & {\nodata} & {\nodata} & {\nodata} & $-4$ & $<13.97$ & $-1$ & {\nodata} & $5.4$\\[2pt]
J101622.60+470643.3 & $0.4322$ & $-62.0$ & $52.0$ & {\nodata} & {\nodata} & {\nodata} & {\nodata} & $<13.51$ & $-1$ & $<13.80$ & $-1$ & $8.7$ & $8.0$\\[2pt]
J101622.60+470643.3 & $0.6647$ & $-77.0$ & $77.0$ & {\nodata} & {\nodata} & {\nodata} & {\nodata} & $<13.84$ & $-1$ & $<14.16$ & $-1$ & $5.1$ & $4.9$\\[2pt]
J101622.60+470643.3 & $0.7278$ & $-55.0$ & $55.0$ & {\nodata} & {\nodata} & {\nodata} & {\nodata} & $<14.06$ & $-1$ & $<14.40$ & $-1$ & $3.1$ & $3.0$\\[2pt]
J104117.16+061016.9 & $0.6544$ & $-20.0$ & $40.0$ & {\nodata} & {\nodata} & {\nodata} & {\nodata} & $<13.26$ & $-1$ & $<13.60$ & $-1$ & $8.6$ & $8.3$\\[2pt]
J104117.16+061016.9 & $0.6551$ & $-50.0$ & $30.0$ & {\nodata} & {\nodata} & {\nodata} & {\nodata} & $<13.38$ & $-1$ & $<13.69$ & $-1$ & $8.6$ & $8.3$\\[2pt]
J104117.16+061016.9 & $0.6554$ & $-36.0$ & $50.0$ & {\nodata} & {\nodata} & {\nodata} & {\nodata} & $<13.39$ & $-1$ & $<13.70$ & $-1$ & $8.5$ & $8.4$\\[2pt]
J105945.24+144143.0 & $0.4658$ & $-60.0$ & $50.0$ & {\nodata} & {\nodata} & {\nodata} & {\nodata} & $<13.21$ & $-1$ & $<13.55$ & $-1$ & $13.2$ & $12.9$\\[2pt]
J105945.24+144143.0 & $0.5765$ & $-60.0$ & $80.0$ & $21.6 \pm 7.7$ & {\nodata} & $95.1 \pm 10.9$ & {\nodata} & $13.91_{-0.10}^{+0.08}$ & $0$ & $<13.74$ & $-1$ & $8.5$ & $8.9$\\[2pt]
J105956.14+121151.1 & $0.3324$ & $-60.0$ & $60.0$ & {\nodata} & {\nodata} & {\nodata} & {\nodata} & $<13.43$ & $-1$ & $<13.79$ & $-1$ & $9.1$ & $8.6$\\[2pt]
J110047.85+104613.2 & $0.4151$ & $-65.0$ & $65.0$ & {\nodata} & {\nodata} & {\nodata} & {\nodata} & $<13.91$ & $-1$ & {\nodata} & $-4$ & $4.1$ & {\nodata}\\[2pt]
J110539.79+342534.3 & $0.2388$ & $-60.0$ & $65.0$ & $1.4 \pm 1.6$ & $0.4 \pm 1.7$ & $105.2 \pm 2.3$ & $77.2 \pm 2.4$ & $14.62_{-0.02}^{+0.02}$ & $-3$ & $14.59_{-0.03}^{+0.03}$ & $0$ & $13.9$ & $13.8$\\[2pt]
J111132.20+554725.9 & $0.6175$ & $-40.0$ & $75.0$ & {\nodata} & {\nodata} & {\nodata} & {\nodata} & $<13.24$ & $-1$ & $<13.62$ & $-1$ & $12.0$ & $11.5$\\[2pt]
J111132.20+554725.9 & $0.6827$ & $-60.0$ & $20.0$ & {\nodata} & {\nodata} & {\nodata} & {\nodata} & $<13.22$ & $-1$ & {\nodata} & $-4$ & $10.6$ & {\nodata}\\[2pt]
J111239.11+353928.2 & $0.2465$ & $-50.0$ & $50.0$ & {\nodata} & {\nodata} & {\nodata} & {\nodata} & {\nodata} & $-4$ & $<13.86$ & $-1$ & {\nodata} & $6.3$\\[2pt]
J111507.65+023757.5 & $0.3692$ & $-40.0$ & $30.0$ & $-7.8 \pm 3.9$ & {\nodata} & $57.2 \pm 5.5$ & {\nodata} & $13.78_{-0.11}^{+0.09}$ & $0$ & $<13.61$ & $-1$ & $8.9$ & $8.9$\\[2pt]
J111754.23+263416.6 & $0.3520$ & $-40.0$ & $50.0$ & $6.4 \pm 6.8$ & $-6.3 \pm 3.2$ & $77.7 \pm 9.7$ & $63.9 \pm 4.5$ & $13.68_{-0.14}^{+0.11}$ & $0$ & $14.37_{-0.07}^{+0.06}$ & $-3$ & $9.3$ & $9.6$\\[2pt]
J111754.23+263416.6 & $0.3523$ & $-30.0$ & $60.0$ & {\nodata} & $5.1 \pm 7.8$ & {\nodata} & $90.0 \pm 11.0$ & $<13.37$ & $-1$ & $13.91_{-0.15}^{+0.11}$ & $0$ & $9.3$ & $9.5$\\[2pt]
J111908.59+211918.0 & $0.1385$ & $-95.0$ & $95.0$ & $7.9 \pm 2.9$ & $6.8 \pm 6.8$ & $91.9 \pm 4.1$ & $114.9 \pm 9.6$ & $13.80_{-0.03}^{+0.02}$ & $0$ & $13.82_{-0.06}^{+0.06}$ & $0$ & $38.6$ & $36.7$\\[2pt]
J112553.78+591021.6 & $0.5575$ & $-60.0$ & $60.0$ & {\nodata} & {\nodata} & {\nodata} & {\nodata} & $<13.54$ & $-1$ & $<13.76$ & $-1$ & $7.3$ & $9.1$\\[2pt]
J112553.78+591021.6 & $0.5582$ & $-60.0$ & $20.0$ & $-19.3 \pm 6.4$ & {\nodata} & $75.8 \pm 9.1$ & {\nodata} & $13.76_{-0.18}^{+0.13}$ & $0$ & $<13.67$ & $-1$ & $6.9$ & $9.3$\\[2pt]
J112553.78+591021.6 & $0.5585$ & $-20.0$ & $60.0$ & {\nodata} & {\nodata} & {\nodata} & {\nodata} & {\nodata} & $-4$ & $<13.72$ & $-1$ & {\nodata} & $8.4$\\[2pt]
J112553.78+591021.6 & $0.6784$ & $-65.0$ & $65.0$ & {\nodata} & {\nodata} & {\nodata} & {\nodata} & $<13.40$ & $-1$ & $<13.72$ & $-1$ & $9.6$ & $9.2$\\[2pt]
J113457.71+255527.8 & $0.4320$ & $-60.0$ & $30.0$ & $-17.0 \pm 4.9$ & {\nodata} & $67.5 \pm 6.9$ & {\nodata} & $13.77_{-0.10}^{+0.08}$ & $0$ & $<13.80$ & $-1$ & $10.6$ & $8.3$\\[2pt]
J113457.71+255527.8 & $0.4323$ & $-40.0$ & $70.0$ & {\nodata} & {\nodata} & {\nodata} & {\nodata} & $<13.34$ & $-1$ & $<13.93$ & $-1$ & $11.0$ & $8.3$\\[2pt]
J113910.70-135044.0 & $0.3195$ & $-75.0$ & $40.0$ & $-18.6 \pm 2.0$ & $-22.3 \pm 3.8$ & $85.7 \pm 2.9$ & $78.9 \pm 5.4$ & $14.02_{-0.03}^{+0.03}$ & $0$ & $14.00_{-0.06}^{+0.05}$ & $0$ & $17.8$ & $17.5$\\[2pt]
J113956.98+654749.1 & $0.3257$ & $-75.0$ & $80.0$ & $7.1 \pm 1.5$ & $3.9 \pm 3.2$ & $85.4 \pm 2.2$ & $78.3 \pm 4.5$ & $13.96_{-0.02}^{+0.02}$ & $0$ & $13.99_{-0.04}^{+0.03}$ & $0$ & $35.2$ & $36.8$\\[2pt]
J113956.98+654749.1 & $0.5279$ & $-60.0$ & $10.0$ & $-19.6 \pm 5.9$ & {\nodata} & $49.0 \pm 8.4$ & {\nodata} & $13.08_{-0.16}^{+0.11}$ & $0$ & {\nodata} & $-4$ & $21.3$ & {\nodata}\\[2pt]
J115120.46+543733.0 & $0.2525$ & $-110.0$ & $140.0$ & $6.0 \pm 1.9$ & $17.3 \pm 4.9$ & $125.2 \pm 2.7$ & $131.2 \pm 6.9$ & $14.34_{-0.01}^{+0.01}$ & $0$ & $14.35_{-0.03}^{+0.03}$ & $0$ & $25.0$ & $18.5$\\[2pt]
J115120.46+543733.0 & $0.6886$ & $-60.0$ & $60.0$ & {\nodata} & {\nodata} & {\nodata} & {\nodata} & $<13.19$ & $-1$ & $<13.44$ & $-1$ & $13.6$ & $13.9$\\[2pt]
J120158.75-135459.9 & $0.4094$ & $-90.0$ & $75.0$ & {\nodata} & {\nodata} & {\nodata} & {\nodata} & $<13.74$ & $-1$ & $<14.08$ & $-1$ & $6.0$ & $5.8$\\[2pt]
J121430.55+082508.1 & $0.3247$ & $-40.0$ & $60.0$ & {\nodata} & {\nodata} & {\nodata} & {\nodata} & $<13.36$ & $-1$ & $<13.72$ & $-1$ & $9.6$ & $9.3$\\[2pt]
J121430.55+082508.1 & $0.3838$ & $-100.0$ & $100.0$ & {\nodata} & {\nodata} & {\nodata} & {\nodata} & $<13.59$ & $-1$ & $<13.93$ & $-1$ & $8.3$ & $7.3$\\[2pt]
J121640.56+071224.3 & $0.3824$ & $-40.0$ & $60.0$ & {\nodata} & $14.9 \pm 5.9$ & {\nodata} & $81.6 \pm 8.4$ & {\nodata} & $-4$ & $14.23_{-0.12}^{+0.09}$ & $0$ & {\nodata} & $6.5$\\[2pt]
J121920.93+063838.5 & $0.2823$ & $-25.0$ & $90.0$ & $15.2 \pm 3.6$ & $16.5 \pm 8.1$ & $81.5 \pm 5.1$ & $67.7 \pm 11.4$ & $13.63_{-0.05}^{+0.04}$ & $0$ & $13.63_{-0.12}^{+0.10}$ & $0$ & $21.8$ & $21.2$\\[2pt]
J122317.79+092306.9 & $0.3795$ & $-50.0$ & $40.0$ & {\nodata} & {\nodata} & {\nodata} & {\nodata} & $<13.32$ & $-1$ & {\nodata} & $-4$ & $9.8$ & {\nodata}\\[2pt]
J122454.44+212246.3 & $0.3784$ & $-80.0$ & $80.0$ & $1.5 \pm 2.6$ & $-6.6 \pm 4.7$ & $88.5 \pm 3.6$ & $121.1 \pm 6.6$ & $14.05_{-0.03}^{+0.03}$ & $0$ & $14.29_{-0.04}^{+0.03}$ & $-3$ & $17.6$ & $17.5$\\[2pt]
J122454.44+212246.3 & $0.4208$ & $-40.0$ & $20.0$ & {\nodata} & {\nodata} & {\nodata} & {\nodata} & {\nodata} & $-4$ & $<12.98$ & $-1$ & {\nodata} & $29.7$\\[2pt]
J122454.44+212246.3 & $0.4214$ & $-60.0$ & $60.0$ & $-9.8 \pm 2.9$ & $-10.4 \pm 4.5$ & $120.1 \pm 4.1$ & $67.3 \pm 6.4$ & $13.89_{-0.02}^{+0.02}$ & $-3$ & $13.75_{-0.06}^{+0.05}$ & $0$ & $29.9$ & $30.3$\\[2pt]
J122512.93+121835.7 & $0.3250$ & $-100.0$ & $100.0$ & {\nodata} & {\nodata} & {\nodata} & {\nodata} & $<13.53$ & $-1$ & $<13.88$ & $-1$ & $9.7$ & $9.2$\\[2pt]
J123304.05-003134.1 & $0.3188$ & $-160.0$ & $100.0$ & $-34.0 \pm 2.5$ & $-44.0 \pm 5.0$ & $159.4 \pm 3.6$ & $158.6 \pm 7.0$ & $14.70_{-0.02}^{+0.02}$ & $0$ & $14.66_{-0.03}^{+0.03}$ & $0$ & $12.1$ & $12.2$\\[2pt]
J123335.07+475800.4 & $0.2849$ & $-60.0$ & $60.0$ & $0.3 \pm 5.3$ & {\nodata} & $54.3 \pm 7.5$ & {\nodata} & $13.93_{-0.09}^{+0.08}$ & $0$ & $<13.73$ & $-1$ & $9.1$ & $9.9$\\[2pt]
J124154.02+572107.3 & $0.2179$ & $-100.0$ & $40.0$ & $4.2 \pm 2.1$ & $6.6 \pm 4.0$ & $64.6 \pm 3.0$ & $71.2 \pm 5.6$ & $14.51_{-0.04}^{+0.03}$ & $0$ & $14.52_{-0.05}^{+0.04}$ & $0$ & $12.3$ & $11.2$\\[2pt]
J124511.26+335610.1 & $0.5567$ & $-60.0$ & $60.0$ & {\nodata} & {\nodata} & {\nodata} & {\nodata} & {\nodata} & $-4$ & $<13.85$ & $-1$ & {\nodata} & $5.6$\\[2pt]
J124511.26+335610.1 & $0.5876$ & $-50.0$ & $50.0$ & {\nodata} & {\nodata} & {\nodata} & {\nodata} & $<13.42$ & $-1$ & $<13.72$ & $-1$ & $9.8$ & $9.2$\\[2pt]
J124511.26+335610.1 & $0.6449$ & $-70.0$ & $70.0$ & {\nodata} & {\nodata} & {\nodata} & {\nodata} & $<13.52$ & $-1$ & $<13.84$ & $-1$ & $7.6$ & $8.2$\\[2pt]
J124511.26+335610.1 & $0.6889$ & $-60.0$ & $25.0$ & {\nodata} & {\nodata} & {\nodata} & {\nodata} & {\nodata} & $-4$ & $<13.73$ & $-1$ & {\nodata} & $7.4$\\[2pt]
J124511.26+335610.1 & $0.6894$ & $-60.0$ & $25.0$ & {\nodata} & {\nodata} & {\nodata} & {\nodata} & {\nodata} & $-4$ & $<13.73$ & $-1$ & {\nodata} & $7.4$\\[2pt]
J124511.26+335610.1 & $0.7130$ & $-40.0$ & $50.0$ & $8.8 \pm 3.4$ & $13.6 \pm 10.8$ & $68.6 \pm 4.8$ & $84.6 \pm 15.3$ & $14.31_{-0.07}^{+0.06}$ & $0$ & $14.55_{-0.29}^{+0.17}$ & $-3$ & $6.1$ & $5.1$\\[2pt]
J125846.65+242739.1 & $0.2262$ & $-100.0$ & $100.0$ & {\nodata} & {\nodata} & {\nodata} & {\nodata} & $<13.54$ & $-1$ & $<13.84$ & $-1$ & $9.5$ & $10.0$\\[2pt]
J130429.02+311308.2 & $0.3099$ & $-11.0$ & $100.0$ & $40.5 \pm 2.4$ & $35.0 \pm 4.0$ & $86.6 \pm 3.4$ & $86.0 \pm 5.7$ & $14.29_{-0.04}^{+0.04}$ & $0$ & $14.30_{-0.06}^{+0.06}$ & $0$ & $11.2$ & $10.8$\\[2pt]
J130451.40+245445.9 & $0.2763$ & $-75.0$ & $75.0$ & {\nodata} & {\nodata} & {\nodata} & {\nodata} & $<13.43$ & $-1$ & $<13.71$ & $-1$ & $11.0$ & $10.7$\\[2pt]
J133045.15+281321.5 & $0.2755$ & $-80.0$ & $90.0$ & $7.9 \pm 4.2$ & $4.1 \pm 7.9$ & $102.6 \pm 6.0$ & $108.7 \pm 11.1$ & $14.29_{-0.05}^{+0.05}$ & $0$ & $14.28_{-0.09}^{+0.07}$ & $0$ & $9.3$ & $9.7$\\[2pt]
J134100.78+412314.0 & $0.3488$ & $-85.0$ & $95.0$ & $-3.2 \pm 1.5$ & $0.7 \pm 2.7$ & $96.7 \pm 2.1$ & $96.3 \pm 3.8$ & $14.52_{-0.02}^{+0.02}$ & $0$ & $14.51_{-0.03}^{+0.03}$ & $0$ & $14.7$ & $14.3$\\[2pt]
J134100.78+412314.0 & $0.6208$ & $-100.0$ & $50.0$ & $-0.5 \pm 3.9$ & $-21.1 \pm 3.3$ & $105.6 \pm 5.6$ & $133.1 \pm 4.6$ & $14.35_{-0.05}^{+0.04}$ & $0$ & $15.06_{-0.06}^{+0.06}$ & $-3$ & $9.5$ & $7.8$\\[2pt]
J134100.78+412314.0 & $0.6214$ & $-80.0$ & $130.0$ & $-7.5 \pm 3.2$ & $-17.6 \pm 2.6$ & $158.1 \pm 4.6$ & $150.7 \pm 3.7$ & $14.95_{-0.05}^{+0.05}$ & $0$ & $14.99_{-0.02}^{+0.02}$ & $0$ & $11.2$ & $12.2$\\[2pt]
J134100.78+412314.0 & $0.6861$ & $-60.0$ & $150.0$ & $46.2 \pm 2.1$ & $53.4 \pm 3.5$ & $101.5 \pm 3.0$ & $101.1 \pm 4.9$ & $14.70_{-0.03}^{+0.03}$ & $-2$ & $14.75_{-0.04}^{+0.03}$ & $-2$ & $11.0$ & $11.1$\\[2pt]
J134447.55+554656.8 & $0.4042$ & $-75.0$ & $50.0$ & {\nodata} & {\nodata} & {\nodata} & {\nodata} & $<13.67$ & $-1$ & {\nodata} & $-4$ & $4.6$ & {\nodata}\\[2pt]
J135726.26+043541.3 & $0.6102$ & $-40.0$ & $75.0$ & $6.5 \pm 2.4$ & $-3.2 \pm 4.0$ & $79.9 \pm 3.4$ & $70.3 \pm 5.6$ & $14.41_{-0.05}^{+0.04}$ & $0$ & $14.89_{-0.11}^{+0.09}$ & $-3$ & $8.6$ & $8.1$\\[2pt]
J140732.25+550725.4 & $0.2433$ & $-75.0$ & $75.0$ & {\nodata} & {\nodata} & {\nodata} & {\nodata} & $<13.36$ & $-1$ & $<13.60$ & $-1$ & $11.8$ & $12.4$\\[2pt]
J140732.25+550725.4 & $0.2469$ & $-50.0$ & $50.0$ & {\nodata} & {\nodata} & {\nodata} & {\nodata} & $<13.29$ & $-1$ & $<13.43$ & $-1$ & $12.3$ & $14.6$\\[2pt]
J140923.90+261820.9 & $0.5749$ & $-50.0$ & $100.0$ & {\nodata} & {\nodata} & {\nodata} & {\nodata} & $<13.05$ & $-1$ & $<13.27$ & $-1$ & $21.5$ & $20.7$\\[2pt]
J140923.90+261820.9 & $0.5996$ & $-50.0$ & $40.0$ & $-8.7 \pm 1.2$ & $4.4 \pm 2.0$ & $80.2 \pm 1.6$ & $95.0 \pm 2.8$ & $13.99_{-0.02}^{+0.02}$ & $0$ & $14.10_{-0.03}^{+0.03}$ & $-3$ & $21.2$ & $22.8$\\[2pt]
J140923.90+261820.9 & $0.6827$ & $-85.0$ & $85.0$ & $3.8 \pm 2.4$ & $-5.6 \pm 4.9$ & $89.1 \pm 3.4$ & $113.9 \pm 6.9$ & $14.09_{-0.02}^{+0.02}$ & $0$ & $14.05_{-0.05}^{+0.04}$ & $0$ & $20.6$ & $20.7$\\[2pt]
J141038.39+230447.1 & $0.3498$ & $-20.0$ & $40.0$ & $11.2 \pm 2.6$ & $14.0 \pm 5.9$ & $45.2 \pm 3.6$ & $76.8 \pm 8.4$ & $13.83_{-0.09}^{+0.07}$ & $0$ & $14.09_{-0.10}^{+0.08}$ & $-3$ & $9.7$ & $9.5$\\[2pt]
J141038.39+230447.1 & $0.3510$ & $-50.0$ & $55.0$ & {\nodata} & {\nodata} & {\nodata} & {\nodata} & $<13.40$ & $-1$ & $<13.71$ & $-1$ & $9.8$ & $9.4$\\[2pt]
J141038.39+230447.1 & $0.5351$ & $-20.0$ & $75.0$ & $23.8 \pm 4.5$ & {\nodata} & $76.7 \pm 6.4$ & {\nodata} & $13.73_{-0.09}^{+0.07}$ & $0$ & $<13.68$ & $-1$ & $12.4$ & $7.7$\\[2pt]
J141910.20+420746.9 & $0.2890$ & $-80.0$ & $80.0$ & $9.3 \pm 3.8$ & $-6.0 \pm 7.6$ & $101.4 \pm 5.3$ & $100.9 \pm 10.8$ & $14.51_{-0.06}^{+0.05}$ & $0$ & $14.46_{-0.10}^{+0.08}$ & $0$ & $8.1$ & $7.1$\\[2pt]
J141910.20+420746.9 & $0.4256$ & $-70.0$ & $10.0$ & $-27.9 \pm 5.3$ & {\nodata} & $60.4 \pm 7.5$ & {\nodata} & $13.73_{-0.13}^{+0.10}$ & $0$ & $<13.74$ & $-1$ & $8.8$ & $8.0$\\[2pt]
J141910.20+420746.9 & $0.5221$ & $-100.0$ & $100.0$ & {\nodata} & {\nodata} & {\nodata} & {\nodata} & $<13.76$ & $-1$ & $<14.08$ & $-1$ & $5.9$ & $5.3$\\[2pt]
J141910.20+420746.9 & $0.5346$ & $-40.0$ & $50.0$ & {\nodata} & {\nodata} & {\nodata} & {\nodata} & $<13.79$ & $-1$ & $<14.15$ & $-1$ & $4.5$ & $4.4$\\[2pt]
J141910.20+420746.9 & $0.6083$ & $-30.0$ & $60.0$ & {\nodata} & {\nodata} & {\nodata} & {\nodata} & $<13.75$ & $-1$ & $<14.04$ & $-1$ & $4.7$ & $4.9$\\[2pt]
J142735.59+263214.6 & $0.3661$ & $0.0$ & $105.0$ & $35.8 \pm 5.9$ & $55.0 \pm 5.4$ & $112.3 \pm 8.3$ & $88.6 \pm 7.6$ & $14.74_{-0.12}^{+0.10}$ & $-3$ & $14.38_{-0.10}^{+0.08}$ & $0$ & $5.1$ & $6.1$\\[2pt]
J142859.03+322506.8 & $0.3823$ & $-20.0$ & $80.0$ & $23.9 \pm 4.9$ & {\nodata} & $56.7 \pm 6.9$ & {\nodata} & $13.79_{-0.09}^{+0.08}$ & $0$ & {\nodata} & $-4$ & $10.0$ & {\nodata}\\[2pt]
J143511.53+360437.2 & $0.3730$ & $-60.0$ & $60.0$ & {\nodata} & {\nodata} & {\nodata} & {\nodata} & $<13.46$ & $-1$ & $<13.81$ & $-1$ & $9.2$ & $8.1$\\[2pt]
J143511.53+360437.2 & $0.3876$ & $-100.0$ & $100.0$ & $-50.9 \pm 15.3$ & {\nodata} & $112.7 \pm 21.6$ & {\nodata} & $14.70_{-0.88}^{+0.27}$ & $-3$ & $<14.12$ & $-1$ & $5.9$ & $5.5$\\[2pt]
J143748.28-014710.7 & $0.2989$ & $-90.0$ & $55.0$ & {\nodata} & {\nodata} & {\nodata} & {\nodata} & $<12.89$ & $-1$ & $<13.15$ & $-1$ & $28.6$ & $28.9$\\[2pt]
J143748.28-014710.7 & $0.6127$ & $-90.0$ & $60.0$ & $-18.3 \pm 2.9$ & $-4.2 \pm 7.0$ & $112.8 \pm 4.0$ & $118.6 \pm 9.9$ & $13.90_{-0.03}^{+0.03}$ & $0$ & $13.83_{-0.08}^{+0.07}$ & $0$ & $22.0$ & $22.1$\\[2pt]
J143748.28-014710.7 & $0.6812$ & $-100.0$ & $100.0$ & {\nodata} & {\nodata} & {\nodata} & {\nodata} & {\nodata} & $-4$ & $<13.41$ & $-1$ & {\nodata} & $20.4$\\[2pt]
J150030.64+551708.8 & $0.3473$ & $-75.0$ & $75.0$ & {\nodata} & {\nodata} & {\nodata} & {\nodata} & $<13.21$ & $-1$ & $<13.59$ & $-1$ & $13.2$ & $12.0$\\[2pt]
J150030.64+551708.8 & $0.3480$ & $-80.0$ & $75.0$ & $-5.8 \pm 4.0$ & {\nodata} & $110.0 \pm 5.7$ & {\nodata} & $14.04_{-0.05}^{+0.04}$ & $0$ & $<13.64$ & $-1$ & $13.0$ & $12.9$\\[2pt]
J152424.58+095829.7 & $0.5185$ & $-70.0$ & $70.0$ & {\nodata} & {\nodata} & {\nodata} & {\nodata} & $<13.06$ & $-1$ & {\nodata} & $-4$ & $18.4$ & {\nodata}\\[2pt]
J152424.58+095829.7 & $0.5718$ & $-55.0$ & $30.0$ & {\nodata} & {\nodata} & {\nodata} & {\nodata} & $<13.20$ & $-1$ & $<13.36$ & $-1$ & $11.1$ & $13.7$\\[2pt]
J152424.58+095829.7 & $0.6754$ & $-70.0$ & $105.0$ & $6.1 \pm 3.2$ & $11.1 \pm 5.6$ & $134.1 \pm 4.5$ & $133.2 \pm 7.9$ & $14.15_{-0.03}^{+0.03}$ & $0$ & $14.20_{-0.06}^{+0.05}$ & $0$ & $15.3$ & $16.1$\\[2pt]
J152424.58+095829.7 & $0.7289$ & $-30.0$ & $50.0$ & $3.3 \pm 4.5$ & {\nodata} & $55.7 \pm 6.3$ & {\nodata} & $13.69_{-0.10}^{+0.08}$ & $0$ & {\nodata} & $-4$ & $10.2$ & {\nodata}\\[2pt]
J154121.48+281706.2 & $0.2825$ & $-120.0$ & $120.0$ & $-9.1 \pm 16.2$ & {\nodata} & $169.9 \pm 22.9$ & {\nodata} & $14.18_{-0.13}^{+0.10}$ & $0$ & $<13.89$ & $-1$ & $6.8$ & $8.4$\\[2pt]
J155048.29+400144.9 & $0.3126$ & $-100.0$ & $100.0$ & {\nodata} & {\nodata} & {\nodata} & {\nodata} & $<13.59$ & $-1$ & $<13.92$ & $-1$ & $8.4$ & $7.8$\\[2pt]
J155048.29+400144.9 & $0.4273$ & $-80.0$ & $110.0$ & $26.9 \pm 4.7$ & $15.3 \pm 7.8$ & $150.2 \pm 6.6$ & $156.6 \pm 11.1$ & $14.37_{-0.04}^{+0.04}$ & $0$ & $14.37_{-0.07}^{+0.06}$ & $0$ & $10.6$ & $10.6$\\[2pt]
J155048.29+400144.9 & $0.4920$ & $-90.0$ & $90.0$ & {\nodata} & $-27.0 \pm 4.5$ & {\nodata} & $113.8 \pm 6.4$ & $<13.46$ & $-1$ & $14.88_{-0.08}^{+0.07}$ & $-3$ & $11.1$ & $10.8$\\[2pt]
J155048.29+400144.9 & $0.4926$ & $-50.0$ & $50.0$ & {\nodata} & {\nodata} & {\nodata} & {\nodata} & $<13.33$ & $-1$ & $<13.61$ & $-1$ & $11.3$ & $10.4$\\[2pt]
J155232.54+570516.5 & $0.3660$ & $-85.0$ & $65.0$ & $18.2 \pm 5.9$ & $20.4 \pm 9.7$ & $89.4 \pm 8.4$ & $95.1 \pm 13.8$ & $14.12_{-0.06}^{+0.05}$ & $0$ & $14.21_{-0.10}^{+0.08}$ & $0$ & $9.9$ & $9.0$\\[2pt]
J155232.54+570516.5 & $0.3665$ & $-50.0$ & $55.0$ & $-5.3 \pm 2.5$ & $-6.1 \pm 5.6$ & $76.5 \pm 3.6$ & $76.0 \pm 7.9$ & $14.25_{-0.05}^{+0.04}$ & $0$ & $14.19_{-0.10}^{+0.08}$ & $0$ & $9.7$ & $8.5$\\[2pt]
J155304.92+354828.6 & $0.2179$ & $-50.0$ & $40.0$ & $-7.5 \pm 3.3$ & $-8.9 \pm 7.4$ & $57.3 \pm 4.6$ & $71.3 \pm 10.5$ & $13.98_{-0.07}^{+0.06}$ & $0$ & $13.94_{-0.16}^{+0.12}$ & $0$ & $8.7$ & $10.4$\\[2pt]
J155304.92+354828.6 & $0.4755$ & $-75.0$ & $75.0$ & {\nodata} & {\nodata} & {\nodata} & {\nodata} & $<13.54$ & $-1$ & $<13.86$ & $-1$ & $9.1$ & $9.2$\\[2pt]
J155504.39+362848.0 & $0.5760$ & $-75.0$ & $75.0$ & {\nodata} & {\nodata} & {\nodata} & {\nodata} & $<13.81$ & $-1$ & $<14.13$ & $-1$ & $4.4$ & $5.0$\\[2pt]
J161916.54+334238.4 & $0.2694$ & $-70.0$ & $70.0$ & $19.4 \pm 3.5$ & $6.4 \pm 10.9$ & $73.0 \pm 5.0$ & $75.3 \pm 15.4$ & $14.46_{-0.07}^{+0.06}$ & $-3$ & $13.86_{-0.16}^{+0.11}$ & $0$ & $9.7$ & $12.2$\\[2pt]
J161916.54+334238.4 & $0.4423$ & $-75.0$ & $100.0$ & {\nodata} & {\nodata} & {\nodata} & {\nodata} & $<13.13$ & $-1$ & $<13.47$ & $-1$ & $19.1$ & $18.6$\\[2pt]
J161916.54+334238.4 & $0.4708$ & $-50.0$ & $100.0$ & $31.2 \pm 0.8$ & $42.7 \pm 2.4$ & $101.9 \pm 1.2$ & $110.6 \pm 3.3$ & $14.93_{-0.02}^{+0.02}$ & $0$ & $15.38_{-0.04}^{+0.04}$ & $-3$ & $19.0$ & $17.4$\\[2pt]
J163201.12+373749.9 & $0.2740$ & $-60.0$ & $25.0$ & $-11.8 \pm 1.1$ & $0.1 \pm 2.3$ & $95.4 \pm 1.6$ & $54.4 \pm 3.2$ & $14.16_{-0.01}^{+0.01}$ & $-3$ & $13.94_{-0.04}^{+0.04}$ & $0$ & $18.1$ & $26.5$\\[2pt]
J163201.12+373749.9 & $0.4177$ & $-60.0$ & $65.0$ & $8.6 \pm 2.5$ & $-2.5 \pm 7.3$ & $94.1 \pm 3.5$ & $83.0 \pm 10.4$ & $13.72_{-0.03}^{+0.03}$ & $0$ & $13.69_{-0.11}^{+0.09}$ & $0$ & $27.1$ & $24.8$\\[2pt]
J215647.46+224249.8 & $0.5928$ & $-30.0$ & $40.0$ & $3.8 \pm 4.7$ & $3.3 \pm 14.1$ & $67.3 \pm 6.6$ & $75.5 \pm 20.0$ & $13.89_{-0.14}^{+0.11}$ & $0$ & $15.00_{-0.57}^{+0.24}$ & $-3$ & $6.7$ & $6.3$\\[2pt]
J215647.46+224249.8 & $0.6200$ & $-50.0$ & $50.0$ & {\nodata} & {\nodata} & {\nodata} & {\nodata} & $<13.60$ & $-1$ & $<13.87$ & $-1$ & $7.0$ & $6.8$\\[2pt]
J215647.46+224249.8 & $0.6210$ & $-50.0$ & $43.0$ & {\nodata} & {\nodata} & {\nodata} & {\nodata} & $<13.57$ & $-1$ & $<13.86$ & $-1$ & $7.1$ & $6.7$\\[2pt]
J215647.46+224249.8 & $0.6214$ & $-27.0$ & $50.0$ & {\nodata} & {\nodata} & {\nodata} & {\nodata} & $<13.53$ & $-1$ & $<13.83$ & $-1$ & $7.0$ & $6.6$\\[2pt]
J225357.75+160853.1 & $0.3206$ & $-40.0$ & $40.0$ & $-0.2 \pm 4.7$ & $-4.4 \pm 6.5$ & $59.4 \pm 6.7$ & $59.1 \pm 9.1$ & $13.90_{-0.12}^{+0.09}$ & $0$ & $14.09_{-0.16}^{+0.12}$ & $0$ & $7.7$ & $7.9$\\[2pt]
J225738.20+134045.4 & $0.4989$ & $-40.0$ & $80.0$ & $-17.3 \pm 5.5$ & $24.8 \pm 6.9$ & $137.6 \pm 7.7$ & $92.4 \pm 9.7$ & $14.42_{-0.08}^{+0.07}$ & $-3$ & $14.29_{-0.11}^{+0.09}$ & $0$ & $7.2$ & $8.0$\\[2pt]
J234500.43-005936.0 & $0.5481$ & $-75.0$ & $100.0$ & $-9.5 \pm 18.0$ & {\nodata} & $128.6 \pm 25.4$ & {\nodata} & $14.65_{-0.68}^{+0.25}$ & $-3$ & $<14.24$ & $-1$ & $4.8$ & $4.8$\\[2pt]
\enddata
\tablecomments{(1) J--Name of the background quasar; (2) Redshift of the absorber; (3) Start velocity of the integration range; (4) End velocity of the integration range; (5) Average velocity centroid of {\ovi} $\lambda$1031; (6) Average velocity centroid of {\ovi} $\lambda$1037; (7) Velocity width of absorption for {\ovi} $\lambda$1031; (8) Velocity width of absorption for {\ovi} $\lambda$1037; (9) Measured column density of the $\lambda$1031 transition; (10) Detection Flag for $\lambda$1031 transition-- 0--detection, $-1$--upper limit, $-2$--saturation, $-3$--contaminated, $-4$--not covered by the spectrum; (11)  Measured column density of the $\lambda$1037 transition; (12) Detection Flag for $\lambda$1037 transition; (13) $S/N$ of the spectrum per pixel for {\ovi} $\lambda$1031; (14) $S/N$ of the spectrum per pixel for {\ovi} $\lambda$1037.}
\end{deluxetable*}
\end{longrotatetable}
\normalsize

\addtolength{\tabcolsep}{3pt}
\startlongtable
\tablewidth{0pc}
\begin{deluxetable*}{lccccccc}
\footnotesize
\tablecaption{Adopted properties of the {\ovi} absorbers in the robust sample\label{tab:ovi-properties}}
\tablehead{
\colhead{Target} & \colhead{$z_{\rm abs}$} & \colhead{$v_{\rm min}$} & \colhead{$v_{\rm max}$} & \colhead{$\langle v \rangle$} & \colhead{$\Delta v_{90}$}  & \colhead{$\log N$} & \colhead{Detection}\\
\colhead{} & \colhead{} & \colhead{({\kms})} & \colhead{({\kms})} & \colhead{({\kms})} & \colhead{({\kms})}  & \colhead{[\cmsq]} & \colhead{Flag}\\
\colhead{(1)} & \colhead{(2)} & \colhead{(3)} & \colhead{(4)} & \colhead{(5)} & \colhead{(6)} & \colhead{(7)}  & \colhead{(8)}
}
\startdata
\hline
\midrule
J000559.23+160948.9 & $0.3058$ & $-100.0$ & $40.0$ & $-52.0 \pm 8.8$ & $46.9 \pm 12.5$ & $13.38_{-0.10}^{+0.08}$ & $0$\\[2pt]
J000559.23+160948.9 & $0.3479$ & $-50.0$ & $20.0$ & $-19.2 \pm 0.9$ & $56.7 \pm 1.2$ & $13.99_{-0.02}^{+0.02}$ & $0$\\[2pt]
J000559.23+160948.9 & $0.3662$ & $-50.0$ & $50.0$ & $-0.3 \pm 1.9$ & $82.5 \pm 2.6$ & $14.02_{-0.03}^{+0.03}$ & $0$\\[2pt]
J004222.29-103743.8 & $0.3161$ & $-60.0$ & $70.0$ & $-4.4 \pm 7.8$ & $72.7 \pm 11.1$ & $13.97_{-0.12}^{+0.09}$ & $0$\\[2pt]
J004705.89+031954.9 & $0.3139$ & $-50.0$ & $50.0$ & {\nodata} & {\nodata} & $<13.39$ & $-1$\\[2pt]
J004705.89+031954.9 & $0.3143$ & $-50.0$ & $100.0$ & {\nodata} & {\nodata} & $<13.48$ & $-1$\\[2pt]
J011013.14-021952.8 & $0.2272$ & $-110.0$ & $100.0$ & $-22.5 \pm 2.2$ & $123.0 \pm 3.2$ & $14.48_{-0.02}^{+0.02}$ & $0$\\[2pt]
J011013.14-021952.8 & $0.5365$ & $-20.0$ & $40.0$ & $4.1 \pm 4.2$ & $34.8 \pm 6.0$ & $13.59_{-0.15}^{+0.11}$ & $0$\\[2pt]
J011016.25-021851.0 & $0.3991$ & $-100.0$ & $80.0$ & $-6.5 \pm 3.3$ & $116.1 \pm 4.6$ & $14.36_{-0.03}^{+0.03}$ & $0$\\[2pt]
J011016.25-021851.0 & $0.5354$ & $-50.0$ & $40.0$ & {\nodata} & {\nodata} & $<13.44$ & $-1$\\[2pt]
J011935.69-282131.4 & $0.3483$ & $-60.0$ & $10.0$ & {\nodata} & {\nodata} & $<13.09$ & $-1$\\[2pt]
J011935.69-282131.4 & $0.3487$ & $-100.0$ & $40.0$ & $-34.4 \pm 3.3$ & $89.8 \pm 4.7$ & $14.07_{-0.04}^{+0.04}$ & $0$\\[2pt]
J012236.76-284321.3 & $0.3650$ & $-45.0$ & $80.0$ & $18.5 \pm 2.2$ & $76.4 \pm 3.1$ & $13.97_{-0.03}^{+0.03}$ & $0$\\[2pt]
J015513.20-450611.9 & $0.2260$ & $-100.0$ & $120.0$ & $7.8 \pm 1.9$ & $108.1 \pm 2.7$ & $14.18_{-0.02}^{+0.02}$ & $0$\\[2pt]
J020157.16-113233.1 & $0.3226$ & $-50.0$ & $50.0$ & {\nodata} & {\nodata} & $<12.89$ & $-1$\\[2pt]
J020157.16-113233.1 & $0.3231$ & $-40.0$ & $25.0$ & $-3.4 \pm 3.1$ & $52.6 \pm 4.4$ & $13.27_{-0.09}^{+0.07}$ & $0$\\[2pt]
J020157.16-113233.1 & $0.3234$ & $-55.0$ & $70.0$ & {\nodata} & {\nodata} & $<13.29$ & $-1$\\[2pt]
J020157.16-113233.1 & $0.3245$ & $-45.0$ & $85.0$ & $16.2 \pm 1.3$ & $99.9 \pm 1.8$ & $14.11_{-0.02}^{+0.02}$ & $0$\\[2pt]
J023507.38-040205.6 & $0.3225$ & $-40.0$ & $40.0$ & $-1.6 \pm 2.9$ & $66.1 \pm 4.1$ & $13.58_{-0.06}^{+0.05}$ & $0$\\[2pt]
J024337.66-303048.0 & $0.3037$ & $-70.0$ & $70.0$ & {\nodata} & {\nodata} & $<13.24$ & $-1$\\[2pt]
J024649.86-300741.3 & $0.3123$ & $-60.0$ & $100.0$ & $25.6 \pm 6.1$ & $106.2 \pm 8.6$ & $14.13_{-0.07}^{+0.06}$ & $0$\\[2pt]
J024649.86-300741.3 & $0.3359$ & $-50.0$ & $70.0$ & {\nodata} & {\nodata} & $<13.44$ & $-1$\\[2pt]
J035128.56-142908.0 & $0.3285$ & $-70.0$ & $100.0$ & $-0.9 \pm 11.6$ & $92.3 \pm 16.3$ & $13.57_{-0.12}^{+0.10}$ & $0$\\[2pt]
J035128.56-142908.0 & $0.3569$ & $-160.0$ & $120.0$ & $-23.2 \pm 4.0$ & $159.5 \pm 5.7$ & $14.69_{-0.03}^{+0.03}$ & $0$\\[2pt]
J035128.56-142908.0 & $0.3572$ & $0.0$ & $70.0$ & {\nodata} & {\nodata} & $<13.34$ & $-1$\\[2pt]
J035128.56-142908.0 & $0.4398$ & $-100.0$ & $100.0$ & $-7.1 \pm 6.0$ & $85.7 \pm 8.5$ & $13.96_{-0.05}^{+0.05}$ & $0$\\[2pt]
J040148.98-054056.5 & $0.2195$ & $-50.0$ & $140.0$ & $40.8 \pm 2.8$ & $134.0 \pm 4.0$ & $14.53_{-0.03}^{+0.03}$ & $0$\\[2pt]
J040148.98-054056.5 & $0.3238$ & $-70.0$ & $70.0$ & {\nodata} & {\nodata} & $<13.45$ & $-1$\\[2pt]
J040748.42-121136.3 & $0.1672$ & $-125.0$ & $95.0$ & $-33.8 \pm 0.4$ & $149.2 \pm 0.5$ & $14.63_{-0.00}^{+0.00}$ & $0$\\[2pt]
J040748.42-121136.3 & $0.3607$ & $-30.0$ & $30.0$ & $0.3 \pm 2.5$ & $45.1 \pm 3.5$ & $12.81_{-0.08}^{+0.07}$ & $0$\\[2pt]
J044011.90-524818.0 & $0.3279$ & $-80.0$ & $80.0$ & {\nodata} & {\nodata} & $<13.18$ & $-1$\\[2pt]
J044011.90-524818.0 & $0.6150$ & $-120.0$ & $135.0$ & $6.9 \pm 2.3$ & $176.7 \pm 3.2$ & $14.86_{-0.02}^{+0.02}$ & $0$\\[2pt]
J044011.90-524818.0 & $0.6157$ & $-20.0$ & $80.0$ & $16.0 \pm 5.3$ & $72.8 \pm 7.5$ & $13.85_{-0.09}^{+0.07}$ & $0$\\[2pt]
J055224.49-640210.7 & $0.3451$ & $-25.0$ & $25.0$ & {\nodata} & {\nodata} & $<12.67$ & $-1$\\[2pt]
J055224.49-640210.7 & $0.4461$ & $-65.0$ & $65.0$ & $-1.9 \pm 3.2$ & $81.8 \pm 4.5$ & $13.66_{-0.04}^{+0.04}$ & $0$\\[2pt]
J063546.49-751616.8 & $0.4175$ & $-50.0$ & $95.0$ & $18.2 \pm 1.3$ & $100.9 \pm 1.9$ & $14.16_{-0.02}^{+0.02}$ & $0$\\[2pt]
J063546.49-751616.8 & $0.4528$ & $-20.0$ & $100.0$ & $41.0 \pm 4.1$ & $80.9 \pm 5.8$ & $13.63_{-0.06}^{+0.05}$ & $0$\\[2pt]
J063546.49-751616.8 & $0.4685$ & $-40.0$ & $80.0$ & $10.1 \pm 3.3$ & $81.5 \pm 4.6$ & $13.69_{-0.04}^{+0.04}$ & $0$\\[2pt]
J071950.89+742757.0 & $0.3777$ & $-40.0$ & $70.0$ & $9.1 \pm 3.4$ & $69.4 \pm 4.8$ & $13.77_{-0.05}^{+0.05}$ & $0$\\[2pt]
J075112.30+291938.3 & $0.4318$ & $-60.0$ & $80.0$ & $13.2 \pm 3.8$ & $89.7 \pm 5.4$ & $13.52_{-0.05}^{+0.04}$ & $0$\\[2pt]
J075112.30+291938.3 & $0.4945$ & $-30.0$ & $50.0$ & $4.4 \pm 6.2$ & $57.2 \pm 8.7$ & $13.02_{-0.14}^{+0.10}$ & $0$\\[2pt]
J080908.13+461925.6 & $0.6191$ & $-90.0$ & $90.0$ & $1.5 \pm 3.2$ & $92.5 \pm 4.5$ & $14.74_{-0.04}^{+0.04}$ & $0$\\[2pt]
J091440.38+282330.6 & $0.2442$ & $-90.0$ & $70.0$ & $-18.3 \pm 4.0$ & $107.1 \pm 5.7$ & $14.42_{-0.05}^{+0.05}$ & $0$\\[2pt]
J092554.44+453544.5 & $0.3096$ & $-50.0$ & $50.0$ & {\nodata} & {\nodata} & $<13.12$ & $-1$\\[2pt]
J094331.61+053131.4 & $0.2290$ & $-50.0$ & $50.0$ & {\nodata} & {\nodata} & $<13.50$ & $-1$\\[2pt]
J100102.64+594414.2 & $0.4156$ & $-60.0$ & $25.0$ & {\nodata} & {\nodata} & $<13.05$ & $-1$\\[2pt]
J100102.64+594414.2 & $0.4160$ & $-80.0$ & $125.0$ & $17.8 \pm 2.0$ & $122.7 \pm 2.8$ & $14.46_{-0.02}^{+0.02}$ & $0$\\[2pt]
J100535.25+013445.5 & $0.4185$ & $-60.0$ & $60.0$ & $-7.9 \pm 7.7$ & $90.7 \pm 10.8$ & $13.39_{-0.11}^{+0.09}$ & $0$\\[2pt]
J100535.25+013445.5 & $0.4197$ & $-40.0$ & $40.0$ & {\nodata} & {\nodata} & $<12.94$ & $-1$\\[2pt]
J100902.06+071343.8 & $0.3372$ & $-65.0$ & $40.0$ & {\nodata} & {\nodata} & $<13.52$ & $-1$\\[2pt]
J101622.60+470643.3 & $0.4322$ & $-62.0$ & $52.0$ & {\nodata} & {\nodata} & $<13.51$ & $-1$\\[2pt]
J104117.16+061016.9 & $0.6544$ & $-20.0$ & $40.0$ & {\nodata} & {\nodata} & $<13.26$ & $-1$\\[2pt]
J104117.16+061016.9 & $0.6551$ & $-50.0$ & $30.0$ & {\nodata} & {\nodata} & $<13.38$ & $-1$\\[2pt]
J104117.16+061016.9 & $0.6554$ & $-36.0$ & $50.0$ & {\nodata} & {\nodata} & $<13.39$ & $-1$\\[2pt]
J105945.24+144143.0 & $0.4658$ & $-60.0$ & $50.0$ & {\nodata} & {\nodata} & $<13.21$ & $-1$\\[2pt]
J105945.24+144143.0 & $0.5765$ & $-60.0$ & $80.0$ & $21.6 \pm 7.7$ & $95.1 \pm 10.9$ & $13.91_{-0.10}^{+0.08}$ & $0$\\[2pt]
J105956.14+121151.1 & $0.3324$ & $-60.0$ & $60.0$ & {\nodata} & {\nodata} & $<13.43$ & $-1$\\[2pt]
J110539.79+342534.3 & $0.2388$ & $-60.0$ & $65.0$ & $0.4 \pm 1.7$ & $77.2 \pm 2.4$ & $14.59_{-0.03}^{+0.03}$ & $0$\\[2pt]
J111132.20+554725.9 & $0.6175$ & $-40.0$ & $75.0$ & {\nodata} & {\nodata} & $<13.24$ & $-1$\\[2pt]
J111132.20+554725.9 & $0.6827$ & $-60.0$ & $20.0$ & {\nodata} & {\nodata} & $<13.22$ & $-1$\\[2pt]
J111507.65+023757.5 & $0.3692$ & $-40.0$ & $30.0$ & $-7.8 \pm 3.9$ & $57.2 \pm 5.5$ & $13.78_{-0.11}^{+0.09}$ & $0$\\[2pt]
J111754.23+263416.6 & $0.3520$ & $-40.0$ & $50.0$ & $6.4 \pm 6.8$ & $77.7 \pm 9.7$ & $13.68_{-0.14}^{+0.11}$ & $0$\\[2pt]
J111754.23+263416.6 & $0.3523$ & $-30.0$ & $60.0$ & {\nodata} & {\nodata} & $<13.37$ & $-1$\\[2pt]
J111908.59+211918.0 & $0.1385$ & $-95.0$ & $95.0$ & $7.8 \pm 2.6$ & $95.4 \pm 3.7$ & $13.80_{-0.02}^{+0.02}$ & $0$\\[2pt]
J112553.78+591021.6 & $0.6784$ & $-65.0$ & $65.0$ & {\nodata} & {\nodata} & $<13.40$ & $-1$\\[2pt]
J113457.71+255527.8 & $0.4320$ & $-60.0$ & $30.0$ & $-17.0 \pm 4.9$ & $67.5 \pm 6.9$ & $13.77_{-0.10}^{+0.08}$ & $0$\\[2pt]
J113457.71+255527.8 & $0.4323$ & $-40.0$ & $70.0$ & {\nodata} & {\nodata} & $<13.34$ & $-1$\\[2pt]
J113910.70-135044.0 & $0.3195$ & $-75.0$ & $40.0$ & $-19.4 \pm 1.8$ & $84.3 \pm 2.5$ & $14.02_{-0.03}^{+0.03}$ & $0$\\[2pt]
J113956.98+654749.1 & $0.3257$ & $-75.0$ & $80.0$ & $6.5 \pm 1.4$ & $84.1 \pm 2.0$ & $13.97_{-0.01}^{+0.01}$ & $0$\\[2pt]
J113956.98+654749.1 & $0.5279$ & $-60.0$ & $10.0$ & $-19.6 \pm 5.9$ & $49.0 \pm 8.4$ & $13.08_{-0.16}^{+0.11}$ & $0$\\[2pt]
J115120.46+543733.0 & $0.2525$ & $-110.0$ & $140.0$ & $7.6 \pm 1.8$ & $126.0 \pm 2.5$ & $14.34_{-0.01}^{+0.01}$ & $0$\\[2pt]
J115120.46+543733.0 & $0.6886$ & $-60.0$ & $60.0$ & {\nodata} & {\nodata} & $<13.19$ & $-1$\\[2pt]
J121430.55+082508.1 & $0.3247$ & $-40.0$ & $60.0$ & {\nodata} & {\nodata} & $<13.36$ & $-1$\\[2pt]
J121430.55+082508.1 & $0.3838$ & $-100.0$ & $100.0$ & {\nodata} & {\nodata} & $<13.59$ & $-1$\\[2pt]
J121920.93+063838.5 & $0.2823$ & $-25.0$ & $90.0$ & $15.4 \pm 3.3$ & $79.2 \pm 4.7$ & $13.63_{-0.04}^{+0.04}$ & $0$\\[2pt]
J122317.79+092306.9 & $0.3795$ & $-50.0$ & $40.0$ & {\nodata} & {\nodata} & $<13.32$ & $-1$\\[2pt]
J122454.44+212246.3 & $0.3784$ & $-80.0$ & $80.0$ & $1.5 \pm 2.6$ & $88.5 \pm 3.6$ & $14.05_{-0.03}^{+0.03}$ & $0$\\[2pt]
J122454.44+212246.3 & $0.4208$ & $-40.0$ & $20.0$ & {\nodata} & {\nodata} & $<12.98$ & $-1$\\[2pt]
J122454.44+212246.3 & $0.4214$ & $-60.0$ & $60.0$ & $-10.4 \pm 4.5$ & $67.3 \pm 6.4$ & $13.75_{-0.06}^{+0.05}$ & $0$\\[2pt]
J122512.93+121835.7 & $0.3250$ & $-100.0$ & $100.0$ & {\nodata} & {\nodata} & $<13.53$ & $-1$\\[2pt]
J123304.05-003134.1 & $0.3188$ & $-160.0$ & $100.0$ & $-36.1 \pm 2.3$ & $159.3 \pm 3.2$ & $14.69_{-0.02}^{+0.02}$ & $0$\\[2pt]
J123335.07+475800.4 & $0.2849$ & $-60.0$ & $60.0$ & $0.3 \pm 5.3$ & $54.3 \pm 7.5$ & $13.93_{-0.09}^{+0.08}$ & $0$\\[2pt]
J124154.02+572107.3 & $0.2179$ & $-100.0$ & $40.0$ & $4.7 \pm 1.8$ & $66.1 \pm 2.6$ & $14.51_{-0.03}^{+0.03}$ & $0$\\[2pt]
J124511.26+335610.1 & $0.5876$ & $-50.0$ & $50.0$ & {\nodata} & {\nodata} & $<13.42$ & $-1$\\[2pt]
J125846.65+242739.1 & $0.2262$ & $-100.0$ & $100.0$ & {\nodata} & {\nodata} & $<13.54$ & $-1$\\[2pt]
J130429.02+311308.2 & $0.3099$ & $-11.0$ & $100.0$ & $39.1 \pm 2.1$ & $86.4 \pm 2.9$ & $14.30_{-0.03}^{+0.03}$ & $0$\\[2pt]
J130451.40+245445.9 & $0.2763$ & $-75.0$ & $75.0$ & {\nodata} & {\nodata} & $<13.43$ & $-1$\\[2pt]
J133045.15+281321.5 & $0.2755$ & $-80.0$ & $90.0$ & $7.1 \pm 3.7$ & $104.0 \pm 5.3$ & $14.29_{-0.04}^{+0.04}$ & $0$\\[2pt]
J134100.78+412314.0 & $0.3488$ & $-85.0$ & $95.0$ & $-2.3 \pm 1.3$ & $96.6 \pm 1.8$ & $14.52_{-0.02}^{+0.02}$ & $0$\\[2pt]
J134100.78+412314.0 & $0.6208$ & $-100.0$ & $50.0$ & $-0.5 \pm 3.9$ & $105.6 \pm 5.6$ & $14.35_{-0.05}^{+0.04}$ & $0$\\[2pt]
J134100.78+412314.0 & $0.6214$ & $-80.0$ & $130.0$ & {\nodata} & {\nodata} & $14.91_{-0.02}^{+0.02}$$^{a}$ & $0$\\[2pt]
J134100.78+412314.0 & $0.6861$ & $-60.0$ & $150.0$ & $53.4 \pm 3.5$ & $101.1 \pm 4.9$ & $14.80_{-0.07}^{+0.07}$$^{b}$ & $0$\\[2pt]
J135726.26+043541.3 & $0.6102$ & $-40.0$ & $75.0$ & $6.5 \pm 2.4$ & $79.9 \pm 3.4$ & $14.41_{-0.05}^{+0.04}$ & $0$\\[2pt]
J140732.25+550725.4 & $0.2433$ & $-75.0$ & $75.0$ & {\nodata} & {\nodata} & $<13.36$ & $-1$\\[2pt]
J140732.25+550725.4 & $0.2469$ & $-50.0$ & $50.0$ & {\nodata} & {\nodata} & $<13.29$ & $-1$\\[2pt]
J140923.90+261820.9 & $0.5749$ & $-50.0$ & $100.0$ & {\nodata} & {\nodata} & $<13.05$ & $-1$\\[2pt]
J140923.90+261820.9 & $0.5996$ & $-50.0$ & $40.0$ & $-8.7 \pm 1.2$ & $80.2 \pm 1.6$ & $13.99_{-0.02}^{+0.02}$ & $0$\\[2pt]
J140923.90+261820.9 & $0.6827$ & $-85.0$ & $85.0$ & $2.0 \pm 2.1$ & $93.8 \pm 3.0$ & $14.08_{-0.02}^{+0.02}$ & $0$\\[2pt]
J141038.39+230447.1 & $0.3498$ & $-20.0$ & $40.0$ & $11.2 \pm 2.6$ & $45.2 \pm 3.6$ & $13.83_{-0.09}^{+0.07}$ & $0$\\[2pt]
J141038.39+230447.1 & $0.3510$ & $-50.0$ & $55.0$ & {\nodata} & {\nodata} & $<13.40$ & $-1$\\[2pt]
J141038.39+230447.1 & $0.5351$ & $-20.0$ & $75.0$ & $23.8 \pm 4.5$ & $76.7 \pm 6.4$ & $13.73_{-0.09}^{+0.07}$ & $0$\\[2pt]
J141910.20+420746.9 & $0.4256$ & $-70.0$ & $10.0$ & $-27.9 \pm 5.3$ & $60.4 \pm 7.5$ & $13.73_{-0.13}^{+0.10}$ & $0$\\[2pt]
J142859.03+322506.8 & $0.3823$ & $-20.0$ & $80.0$ & $23.9 \pm 4.9$ & $56.7 \pm 6.9$ & $13.79_{-0.09}^{+0.08}$ & $0$\\[2pt]
J143511.53+360437.2 & $0.3730$ & $-60.0$ & $60.0$ & {\nodata} & {\nodata} & $<13.46$ & $-1$\\[2pt]
J143748.28-014710.7 & $0.2989$ & $-90.0$ & $55.0$ & {\nodata} & {\nodata} & $<12.89$ & $-1$\\[2pt]
J143748.28-014710.7 & $0.6127$ & $-90.0$ & $60.0$ & $-16.3 \pm 2.6$ & $113.6 \pm 3.7$ & $13.89_{-0.03}^{+0.03}$ & $0$\\[2pt]
J143748.28-014710.7 & $0.6812$ & $-100.0$ & $100.0$ & {\nodata} & {\nodata} & $<13.41$ & $-1$\\[2pt]
J150030.64+551708.8 & $0.3473$ & $-75.0$ & $75.0$ & {\nodata} & {\nodata} & $<13.21$ & $-1$\\[2pt]
J150030.64+551708.8 & $0.3480$ & $-80.0$ & $75.0$ & $-5.8 \pm 4.0$ & $110.0 \pm 5.7$ & $14.04_{-0.05}^{+0.04}$ & $0$\\[2pt]
J152424.58+095829.7 & $0.5185$ & $-70.0$ & $70.0$ & {\nodata} & {\nodata} & $<13.06$ & $-1$\\[2pt]
J152424.58+095829.7 & $0.5718$ & $-55.0$ & $30.0$ & {\nodata} & {\nodata} & $<13.20$ & $-1$\\[2pt]
J152424.58+095829.7 & $0.6754$ & $-70.0$ & $105.0$ & $7.4 \pm 2.8$ & $133.9 \pm 3.9$ & $14.16_{-0.03}^{+0.03}$ & $0$\\[2pt]
J152424.58+095829.7 & $0.7289$ & $-30.0$ & $50.0$ & $3.3 \pm 4.5$ & $55.7 \pm 6.3$ & $13.69_{-0.10}^{+0.08}$ & $0$\\[2pt]
J155048.29+400144.9 & $0.3126$ & $-100.0$ & $100.0$ & {\nodata} & {\nodata} & $<13.59$ & $-1$\\[2pt]
J155048.29+400144.9 & $0.4273$ & $-80.0$ & $110.0$ & $23.8 \pm 4.0$ & $151.9 \pm 5.7$ & $14.37_{-0.04}^{+0.04}$ & $0$\\[2pt]
J155048.29+400144.9 & $0.4920$ & $-90.0$ & $90.0$ & {\nodata} & {\nodata} & $<13.46$ & $-1$\\[2pt]
J155048.29+400144.9 & $0.4926$ & $-50.0$ & $50.0$ & {\nodata} & {\nodata} & $<13.33$ & $-1$\\[2pt]
J155232.54+570516.5 & $0.3660$ & $-85.0$ & $65.0$ & $18.8 \pm 5.1$ & $90.9 \pm 7.2$ & $14.14_{-0.05}^{+0.05}$ & $0$\\[2pt]
J155232.54+570516.5 & $0.3665$ & $-50.0$ & $55.0$ & $-5.4 \pm 2.3$ & $76.4 \pm 3.3$ & $14.24_{-0.04}^{+0.04}$ & $0$\\[2pt]
J155304.92+354828.6 & $0.2179$ & $-50.0$ & $40.0$ & $-7.8 \pm 3.0$ & $59.6 \pm 4.2$ & $13.98_{-0.06}^{+0.06}$ & $0$\\[2pt]
J155304.92+354828.6 & $0.4755$ & $-75.0$ & $75.0$ & {\nodata} & {\nodata} & $<13.54$ & $-1$\\[2pt]
J161916.54+334238.4 & $0.2694$ & $-70.0$ & $70.0$ & $6.4 \pm 10.9$ & $75.3 \pm 15.4$ & $13.86_{-0.16}^{+0.11}$ & $0$\\[2pt]
J161916.54+334238.4 & $0.4423$ & $-75.0$ & $100.0$ & {\nodata} & {\nodata} & $<13.13$ & $-1$\\[2pt]
J163201.12+373749.9 & $0.2740$ & $-60.0$ & $25.0$ & $0.1 \pm 2.3$ & $54.4 \pm 3.2$ & $13.94_{-0.04}^{+0.04}$ & $0$\\[2pt]
J163201.12+373749.9 & $0.4177$ & $-60.0$ & $65.0$ & $7.5 \pm 2.3$ & $93.0 \pm 3.3$ & $13.72_{-0.03}^{+0.03}$ & $0$\\[2pt]
J225738.20+134045.4 & $0.4989$ & $-40.0$ & $80.0$ & $24.8 \pm 6.9$ & $92.4 \pm 9.7$ & $14.29_{-0.11}^{+0.09}$ & $0$\\[2pt]
\enddata
\tablecomments{$^{a}$ For the $z = 0.6214$ absorber towards J134100.78$+$412314.0, we found that both the {\ovi} transitions were affected by contamination to different extents in the velocity range of integration. For this system, we have summed the column densities measured over different velocity ranges for both transitions, avoiding regions of contamination. We have also not measured $\langle v \rangle$ and $\Delta v_{90}$ for this system because both transitions are affected by contamination in different velocity ranges. $^{b}$ For the $z = 0.6861$ absorber towards J134100.78$+$412314.0, we found evidence for saturation in the {\ovi}$\lambda$1031 line, which we corrected for (see text for more detail).}
\end{deluxetable*}

For {\hi}, {\cii}, and {\ciii}, we adopt the column densities, kinematics, and integration ranges from \citetalias{Lehner2018}. For multiple transitions of a given species detected at more than $2\sigma$ and uncontaminated (e.g., \cii, \hi), \citetalias{Lehner2018} used a variance-weighted mean of the measurements when no saturation was present. Saturation was assessed using atoms or ions with multiple transitions (see Section~4.2 in \citetalias{Lehner2018} for more detail). For single transitions (such as \ciii), saturation was evaluated by determining at which peak optical depth there is evidence for saturation in species with multiple transitions; if saturation is present, the column density is reported as a lower limit with a 0.15 dex correction applied (see explanation in \S4.2 of \citetalias{Lehner2018}). For the \hi\ transitions, only the weaker transitions were considered (i.e., Ly$\alpha$, Ly$\beta$, and typically Ly$\gamma$ were excluded). The \hi\ column densities were derived from the mean of the values calculated using two to three methods for the SLFSs and pLLSs, including the AOD method, Voigt profile fitting, and a curve-of-growth analysis (see \citetalias{Lehner2018} for more details).

We remind the reader that the velocity ranges over which the absorption lines were integrated in \citetalias{Lehner2018} were selected using the weakest \hi\ absorption transitions so that the absorbers consist as much as possible of a single \hi\ absorbing complex at the COS resolution and in these transitions. However, in a few cases for the pLLSs ($\la10\%$ of the present sample) and LLSs, stronger \hi\ transitions may show an additional,  weaker component(s). Such an example is the absorber toward J035128.56--142908.0 at $z=0.439804$ where \hi\ $\lambda\lambda$917, 918 show little evidence of an additional component (see \citetalias{Lehner2018}), but a blended component is evident in the stronger \hi\ transitions shown in the figures of the Appendix (which is located at $\approx -50$\,\kms\ relative to the stronger absorption component). In that case, the additional component adds $\approx 0.1$\,dex to the \nhi\ value of the pLLS. 
 
While we adopt the integration ranges from \citetalias{Lehner2018} for the low ions and \hi, we cannot apply the same integration ranges for \ovi. The \ovi\ profiles can be more extended than \hi\ or shifted in velocity relative to the {\hi}. A re-analysis of the velocity integration ranges for \ovi\ is therefore required (see also Footnote~10 in \citetalias{Lehner2018}). Our methodology to select $[\vmin, \vmax]$ for integrating the {\ovi} $\lambda\lambda$1031, 1037 profiles is as follows: 1) if the stronger transitions of \hi\ indicate no blending with a weaker \hi\ absorber, we consider the entire velocity range over which the \ovi\ absorption is observed; 2) if there is some evidence for an additional blend in the \hi\ transitions with a weaker \hi\ absorber (or absorbers), \vmin\ and/or \vmax\ are set to the values determined from the \hi. This approach minimizes the inclusions of other  \hi\ absorbers in the \ovi\ absorption. In both cases,  \vmin\ and \vmax\ may be somewhat adjusted if there is some evidence of contamination in one of the \ovi\ transitions.  In Fig.~\ref{fig:example}, we show as an example the {\nav} profiles of {\ovi}, {\cii}, {\ciii}, and {\hi} of a blended system with two components that are treated each as an individual absorber. 

As part of revisiting the integration range for \ovi, we also conducted a visual inspection of all the continua near the two \ovi\ transitions. This reexamination was needed, as the \citetalias{Lehner2018} work focused on lower ionization gas and did not emphasize \ovi\ absorption in their analysis. The continuum regions used for the continuum fit with low-order Legendre polynomials were automatically selected following the method described in  \citetalias{Lehner2018}. In $\sim$20\% of the cases, we determined that the continua could be improved and adjusted the continua by manually selecting the continuum regions to be fit with the Legendre polynomials. In 92\% of these manually adjusted cases, the resulting column densities were similar (within $\la 0.05$\,dex) to those derived in \citetalias{Lehner2018} using the same velocity intervals for the integration. For 8\% of these cases (3 absorbers), the new continua led to a larger difference of about 0.10--0.15 dex (the largest value being for an absorber with unresolved saturation that was not assessed as such in \citetalias{Lehner2018}; see below). These continuum readjustments also resulted in improved agreement (all within 1$\sigma$, see below) between the column density measurements derived from the two {\ovi} transitions.

When deriving the kinematics and column densities of the \ovi\ $\lambda\lambda$1031, 1037 doublet, we encounter various scenarios, which we detail below. The simplest case is when {\ovi} $\lambda$1031 is not detected at the $3\sigma$ level. In this case, we adopt the 2$\sigma$ column density upper limit derived from the equivalent width uncertainty, assuming the line lies on the linear portion of the curve of growth, and adopting the same integration range as that of {\hi}. 

In the case where both transitions of the \ovi\ doublet are available, we determine if the absorption is saturated by comparing their integrated column densities and $N_a(v)$ profiles. Saturation can cause the integrated column density derived from the weaker transition to be larger than that derived from the stronger transition. However, contamination can also produce a similar effect. To distinguish between these two possibilities, we visually inspected the profiles to determine if any discrepancy occurs at or near the peak optical depth, where saturation effects are most prominent. We found consistency between the two \ovi\ transitions at the $\approx 1\sigma$ confidence level and column density profiles (see an example in Fig.~\ref{fig:example}), implying no saturation, even for {\ovi} absorbers with \coldenovi $\gtrsim 14.5$ with the exception of one absorber at redshift $z=0.6861$ toward J134100.78+412314.0. For this system, the weaker transition exhibits a marginally higher column density (by 0.05 dex) compared to the stronger transition, although still within 1$\sigma$. The {\nav} profile comparison reveals that this discrepancy is localized in the absorption core, strongly suggesting saturation, which we corrected for using the method outlined in \citet{Savage1991}. 

To identify contamination, we employ a similar procedure, but in this case, either transition may yield a larger integrated column density. The appendix figures illustrate several instances of partial or complete contamination. Some examples of unambiguous contamination include absorbers at $z=0.356924$ toward J035128.56$-$142908.0 and at $z=0.619130$ toward J080908.13+461925.6, where \ovi\ $\lambda$1031 is affected. A more nuanced case is the absorber at $z=0.378447$ toward J122454.44+212246.3, where \ovi\ $\lambda$1037 shows contamination. In this instance, the discrepancy is evident at peak optical depth and persists at low optical depths in the negative velocity wing, where the \nav\ profiles should match (as they do at positive velocities), leading to its classification as contaminated. For the $z = 0.6214$ absorber towards J134100.78+412314.0, both \ovi\ transitions exhibit contamination at different velocities. To address this, we have aggregated the column densities measured over distinct velocity ranges for both transitions, excluding contaminated regions.

Among the 73 detected absorbers in our robust sample (see Section~\ref{sec:detectionrate}), six cases have only the \ovi\ $\lambda$1031 transition. For these instances, we classify a detection as clean when it is neither affected by nor is an interloper after visually inspecting the absorption velocity profile relative to those observed in \hi\ and other ions. For these single-transition systems, we found no evidence of saturation in the {\ovi} absorption, as the peak optical depth consistently remains $\la 1$.

We summarize in Table~\ref{tab:raw-ovi-properties} the results ($\langle v \rangle$, $\Delta v_{90}$, $\log N$) for each transition of the \ovi\ doublet. All listed uncertainties are 1$\sigma$ values while upper limits are given at the $2\sigma$ level. The {\nav} and normalized flux profiles for all the systems with {\ovi} coverage from the CCC are presented as supplementary materials in the appendix.  In Table~\ref{tab:ovi-properties}, we tabulate the adopted {\ovi} measurements for the robust sample. All the listed uncertainties correspond to 1$\sigma$ values. Upper limits are again given at the $2\sigma$ level. When both transitions of the doublet are detected at $>3\sigma$ with no evidence of saturation or contamination (see above), the adopted column densities and kinematics are a weighted mean between the two transitions. The uncertainty on the logarithmic of the column density is obtained following the prescription in~\S2.8 of~\citet{Barlow2003} to account for the possible asymmetric uncertainties in the measurements. In the one case of saturation, we adopt the column density from the weak transition after correcting for saturation following the procedure above; in that case, the kinematics are adopted from the weak transition. If one of the transitions is contaminated, we adopt the results from the uncontaminated transition.

\section{Results}\label{s-results}

Below, we present the empirical properties of {\ovi} absorption associated with {\hi}-selected absorbers whose metallicities have been systematically determined. As previously discussed, we adopt the metallicities of the cool photoionized gas from \citetalias{Lehner2019}. Following the definitions in \citetalias{Wotta2019} and \citetalias{Lehner2019}, we classify absorption systems as low-metallicity (LM) when {\xh} $\leq -1.4$ and high-metallicity (HM) when {\xh} $> -1.4$ (see Section~\ref{s-data}).

\subsection{Frequency of \ovi}\label{sec:detectionrate}

The frequency or detection rate is a sample-dependent heuristic that represents the fraction of detected {\ovi} absorption systems within a given {\hi} absorber sample. We compute these rates assuming a binomial distribution, following the method of \citet{Cameron2011}. This approach assesses the likelihood function for detection rate values, given the number of detections relative to the total number in a chosen subsample. \citeauthor{Cameron2011} demonstrates that the normalized likelihood function used for estimating Bayesian confidence intervals on a binomial distribution with a uniform prior follows a $\beta$ distribution. For a detailed discussion of this method, which is particularly suitable for robust determination of detection rates, we refer the reader to \citet{Cameron2011} (see also \citealt{Howk2017}). We report the median of the distribution along with asymmetric uncertainties corresponding to the 16th and 84th percentiles.

Our robust sample includes 100 SLFSs and 26 pLLSs (see Table~\ref{ovi-sample}). We remind that our robust sample has ${\sn} \geq 8$ per resolution element yielding sensitivity to absorbers with {\coldenovi}$\,\geq 13.6$ at the $2\sigma$ level (see Section~\ref{s-data}); the interquartile range (IQR) of {\sn} is [10, 19] with a median of 12. We summarize the detection rates as a function of {\hi}-selected and metallicity-selected subsamples in Table~\ref{detection_rates}. In the 43 SLFSs with {\ovi} non-detections, the $2\sigma$ upper limits on the {\ovi} column density range from {\coldenovi}$\,<13.6$ to {\coldenovi}$\,< 12.9$; in the 9 pLLSs with {\ovi} non-detections, the $2\sigma$ upper limits on the {\ovi} column density range from {\coldenovi}$\,<13.6$  to {\coldenovi}$\,<12.7$.  The detection rates are therefore determined by treating measured values below the survey sensitivity of {\coldenovi}$\,<13.6$ as non-detections.

\begin{deluxetable}{lcc}
\tabcolsep=10pt
\tablecolumns{3}
\tablewidth{0pc}
\tablecaption{Detections rates of {\ovi} for different subsamples\label{detection_rates}}
\tabletypesize{\normalem}
\tablehead{\colhead{Sample} & \colhead{Size} &\colhead{Detection Rate (\%)}
}
\startdata
SLFSs $+$ pLLSs & 126 & $51 \pm 4$ \\
SLFSs & 100 & $48 \pm 5$ \\
pLLSs & 26 & $61 \pm 9$ \\
LM & 44 & $28_{-6}^{+7}$\\
HM & 82 & $63 \pm 5$ \\
SLFSs--LM & 36 & $21_{-6}^{+7}$\\
SLFSs--HM & 64 & $64 \pm 6$\\
pLLSs--LM & 8 & $61_{-16}^{+15}$\\
pLLSs--HM & 18 & $60 \pm 11$\\
\hline
\enddata
\tablecomments{LM refers to the low-metallicity subsample, and HM refers to the metal-enriched subsample. The median value and the uncertainties of the distribution corresponding to the 16th and 84th percentiles are given. The detection rates are calculated at the survey sensitivity of {\coldenovi} $\geq$ 13.6.}
\end{deluxetable}

The detection rates of {\ovi} in SLFSs and pLLSs are found to be $48 \pm 5$\% and $61 \pm 9$\%, respectively. Although \ovi\ detection rates overlap between  SLFSs and pLLSs, there is a suggestion of a somewhat lower frequency of \ovi\ with {\coldenovi}$\,\geq 13.6$ in SLFSs than in pLLSs (but see below). For comparison, \citet{Fox2013}, who studied the properties of the intermediate ions and high ions in 23 pLLSs, reported a {\ovi} detection rate of $83_{-9}^{+8}$\% in their survey of pLLSs at the $3\sigma$ level, calculated accounting for the detection sensitivity of their survey, consistent with our value for pLLSs within $1.75\sigma$.

We find that the HM sample has a significantly higher ($> 4\sigma$) \ovi\ detection rate of  $63 \pm 5$\% compared to $28_{-6}^{+7}$\% for the LM sample. However, this difference is driven by the low frequency of \ovi\ in LM SLFSs since the detection rates in the HM and LM pLLSs and HM SLFSs are similar $\sim$60\% (see Table~\ref{detection_rates}). While there are only 8 systems in the LM pLLS sample, there is the same balance of HM and LM absorbers in the pLLSs and SLFSs, with $\sim$\,2/3 HM and $\sim$\,1/3 LM. Therefore the high frequency of \ovi\ with LM pLLSs compared to LM SLFSs is unlikely to arise owing to small statistical numbers and the smaller frequency of \ovi\ with  {\coldenovi}$\,\geq 13.6$ in SLFSs than pLLSs is driven by the LM SLFS sample.

\begin{figure}[tbp]
\centering
\includegraphics[scale=0.7]{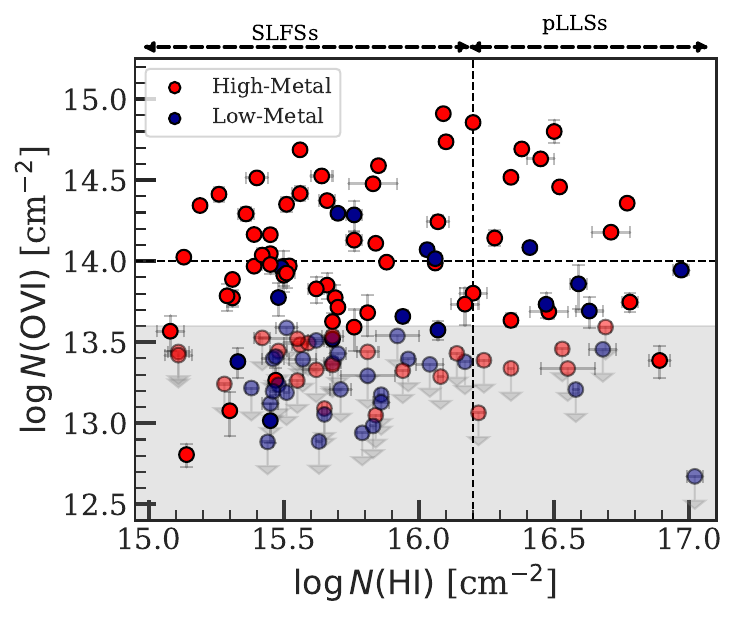}
\caption{{\ovi} column density as a function of the \hi\ column density. The vertical line divides the sample into SLFSs and pLLSs. The horizontal line separates the sample based on {\coldenovi}$\,=14$. The data points are color-coded by the metallicity of lower ionization gas. The HM sample is shown in red and the LM sample is shown in blue. Upper limits on {\ovi} column density measurements are shown as grey downward pointing arrows. The light-shaded region highlights the sensitivity limit of our survey of {\coldenovi}$\,\geq 13.6$.
\label{fig:NOVI_NHI}}
\end{figure}

\begin{figure*}[tbp]
\centering
\includegraphics[scale=0.5]{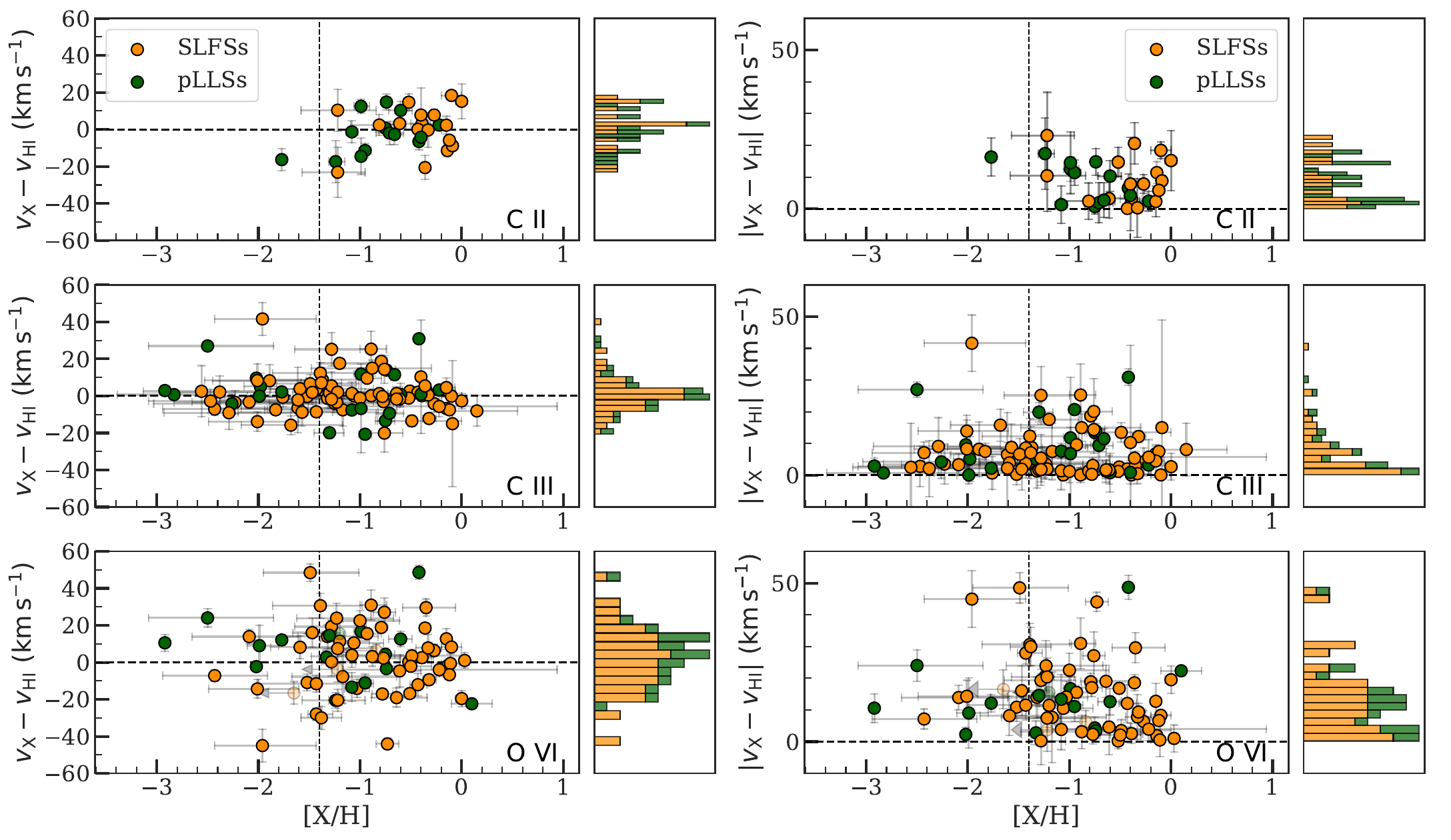}
\caption{Comparison of (\textit{left}) velocity centroid offset  and (\textit{right}) absolute velocity centroid offset, with metallicity for {\cii}, {\ciii}, and \ovi.  The vertical line marks the separation between low metallicity and HM absorbers. The horizontal line marks the mean of the sample. SLFSs are colored in orange, while pLLSs are colored in green. The stacked histograms to the right are presented to highlight the dispersion in the velocity centroid offsets}.\label{fig:dv90_Z}
\end{figure*}

\subsection{{\coldenovi}--{\colden} Relationship}\label{ovi-hi}

In Fig.~\ref{fig:NOVI_NHI}, we plot the relationship between {\coldenovi} and {\colden}, and color code it by the metallicity of the lower-ionization gas. Visually we do not see an apparent correlation between the parameters of {\coldenovi}--{\colden} for either the LM or the HM sample. \citet{Danforth2005}, using an {\ovi}-selected sample, found a very mild relationship between $N($\ovi$)$ and $N($\hi$)$, and fit the relation with a $N($\hi$)^{0.1}$ power law over the {\hi} column density of $13 \lesssim$\,{\colden}\,$\lesssim 16$. We test for a correlation between the {\ovi} and {\hi} column densities by performing the ATS test. This test performs linear regression for bivariate data accounting for the upper limits, and quantifies the null-hypothesis probability. We test the null hypothesis of no correlation between {\coldenovi} and {\colden}. For the robust, HM, and LM samples, we obtain $p$-values $\gg$ 0.05, i.e., the null hypothesis cannot be rejected. In contrast, \citetalias{Lehner2018} found evidence for a significant correlation between the column densities of the singly and doubly ionized species of carbon with {\hi} in the $15 \leq$ {\colden} $\leq 17$ range (see their Figure 11). The strong association between low- and intermediate ions, and {\hi} indicates that these ions likely exist in the same gas phase, and indeed single-phase photoionization models reproduce quite well the column densities of these ions \citepalias{Wotta2019,Lehner2019}. Conversely, the absence of substantial correlation between {\coldenovi} and {\colden} suggests that {\ovi} is likely located in a separate phase from the lower-ionization gas. 

\subsection{{\ovi} Kinematics}\label{ovi-kinematics}
\subsubsection{Average Central Velocities}

\begin{deluxetable}{lcccc}
\tabcolsep=5pt
\tablecolumns{5}
\tablewidth{0pc}
\tablecaption{Summary statistics of \hbox{$|v_{\rm X} - {\vhi}|$} values for different ions\label{table-dv-kinematics}}
\tabletypesize{\normalem}
\tablehead{\colhead{Ion} & \colhead{Sample} & \colhead{IQR} & \colhead{\mean\,$\pm$\,\sdmu} & \colhead{\std}\\
\colhead{(X)} & \colhead{Size} & \colhead{(\kms)} & \colhead{(\kms)} & \colhead{(\kms)}}
\startdata
{\cii}   & 31 & 3--15  & 9 $\pm$ 1 & 7\\
{\ciii}  & 94 & 2--10  &  7 $\pm$ 1 & 8\\
{\ovi}  & 73 & 6--19  &  14 $\pm$ 1 & 11\\
{\cii}--SLFSs  &  17 &  3--15  & 9 $\pm$ 2 & 7\\
{\cii}--pLLSs  &  14 &  2--14  & 8 $\pm$ 2 & 6\\
{\ciii}--SLFSs &  72 &  2--9  &  7 $\pm$ 1 & 7\\
{\ciii}--pLLSs &  22 &  3--13  &  9 $\pm$ 1 & 9\\
{\ovi}--SLFSs &  56 & 6--19  &  14 $\pm$ 1 & 11\\
{\ovi}--pLLSs &  17 & 9--17  &  14 $\pm$ 1 & 11\\
{\ciii}--LM &  32 & 2--8  &  7 $\pm$ 1 & 8\\
{\ciii}--HM & 62 & 2--12  &  8 $\pm$ 1 & 7\\
{\ovi}--LM  & 16 & 10--18  &  17 $\pm$ 1 & 13\\
{\ovi}--HM  & 57 & 4--19  &  13 $\pm$ 1& 11\\
\enddata
\tablecomments{We do not estimate the summary statistics for {\cii}-LM since there is only one data point. } 
\end{deluxetable}

We consider the statistic of kinematic velocity centroid offset, $v_{\rm X} - {\vhi}$, between the absorption component in a metal ion, $X$, from {\hi}.\footnote{For saturation-affected features, we determine the velocity centroid using a treatment similar to that for unsaturated profiles. The primary effect of saturation is to decrease the true weight of the line core velocities relative to the wings in the centroid calculation. In our analysis, {\ciii} is the ion most affected by saturation effects: 37/94 {\ciii}, 1/31 {\cii}, and 1/73 \ovi\ absorbers in our robust sample exhibit saturation. As discussed in Section~\ref{s-intro}, we employ a Monte Carlo sampling approach to account for measurement errors. For saturated absorption features, we modify this approach to mitigate the influence of less robustly constrained velocity centroids. Instead of drawing from a normal distribution, we sample from a uniform distribution ranging between the minimum and maximum values corresponding to one standard deviation below and above the mean velocity centroid value. This uniform distribution sampling for saturated absorbers aims to capture a broader range of possible centroid values and reduce the impact of less reliable saturated core velocities on our statistical analysis.} It provides information on how absorbing ions are distributed in velocity space within each absorption system.\footnote{\citetalias{Lehner2018} found that the absolute COS wavelength calibration to be accurate within one COS resolution element ($\approx 15$ \kms; see also \citealt{Holland2012}) using the Lyman series transitions. Consequently, any velocity shifts larger than $\ga 15$ \kms\ likely represent real velocity differences rather than data artefacts.} Such velocity offsets can indicate a multiphase structure of the absorption system~\citep{Fox2013}. \citetalias{Savage2014} noted that velocity centroid offsets between {\ovi} and {\lya} absorption components, $|{\vovi} - {\vhi}|>10$\,\kms, often indicate that these ions reside in separate phases based on a detailed comparison of the \hi\ and \ovi\ velocity profiles observed in the highest--S/N {\it HST}/COS G130M/G160M spectra.

In the left panels of Fig.~\ref{fig:dv90_Z}, we plot the velocity centroid offset, $v_{\rm X} - {\vhi}$, of low ion (\cii), intermediate ion (\ciii), and high ion (\ovi) species, as a function of metallicity. To the right of each scatter plot, we present a histogram showing the distribution of velocity centroid offsets for the corresponding ion. The {\cii} histogram displays a relatively flat distribution in the range of $[-20,20]$ {\kms}, though the sample size is comparatively small. For {\ciii}, the histogram exhibits a symmetric distribution centered near zero {\kms}, with a pronounced peak at $\approx0$\,\kms and fewer data points extending to large positive or negative offsets. In contrast, the {\ovi} histogram reveals a much broader distribution compared to {\cii} and {\ciii}. While its peak is close to $\approx0$\,\kms, the distribution shows considerable dispersion, with many data points well beyond $\pm20$\,\kms---far exceeding offsets that could be attributed to COS wavelength calibration.

In Fig.~\ref{fig:dv90_Z} (right), we show the absolute velocity centroid offsets, $|v_{\rm X} - {\vhi}|$, and their distributions for {\cii}, {\ciii}, and {\ovi}. The IQR, mean, and 1$\sigma$  dispersion values for $|v_{\rm X} - {\vhi}|$ are tabulated in Table~\ref{table-dv-kinematics}. \citet{Fox2013} in their sample of pLLSs find the absolute velocity centroid offsets to be  $6\,\pm\,1$ {\kms}, $10\,\pm\,2$  {\kms} and $14\,\pm\,4$ {\kms} for {\cii}, {\ciii} and {\ovi} ions, respectively, consistent with our findings for pLLSs in our sample. We note that there is a large dispersion in the velocity centroid offsets for the various ions as tabulated in the last column of Table~\ref{table-dv-kinematics}.

To investigate any differences in the $|v_{\rm X} - {\vhi}|$ distributions of the LM and HM groupings of {\ciii} and {\ovi}, we perform a two-sample Anderson-Darling (AD) test for the null hypothesis that the two samples are identical. AD test is a non-parametric test and can be more powerful than the  KS test for smaller sample sizes ($\lesssim\,25$)~\citep{epps}. We obtain $p$-values $\gg 0.05$ for both {\ciii} and {\ovi}, implying that the null hypothesis cannot be rejected.  We do not perform the AD test for {\cii} because of only one data point in LM grouping for {\cii}.

We also perform the AD test to compare the $|v_{\rm X} - {\vhi}|$ distributions of low, intermediate, and high ions irrespective of the metallicity. We find a statistically significant difference in the $|v_{\rm X} - {\vhi}|$ distributions of {\ciii} compared with {\ovi} with $p$-value of 0.001 $\pm$ 0.001. But, {\cii} does not show a statistically significant difference compared with {\ovi} ($p$-value = 0.15 $\pm$ 0.18). {\cii} also does not a statistically significant difference in the velocity centroid offsets compared with {\ciii} ($p$-value $\gg$ 0.05).  We note that the results for {\cii} could be affected by the small sample sizes and, hence, may not be statistically meaningful. These results suggest that the intermediate ion of {\ciii} and {\ovi} do not always arise in a single phase and could imply a multiphase medium. We further compare the $|v_{\rm X} - {\vhi}|$ distributions of SLFSs and pLLSs for each of the ions of {\cii}, {\ciii} and {\ovi}, but do not find evidence for a statistically significant difference in their distributions with $p$-values $\gg 0.05$.

\subsubsection{Velocity Widths}
\begin{figure}[htbp!]
\epsscale{1.}
\plotone{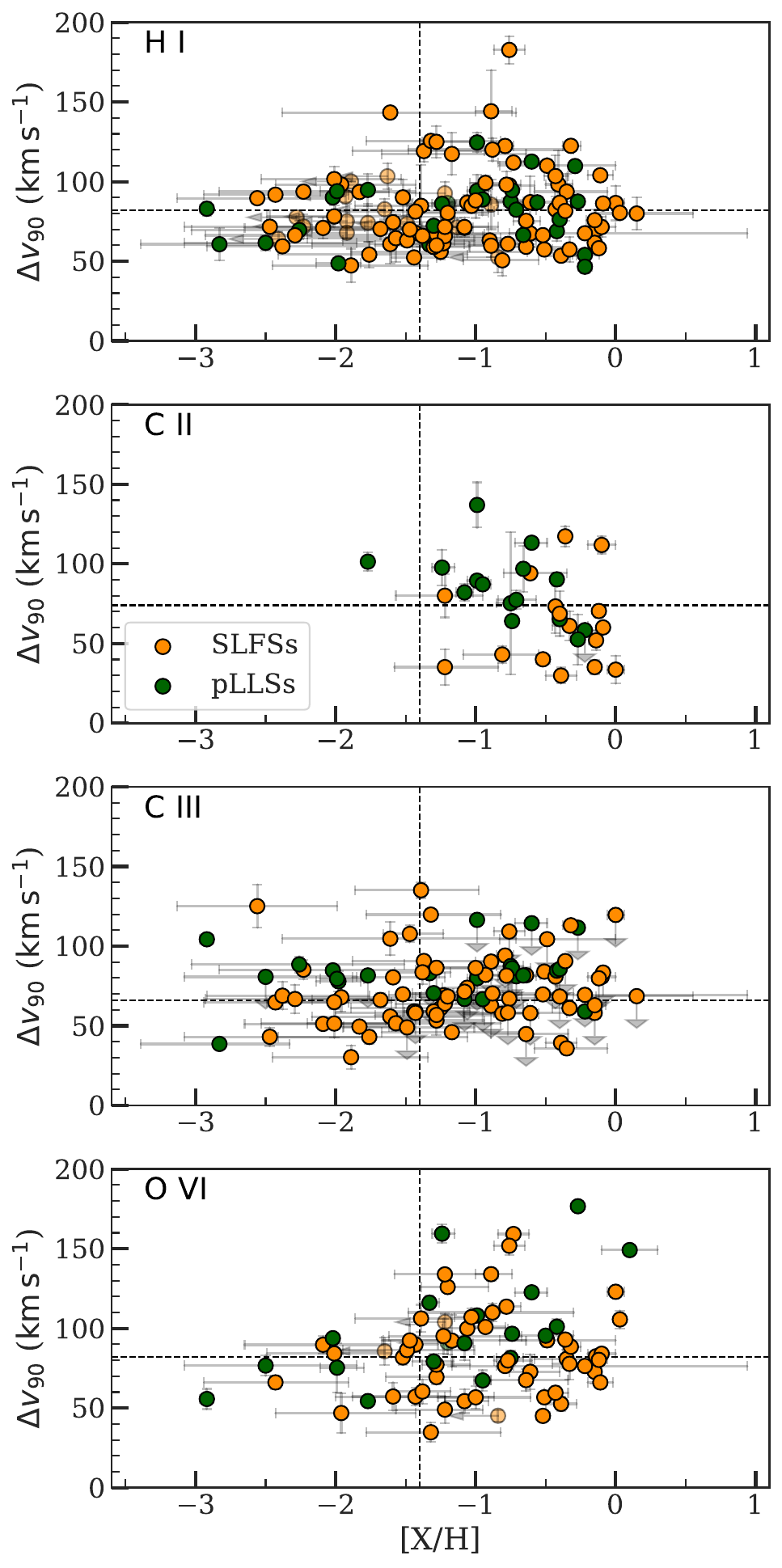}
\caption{Comparison of velocity width, $\Delta v_{90}$, with metallicity for {\hi}, {\cii}, {\ciii}, and \ovi. The vertical line marks the separation between LM and HM absorbers. The horizontal line marks the mean value of velocity width. SLFSs are colored in orange, while pLLSs are colored in green. Saturated absorption is seen mainly in {\ciii} and is presented as upper limits. \label{fig:Deltav90_Z}}
\end{figure}

Another indicator of kinematic differences is the velocity width, $\Delta v_{90}$ (see Section~\ref{s-measurements}). We measure it for four species: {\hi}, {\cii}, {\ciii}, and {\ovi}. In Fig.~\ref{fig:Deltav90_Z}, we plot the velocity width, $\Delta v_{90}$, as a function of metallicity. The IQR, mean, and $1\sigma$ dispersion values for $\Delta v_{90}$ are tabulated in Table~\ref{table-Dv90-kinematics}. To handle the measurements associated with saturated {\ciii} absorbers, the velocity width was randomly drawn from a uniform distribution spanning the range from $30$\,\kms\ to the upper limit value when generating 1000 replicated datasets. We chose the lower bound of 30 \kms\ because it is about the smallest velocity width observed for unsaturated {\ciii} absorption in our sample. The {\ovi} sample exhibits a noticeably broader velocity width when compared to the low and intermediate metal ions.

\begin{deluxetable}{lcccc}
\tabcolsep=5pt
\tablecolumns{5}
\tablewidth{0pc}
\tablecaption{Summary statistics of $\Delta v_{90}$ values for different ions\label{table-Dv90-kinematics}}
\tabletypesize{\normalem}
\tablehead{\colhead{Atom/Ion} & \colhead{Sample} & \colhead{IQR} & \colhead{\mean\,$\pm$\,\sdmu} & \colhead{\std}\\
\colhead{} & \colhead{Size} & \colhead{(\kms)} & \colhead{(\kms)} & \colhead{(\kms)}}
\startdata
{\hi}   & 126 & 66--93  & 82 $\pm$ 1 & 22\\
{\cii}   & 31 & 56--92  & 74 $\pm$ 2 & 27\\
{\ciii}  & 94 & 61--89  &  67 $\pm$ 1 & 21\\
{\ovi}  & 73 & 67--101  &  89 $\pm$ 1 & 32\\
{\hi}--LM  & 44 &  66--90 &  78 $\pm$ 1 & 18\\
{\hi}--HM  & 82 &  66--95 &  84 $\pm$ 1 & 24\\
{\ciii}--LM  & 32 &  53--83 &  66 $\pm$ 1 & 22\\
{\ciii}--HM  & 62 &  65--91 &  67 $\pm$ 2 & 26\\
{\ovi}--LM  & 16 &  57--87 &  75 $\pm$ 2 & 16\\
{\ovi}--HM  & 57 &  73--107 &  92 $\pm$ 1 & 31\\
\enddata
\tablecomments{We do not estimate the summary statistics for {\cii}-LM since there is only one data point. The IQR for {\ciii} is determined by treating the upper limits on $\Delta v_{90}$ as measured values.} 
\end{deluxetable}

We observe that the mean value of $\Delta v_{90}$ for the {\ovi} HM sample is greater than that of {\ovi} LM sample. We perform the MWU test, a non-parametric test, to compare the {\ovi} $\Delta v_{90}$ distribution associated with HM lower-ionization gas and the {\ovi} $\Delta v_{90}$ distribution associated with LM lower-ionization gas. This statistical test is used to determine if there is a significant difference between the average values of two distributions. We find that {\ovi} $\Delta v_{90}$ distribution associated with HM population is statistically greater than the {\ovi} $\Delta v_{90}$ distribution associated with LM population ($p$-value\,$= 0.03\,\pm\,0.02$). We do not find evidence for a statistically significant difference in the $\Delta v_{90}$ distributions of LM and HM gas for {\hi} and {\ciii} with $p$-values $\gg$ 0.05.
 
The primary conclusion from our kinematic analysis is that low and intermediate ions exhibit relatively smaller centroid offsets and line widths compared to the {\ovi} ion. From these results, we draw two main inferences: i) the lower-ionization gas and {\ovi} do not always trace the same plasma, and ii) there are distinct differences in the kinematic properties of {\ovi} associated with low-metallicity (LM) and high-metallicity (HM) lower-ionization gas.

\begin{figure}[htbp!]
\includegraphics[scale=0.625]{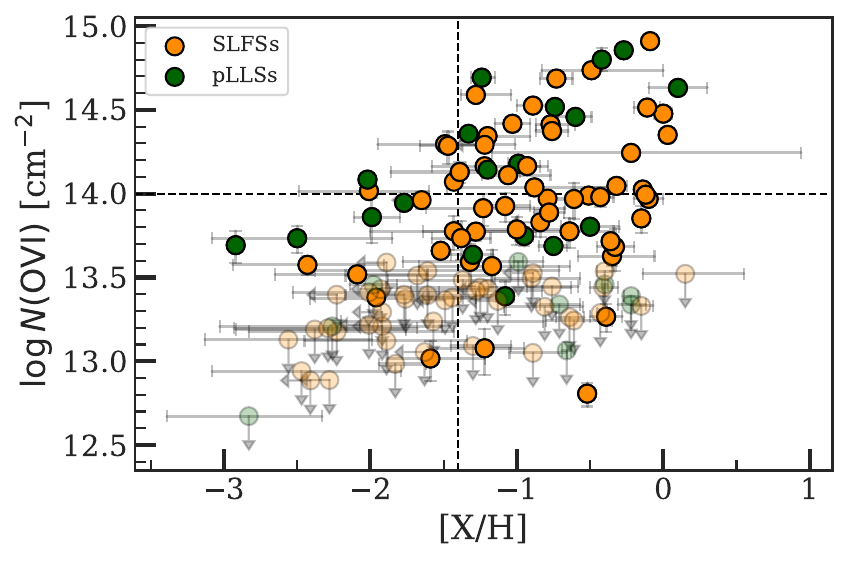}
\caption{The {\ovi} column density as a function of metallicity for our sample. {\ovi} detections are shown as dark-filled circles, the non-detections (upper limits) are shown as filled circles with attached downward-pointing arrows and shaded in a lighter tone. The vertical line divides the sample into LM and HM subsamples. The horizontal line at {\coldenovi}\,$=14$ divides the sample based on the {\ovi} absorption strength. The data points are color-coded to differentiate SLFSs (orange) and pLLSs (green). \label{fig:logNOVI_Z}}
\end{figure}

\subsection{{\ovi} Column Densities}\label{sec:columns}

We measure {\ovi} absorption at a significance level of 3$\sigma$ or above in 73 of the 126 associated {\hi} components. Our upper limits for non-detected absorption are reported at the 2$\sigma$ level\footnote{To determine if a line is detected or not detected, we compare the measured equivalent width within the velocity range of integration, $EW$,  to the total uncertainty on the equivalent width within the velocity range of integration, $\rm EW_{\rm err}$, (accounting for both the statistical uncertainty and continuum uncertainty). If the condition $\rm EW \geq 3\,\rm EW_{\rm err}$ is met, we consider the line to be detected at the 3$\sigma$ level. If not, we report the linear column density measurement corresponding to 2$\rm EW_{\rm err}$ as the upper limit value.}. The IQR of the {\coldenovi} in our robust sample is 12.81--14.00, accounting for the upper limits. The mean of the {\ovi} column density distribution and the standard error on the mean are $\meancoldenovi = 13.98 \pm 0.01$, estimated using the KM estimator accounting for the upper limits.\footnote{Hereafter, we use the notation of {\meancoldenovi} and {\geomeancoldenovi} to imply the mean of the {\ovi} column density and the error on the mean, and the geometric mean of the {\ovi} column density and the error on the geometric mean, respectively.} The geometric mean and the standard error on the geometric mean are $\geomeancoldenovi = 13.52 \pm 0.02$. Considering just the detections, we obtain mean and geometric mean values of $\meancoldenovi = 14.21 \pm 0.01$ and $\geomeancoldenovi = 14.01 \pm 0.01$. 

Motivated by the inference from the {\ovi} kinematics that the {\ovi} and lower ionization gas likely reside in different gas phases, we next investigate the trend between the {\ovi} column density and metallicity. We would expect a correlation if the properties of the different phases are linked. In Fig.~\ref{fig:logNOVI_Z}, we plot the column density of {\ovi} as a function of the cool gas metallicity derived in \citetalias{Lehner2019}. On average, the HM systems ($\xh>-1.4$) show stronger {\ovi} absorption when compared to the LM systems. We find that when {\coldenovi} $\geq$ 14, only $16_{-5}^{+7}$\% of the systems are associated with LM gas. But strikingly, $84_{-7}^{+5}$\% of the systems with  {\coldenovi} $\geq$ 14 are associated with HM gas. The IQR for the HM and LM subsamples spans between 13.08--14.24 dex and 12.89--13.68 dex, respectively.  The geometric means of \ovi\ column density for HM and LM subsamples are $\geomeancoldenovi = 13.73 \pm 0.02 $ and $13.22 \pm 0.05$, respectively. The means of \ovi\ column density for HM and LM subsamples are $\meancoldenovi = 14.11 \pm 0.01 $ and $13.46 \pm 0.02$, respectively.

We observe a positive correlation between the {\ovi} column density and metallicity of the cool gas. We perform the ATS test to investigate the null hypothesis of no correlation between these parameters, and find a $p$-value\,$= (0.6 \pm 0.9)\times10^{-3}$, rejecting the null hypothesis. This suggests that the {\ovi} column density and the metallicity of lower-ionization gas are correlated. The correlation between the column density of {\ovi} and the metallicity of the cool photoionized gas is remarkable, given that the {\ovi} and low-ionization gas were found to show statistically significant differences in kinematics and reside in different gas phases.
 
\citet{Fox2013} found similarly a trend in which metal-rich pLLSs showed higher {\ovi} columns than metal-poor pLLSs. Consistent with their finding, we determine that the mean {\coldenovi} for HM pLLSs is higher than that of LM pLLSs with mean values of $\meancoldenovi = 14.20 \pm 0.02$ and  $13.67 \pm 0.03$, respectively. The geometric mean for the HM pLLSs and LM pLLSs are $\geomeancoldenovi = 13.87 \pm 0.03$ and $13.57 \pm 0.03$, respectively. We find that the \ovi\ frequency with {\coldenovi}$\,\geq 14$ in LM pLLSs is $18_{-10}^{+14}$\% compared with $45 \pm 11$\% in HM pLLSs. 

\begin{figure}[tbp]
\epsscale{1.0}
\plotone{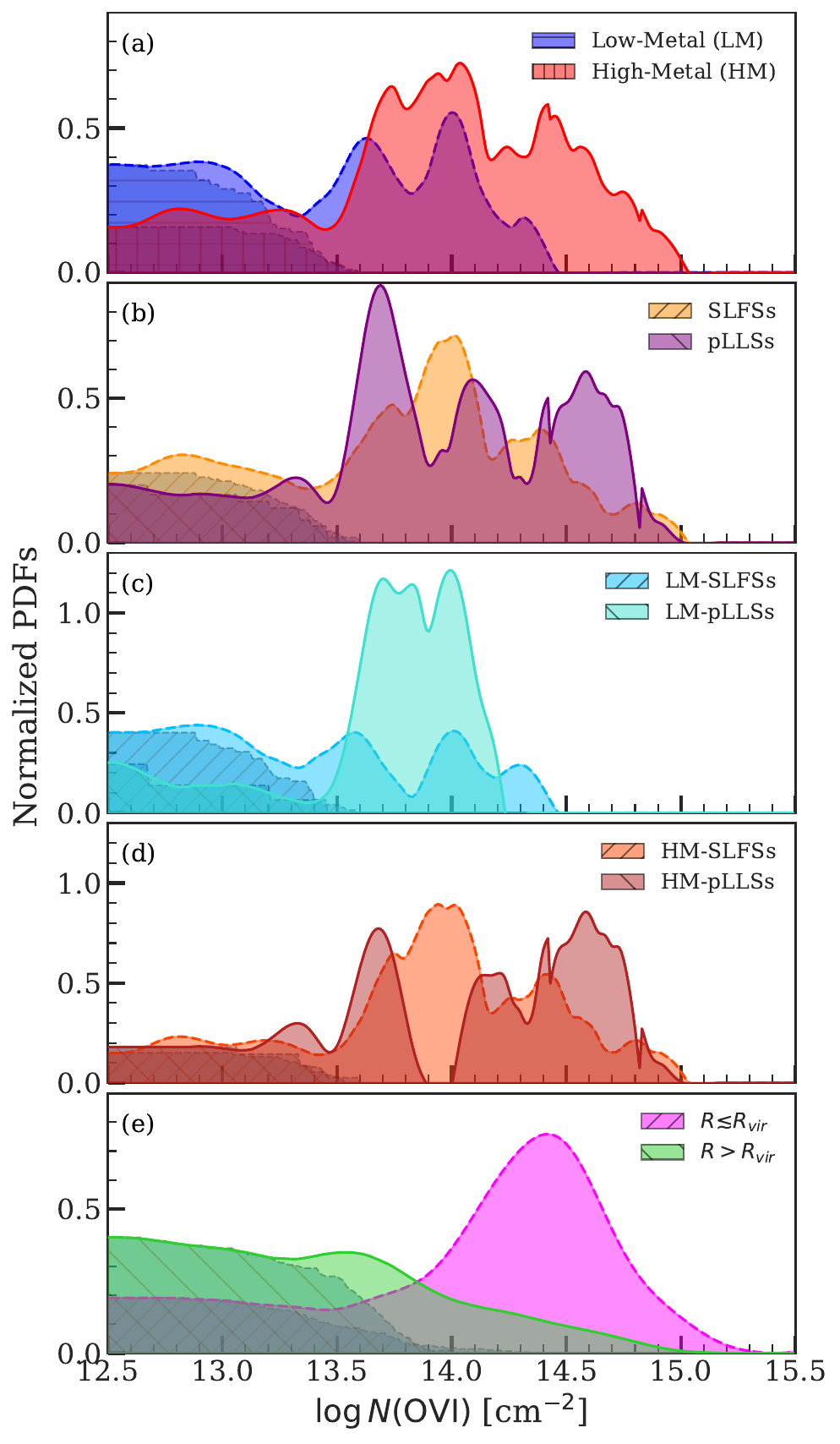}
\caption{Normalized PDFs of {\ovi} column density for the (a) LM subsample and the HM subsample; (b) for the SLFSs and pLLSs sub-samples; (c) for the LM-SLFSs and LM-pLLSs sub-samples; (d) for the HM-SLFSs and HM-pLLSs sub-samples; and (e) for absorbers within $\la \rvir$ (103 absorbers) and absorbers beyond $>\rvir$ (146 absorbers) taken from \citetalias{Tchernyshyov2022}. The contribution from the lower limits is shown with hatchings. \label{fig:logNOVI_Z_NHI_histograms}}
\end{figure}

So far, we have considered the discrete distributions of the {\ovi} column density; in Fig.~\ref{fig:logNOVI_Z_NHI_histograms}, we display the normalized PDFs of {\ovi} column density for various subsamples. The detections are modeled by approximating them as a Fechner distribution accounting for asymmetric uncertainties in the {\ovi} column density measurements. The upper limits are modeled as uniform distributions spanning a {\coldenovi} range of 11.5 dex to the estimated upper limit value. Our choice of the lower bound of {\coldenovi} $=11.5$ is motivated by the lowest {\ovi} column densities predicted by \textsc{EAGLE} cosmological simulations~\citep{Oppenheimer2016}. Figure 2 in their paper shows that the predicted {\ovi} column densities extend down to {\coldenovi} $\approx 11.5$ for halos with masses between $10^{11.1} M_{\odot} < M_{200} < 10^{13.2} M_{\odot}$, with less prevalence of lower column density {\ovi} gas at higher halo masses.

In Fig.~\ref{fig:logNOVI_Z_NHI_histograms}a, we show the \coldenovi\ distributions of LM and HM lower-ionization gas. The LM PDF has two main peaks at {\coldenovi}\,$\approx 13.6$ and 14.0, and an important tail at {\coldenovi}\,$<13.2$ dex; the detection rate for this subsample is only $28 \pm 7$\%, which is driven by the SLFS sample. The PDF of the HM subsample shows two apparent peaks at {\coldenovi}\,$\approx 13.9$ and 14.4, and a relatively weaker tail extending to {\coldenovi}\,$\lesssim 13.5$. Visually, the  \coldenovi\ distributions of the LM and HM absorbers are quite different. Statistically, we compare the \coldenovi\ distributions of the LM and HM groupings to test the null hypothesis that they belong to the same underlying parent population, while accounting for the upper limits using the log-rank test (implemented using \textit{survdiff} package in R language). The log-rank test does not assume the normality of the distributions and is a non-parametric test. We find a statistically significant difference in the distributions of LM and HM subsamples ($p$-value\,$= (4 \pm 1) \times 10^{-4}$). This result implies that the column densities of {\ovi} absorbers arising in LM and HM systems are significantly different. As a reminder, we also found statistically significant differences ($p$-value $= 0.03 \pm 0.02$) in {\ovi} velocity widths between LM and HM subsamples in Section~\ref{ovi-kinematics}. 
 
We compare the \coldenovi\ distributions for the SLFSs and pLLSs in our sample in Fig.~\ref{fig:logNOVI_Z_NHI_histograms}b. We observe that the SLFSs show a main peak at \coldenovi$\,\approx 14.0$ with a broad distribution below and above this value and several smaller peaks. The pLLSs, on the other hand, show three peaks - a dominant one and two smaller peaks at \coldenovi\,$\approx 13.7$, 14.1, and 14.6, respectively. We again perform the log-rank test for the null hypothesis that SLFSs and pLLSs are probing the same parent population of {\ovi} absorbers but cannot reject the null hypothesis ($p$-value $\gg$ 0.05).

\begin{deluxetable*}{lcccc}
\tabcolsep=5pt
\tablecolumns{5}
\tablewidth{0pc}
\tablecaption{$p$-values from log-rank tests comparing {\ovi} column density distributions: metallicity-selected and {\hi}-selected groups vs. absorbers within ($R \la \rvir$) and beyond ($R > \rvir$) the virial radius of associated galaxies. \label{t-2samp-abstype_galaxy}}
\tabletypesize{\normalem}
\tablehead{\colhead{Environment} & \colhead{LM-SLFSs} & \colhead{HM-SLFSs} & \colhead{LM-pLLSs} & \colhead{HM-pLLSs}}
\startdata
$R \la \rvir$   & (12 $\pm$ 1)$\times 10^{-5}$ & 0.32 $\pm$ 0.04 & 0.47 $\pm$ 0.03 & 0.78 $\pm$ 0.02\\
$R > \rvir$  & 0.91 $\pm$ 0.03  &  (2 $\pm$ 1) $\times 10^{-11}$ & 0.002 $\pm$ 0.001 & (3 $\pm$ 1)$\times 10^{-5}$\\
\enddata
\tablecomments{For the log-rank test, if $p\ge 0.05$, then the null-hypothesis is retained and both samples have identical distributions; if $p<0.05$, then the null-hypothesis is rejected and both samples have different distributions.} 
\end{deluxetable*}

\citet{Berg2023} found that the HM pLLSs are typically associated with galaxies, whereas LM pLLSs trace more diverse locations, including the IGM at the periphery of galaxies. To explore the relationship of {\ovi} linked with LM and HM lower-ionization gas with the environments, we divide our sample into four groups based on metallicity-selection and {\hi}-selection - LM-SLFSs, LM-pLLSs, HM-SLFSs, and HM-pLLSs. Figs.~\ref{fig:logNOVI_Z_NHI_histograms}c and \ref{fig:logNOVI_Z_NHI_histograms}d show these four groupings. The {\ovi} column density distribution of LM-SLFSs shows two small peaks at {\coldenovi} $\approx$ 13.5 and 14, and it is dominated by upper-limits extending to lower column densities. This subsample has a detection rate of only $21 \pm 6$\% (see Table~\ref{detection_rates}). The {\ovi} column density distribution of LM-pLLSs shows a broad peak centered at \coldenovi\,$\approx 13.9$, with a weaker tail arising from the upper-limits extending to lower column densities. The {\ovi} column density distribution of HM-SLFSs shows a peak centered at \coldenovi\,$\approx 13.9$, with contribution from {\ovi} column densities in the range \coldenovi\,$\approx$ 13.9--14.7, and a weaker tail arising from the upper-limits extending to lower column densities.  The {\ovi} column density distribution of HM-pLLSs shows two peaks centered at \coldenovi\,$\approx 13.8$ and 14.5, and a relatively smaller peak at \coldenovi\,$\approx 14.2$, and a weaker tail arising from the upper-limits extending to lower column densities. The LM-pLLSs, HM-SLFSs, and HM-pLLSs, all have similar detection rates of $\approx$ 60\%. However, Figs.~\ref{fig:logNOVI_Z_NHI_histograms}c and \ref{fig:logNOVI_Z_NHI_histograms}d show that LM-pLLSs have {\ovi} column densities well under {\coldenovi}\,$\la 14.2$, while the {\ovi} column density PDF of HM-SLFSs and HM-pLLSs show significant contributions from {\coldenovi}\,$\geq 14.2$. 

We finally compare the {\ovi} column density distributions of the above four groupings to the {\ovi} column density distribution of absorbers within the virial radius ($\la \rvir$) of their associated galaxy and those beyond $>\rvir$ from \citetalias{Tchernyshyov2022}. The {\ovi} column density distributions of absorbers from \citetalias{Tchernyshyov2022} segregated by {\rvir} is shown in  Fig.~\ref{fig:logNOVI_Z_NHI_histograms}e. The {\ovi} absorbers within $\la \rvir$ have an IQR that spans 13.36--14.44 dex, and a higher mean {\ovi} column density,  $\meancoldenovi = 14.27 \pm 0.05$ ($\geomeancoldenovi = 13.96 \pm 0.06$) than that of the $> \rvir$ population for the which the IQR spans 13.10--13.47 dex and has a mean of $\meancoldenovi = 13.67 \pm 0.07$ ($\geomeancoldenovi = 13.41 \pm 0.07$). The {\ovi} column density PDF of the absorbers at $R \la \rvir$ from galaxies shows a strongly peaked mode at {\coldenovi}\,$\approx 14.4$, with a long tail extending to lower column densities. The {\ovi} column density PDF of absorbers at $R > \rvir$ from galaxies does not show any particular peak and is dominated by non-detections. To gain further insights into the association between {\ovi} gas and the environments it may be tracing, we perform the log-rank to compare the {\coldenovi} distributions associated with metallicity-selected and {\hi}-selected groupings with {\coldenovi} distributions of {\ovi} absorbers segregated based on \rvir. We tabulated the $p$-values from the log-rank test in Table~\ref{t-2samp-abstype_galaxy}. Based on the $p$-values, we find that the {\ovi} associated with LM-SLFSs is significantly different from the {\ovi} arising within the $\rvir$ of associated galaxies ($p$-value\,$ = (12 \pm 1) \times 10^{-5}$), but no significant differences are seen between {\ovi} associated with LM-pLLSs and HM lower ionization gas, and {\ovi} arising within $\rvir$ of associated galaxies. On the other hand, {\ovi} associated with LM-pLLSs and HM lower ionization gas is found to be significantly different from the {\ovi} arising beyond $\rvir$ of associated galaxies, but no significant differences are seen between {\ovi} associated with LM-SLFSs and {\ovi} beyond $\rvir$ of associated galaxies. 

\begin{figure}[htbp!]
\includegraphics[scale=0.65]{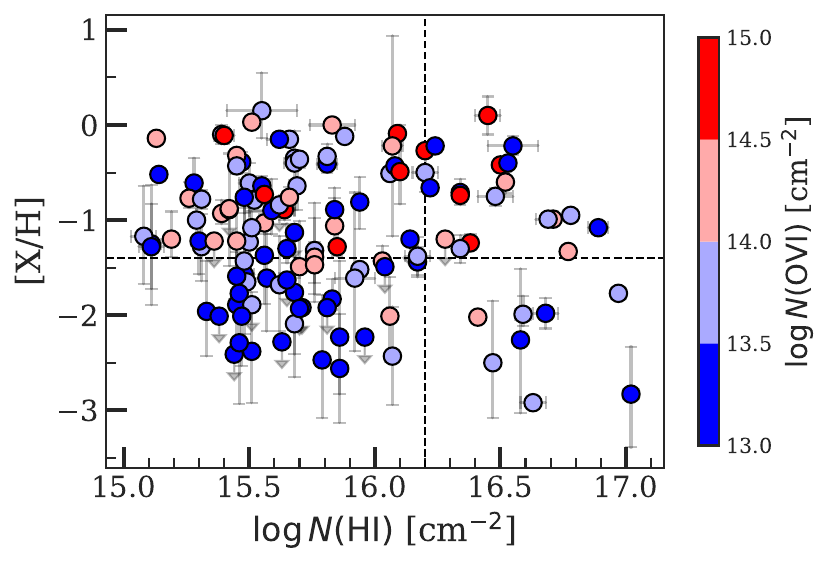}
\caption{The metallicity of the cool gas as a function of \hi\ column density for our sample. The data points are colored by their {\coldenovi}. The horizontal line divides the sample into low-metal and high-metal subsamples. The vertical line divides the sample into SLFSs and pLLSs. About 80\% of the absorbers with {\coldenovi}\,$< 13.6$ have no \ovi\ detection.\label{fig:logNOVI_logNHI_cbarOVI}}
\end{figure}

\begin{figure*}[tbp]
\centering
\includegraphics[scale=0.525]{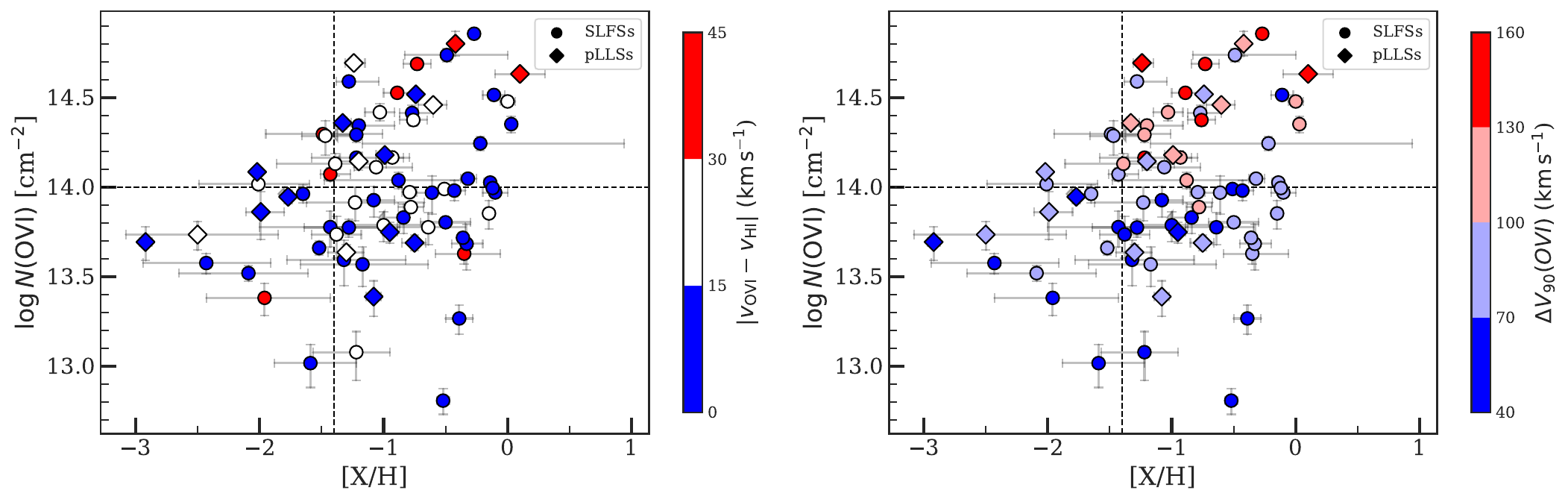}
\caption{The {\ovi} column density as a function of metallicity of the cool gas for absorbers with detected \ovi\ in our sample colored by their kinematic properties. {\it Left:}\ colored by absolute velocity centroid offset, $|{\vovi} - {\vhi}|$; {\it right:}\  velocity width, $\Delta v_{90} (\ovi)$. \label{fig:logNOVI_Z_cbar}}
\end{figure*}

\section{Discussion}\label{s-discussion}
\subsection{Preamble}
We have studied a sample of 126 {\hi}-selected absorbers at $z \lesssim 1$  from the CCC survey with {\ovi} $\lambda$1031 and/or $\lambda$1037 coverage in COS G130M/G160M spectra with $\sn \ga 8$. The {\hi}-selected absorbers span the \hi\ column density range $15 \lesssim$\,{\colden}\,$\lesssim 17.2$. This \hi\ column density range allows for these absorbers to be sensitive to both HM and LM gas at $z\la 1$ \citepalias{Lehner2018}, so this sample has no bias with regard to the metallicity of the cool gas. There is also no bias with regard to the strength of the \ovi\ absorption down to the sensitivity level of our survey, {\coldenovi}\,$\la 13.6$. This contrasts remarkably from blind \ovi\ surveys, which requires detection of \ovi\ and probe mostly absorbers with $13 \la$\,{\colden}\,$\la 14.5$ (e.g., \citealt{Danforth2005,Tripp2008,Thom2008b}; \citetalias{Savage2014}). The \hi-selection is also different from galaxy-centric surveys (e.g., \citealt{Tumlinson2011,Werk2012,Stocke2013,Johnson2015,Keeney2017,Wilde2021, Tchernyshyov2023}) because even though our survey probes overdensities of $\delta \approx 15$--200 at $z\la 1$ (using $N($\hi)$ \propto (1 + \delta)^{1.5}$ where $\delta \equiv n_{\rm H}/\bar{n}_{\rm H}$ is the overdensity, see \citealt{Schaye2001}), the \hi-selection is agnostic in terms of the type and mass of the galaxies or the environments that these absorbers may probe (e.g., denser regions of the IGM instead of the CGM, intragroup gas, etc.). Therefore, the \hi-selection provides a different perspective relative to blind \ovi\ or galaxy-centric surveys. It is instructive therefore to compare our results to these surveys to understand the origin(s) of these absorbers.

\begin{figure}[htbp!]
\includegraphics[scale=0.5]{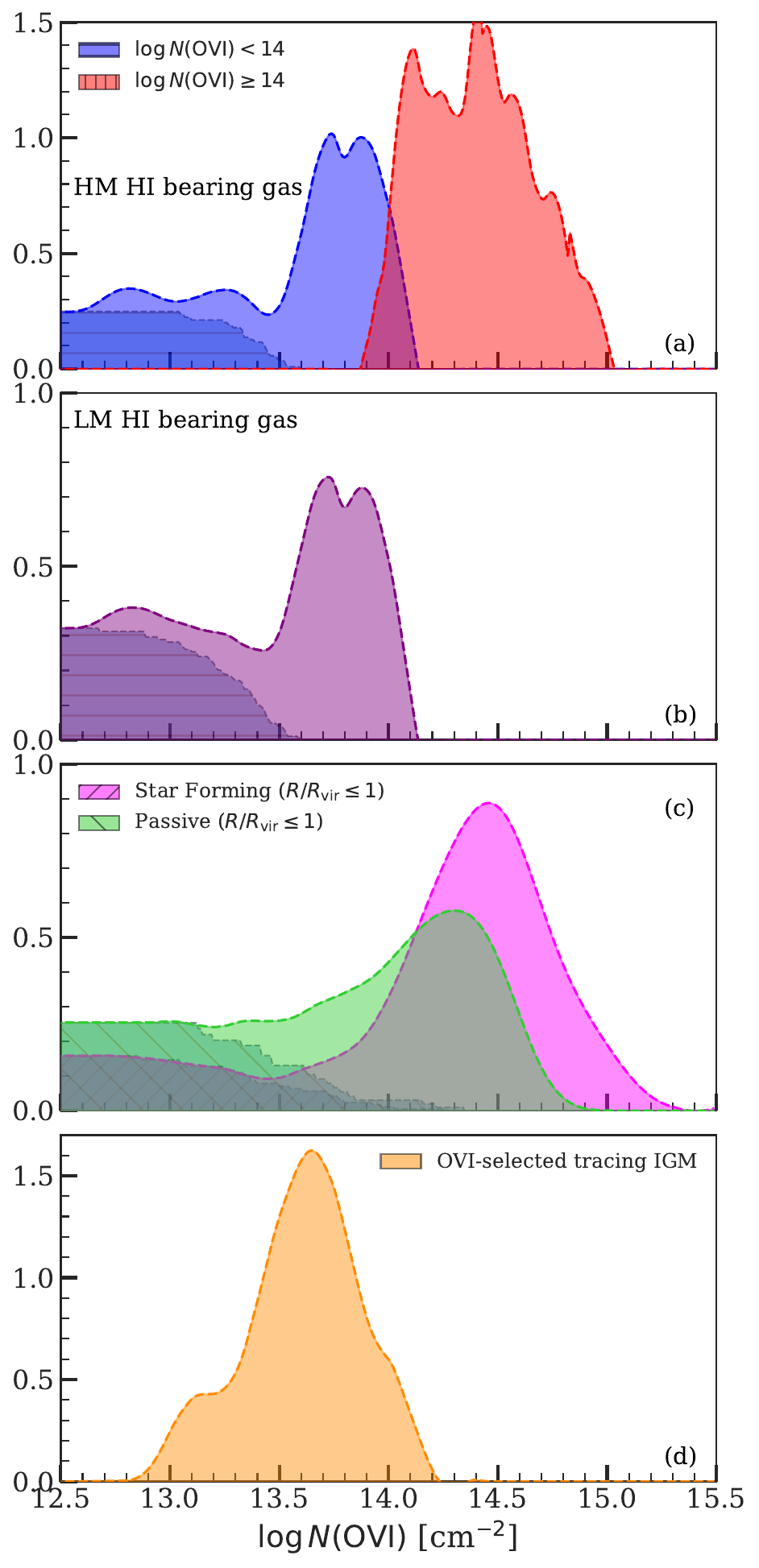}
\caption{Normalized PDFs of {\ovi} column density for the (a) HM lower-ionization gas comprising of strong {\ovi} absorbers, {\coldenovi} $\geq$ 14 (in red) and weaker {\ovi} absorbers, {\coldenovi} $<$ 14 (in blue); (b) LM lower-ionization gas excluding the one data point with {\coldenovi} $>$ 14; (c) star-forming (67 in number) and passive (36 in number) galaxies from \citetalias{Tchernyshyov2022} within {\rvir}; d) {\ovi}-selected absorbers (22 in number) extending 200 to 600 kpc beyond the closest associated galaxies from the survey of \citetalias{Savage2014}. The contribution from the upper limits are shown with hatchings.\label{fig:logNOVI_sf_q_hi_strong_weak}}
\end{figure}

From Section~\ref{s-results}, we categorize the \ovi\ absorbers into: (1) non-detections with {\coldenovi}$\,< 13.6$; (2) detections with {\coldenovi}$\,\lesssim 14$; and (3) detections with {\coldenovi}$\,\gtrsim 14$. To motivate this, Fig.~\ref{fig:logNOVI_logNHI_cbarOVI} shows the \xh--$N($\hi$)$ distribution colored by {\coldenovi}. A majority ($84_{-7}^{+5}$\%) of the absorbers with {\coldenovi}\,$\ge 14$ have high metallicities (100\% for {\coldenovi}\,$\ga 14.5$), while those with {\coldenovi}\,$< 14$ span all metallicities and \hi\ columns. As Fig.~\ref{fig:logNOVI_Z_cbar} shows, category (3) absorbers tend to be also broader ($\Delta v_{90} \ga 100$ \kms) and/or have large \ovi--\hi\ velocity offsets ($|{\vovi} - {\vhi}|  \gtrsim 20$ \kms), possibly suggesting a galaxy feedback origin for these (see Section~\ref{s-ovi-strong}). 

\subsection{\hi-Selected Absorbers with no \ovi: Single-Phase Gas}\label{s-noovi}

While blind \ovi\ surveys found very little evidence of \ovi\ without associated \hi\ (e.g., \citealt{Thom2008a,Savage2014,Keeney2017}), $\sim$$40\%$ of our \hi-selected sample has no \ovi\ detected down to {\coldenovi}\,$\la 13.6$ (and even {\coldenovi}\,$\la 13$ in several cases). Although previous blind surveys were sensitive to {\coldenovi}\,$\ga 12.8$--13.1, most \ovi\ systems are found with {\coldenovi}\,$\ga 13.5$ (e.g., \citealt{Tripp2008}; \citetalias{Savage2014}). Hence, the lack of \ovi\ absorption in our sample indicates a real deficit of highly ionized gas. Indeed, using spectra from \citetalias{Lehner2018}, these absorbers show no other high ions (e.g., \neviii), and all remaining ions (excluding N and Fe ions, which were not typically included in these models) can be modeled by a single photoionized phase \citepalias{Lehner2019}.

The absence of {\ovi} absorption in $\sim 40\%$ of absorbers therefore implies that these systems may probe a single-phase gas, in contrast with the frequent presence of multiphase gas in CGM absorbers (e.g., \citealt{zahedy2019,sameer2021}). The frequency of detected \ovi\ with {\coldenovi}\,$\ga 13.6$ is similar around 60\% in HM SLFSs, LM and HM pLLSs (see Table~\ref{detection_rates}), but it is substantially lower in LM SLFSs at only $21 \pm 6\%$. This implies that pLLSs have, in general, a more complex gas-phase structure of the metallicities. Hence, LM SLFSs mostly probe single-phase gas, while HM SLFSs and pLLSs typically trace multiphase gas.\footnote{Although they could also probe more complex regions where the \ovi\ and lower ions may not be cospatial, this is unlikely to be systematic owing to the close velocity proximity of the \hi\ and \ovi\ velocities that are separated by $\la 40$ \kms, see Fig.~\ref{fig:dv90_Z}.}

In their survey of galaxies at $z \sim 0.2$, which is combined with the COS-Halos survey \citep{Tumlinson2013}, \citet{Johnson2015} found that the vast majority of the \ovi\ non-detections were beyond the virial radius of galaxies, but these absorbers have \hi\ column densities ${\rm \log}\,N($\hi$)<15$, i.e., much lower than those in our survey. A handful of \ovi\ non-detections with {\coldenovi}\,$\la 13.6$ are found within $\la \rvir$ where $\log N($\hi$) \ga 15$; the majority of these galaxies are early-type galaxies (see also \citetalias{Tchernyshyov2022} and Fig.~\ref{fig:logNOVI_sf_q_hi_strong_weak}). In the CGM$^2$ galaxy survey, the population of \ovi\ non-detections is also predominantly found beyond \rvir\ (see Section~\ref{sec:columns} and Fig.~\ref{fig:logNOVI_Z_NHI_histograms}c) \citepalias{Tchernyshyov2022}. This combined with the fact that the highest fraction of non-detections of \ovi\ are found in LM absorbers, they may instead trace the overdense IGM, like some of the LM pLLSs \citep{Berg2023}. In addition, a fraction of these might be found in CGM ($\la \rvir$) of dwarf galaxies with stellar masses of $\la 10^8$--$10^9$ M$_\sun$ since 3/8 of the dwarfs in the sample of \citet{Johnson2017} have no \ovi\ detection but have relatively strong \hi\ absorption (\colden$\,\ga 14.5$).

\subsection{\hi-Selected Absorbers with \ovi\ Detections: Multiple-Phase Gas}\label{s-ovi-detected}

About 20\% of the LM SLFSs and 60\% of the LM/HM pLLSs and HM SLFSs have associated \ovi\ absorption (see Table~\ref{detection_rates}). As shown by many independent works studying the detailed ionization processes at play, photoionization models can generally match the \hi, low and intermediate ions, but rarely \ovi\ in the same gas phase at $z\la1$ as the EUVB is typically not hard enough and the densities too high to produce coexisting \ovi\ and low ions (e.g., \citealt{Cooksey2008,Lehner2009,Lehner2013,Kacprzak2012,Crighton2013}; \citetalias{Savage2014}; \citealt{Werk2016,Stocke2017,Rosenwasser2018,zahedy2019,Cooper2021,haislmaier2021,sameer2021,Sameer2024CMBM}). Hence, detected \ovi\ likely traces a separate hot, collisionally ionized phase, cooler collisionally ionized gas in non-equilibrium, or a very low-density photoionized phase distinct from the majority of \hi\ gas at $z\la 1$.

When we compare the velocity profiles between \ovi, \hi, \cii, and \ciii, we find larger centroid offsets for \ovi\ relative to \hi\ compared to the lower ions  (see Figs.~\ref{fig:dv90_Z} and \ref{fig:logNOVI_Z_cbar}), indicating \ovi\ traces a distinct phase \citepalias[e.g.,][]{Savage2014}. While the low/intermediate ions show consistent velocity width distributions between metallicity subsamples, \ovi\ widths differ significantly, with HM absorbers being broader than LM absorbers (Fig.~\ref{fig:Deltav90_Z}). This suggests intrinsic differences in the \ovi-bearing gas related to metallicity, possibly owing to outflows driven or recycling material produced by high stellar activity within star-forming galaxies where higher metallicities and broader \ovi\ absorbers are expected \citep[e.g.,][]{Lehner2014,Lehner2017,Werk2016,Qu2024}.

\subsection{\hi-Selected Absorbers with \coldenovi$\,\la 14$}\label{s-ovi-weak}

Weak \ovi\ absorption (\coldenovi$\,\la 14$) in \hi-selected absorbers shows little connection between the properties of \ovi\ and the cool, lower ionization gas. This \ovi\ can be found over the whole range of \hi\ column densities and metallicities of our sample (Fig.~\ref{fig:logNOVI_logNHI_cbarOVI}). However, in contrast to their stronger analogs, the kinematics of the weak \ovi\ in these absorbers are not as extreme; their velocity widths are typically $\Delta v_{90}($\ovi$)\la 80$ \kms\ (Fig.~\ref{fig:logNOVI_Z_cbar}).

In Figs.~\ref{fig:logNOVI_sf_q_hi_strong_weak}a, b, we show the \ovi\ column density distributions for the HM and LM portions of our sample, respectively, splitting the \ovi\ distributions for absorbers on either side of the \coldenovi$\,=14$ threshold, which corresponds to the geometric mean of the sample with \ovi\ detections (see \S\ref{sec:columns}). For comparison, we show the \ovi\ column density distributions for absorbers projected within $\rvir$ of star-forming and passive galaxies from \citetalias{Tchernyshyov2022} in Fig.~\ref{fig:logNOVI_sf_q_hi_strong_weak}c, while we show the distribution of \ovi\ column densities for gas projected at large distances from the closest galaxies ($200 \le R \le 600$\,kpc) from \citetalias{Savage2014} in Fig.~\ref{fig:logNOVI_sf_q_hi_strong_weak}d.  Fig.~\ref{fig:logNOVI_sf_q_hi_strong_weak} shows the weak \ovi\ absorbers in the HM and LM groupings are not statistically different from one another ($p$-value $\gg$ 0.05), and those are similar in many ways to that of the IGM sample at $R > \rvir$ \citepalias{Savage2014}. A log-rank test comparing the weak LM and weak HM groupings with the IGM sample yields $p$-values $\gg$ 0.05, in both cases, suggesting that these distributions are not significantly different from one another. Thus, the \hi-selected absorbers with weak \ovi\ may predominantly trace the IGM and extended CGM (beyond \rvir), although a small population of weak absorbers is also seen in the tail of the galaxy-centered samples (and some of these may include dwarf galaxies, see \citealt{Johnson2017,Qu2024}). 

\subsection{\hi-Selected Absorbers with \coldenovi$\,\ga 14$}\label{s-ovi-strong}

Similar to their weaker counterparts, \hi-selected absorbers with strong \ovi\ absorption (\coldenovi$\,\ga 14$) show no correlation between their \ovi\ and \hi\ column densities. However, they exhibit a strong connection with the metallicity of the cool photoionized gas, as they are found nearly exclusively at $\xh>-1.4$. Out of 35 absorbers with \coldenovi$\,\ga 14$, only 5 are found with $\xh \la -1.4$. Using the same criterion to define the threshold between HM and LM gas at $z\sim2.4$--3.5 places the split at $\xh=-2.4$ \citep{Lehner2022}, above which strong \ovi\ also predominates, implying that this relation is observed at any $z$. This is striking since the metallicities of the cool, lower ionization phase and the \ovi-bearing gas can differ in multiphase absorbers \citep[e.g.,][]{Savage2011a,Muzahid2015,Cooper2021}. The pronounced correlation suggests strong \ovi\ and HM cool gas are likely produced by similar phenomena, such as galaxy feedback.

As shown in Figs.~\ref{fig:logNOVI_Z_NHI_histograms} and \ref{fig:logNOVI_sf_q_hi_strong_weak}, strong \ovi\ dominates in the CGM ($<R_{\rm vir}$) of star-forming galaxies (\citealt{Tumlinson2011,Johnson2015}; \citetalias{Tchernyshyov2022}), including those of some dwarf galaxies \citep{Johnson2017,Qu2024}. \citet{Tchernyshyov2023} find higher \ovi\ covering fractions around star-forming galaxies across three galaxy mass ranges than about passive, quiescent galaxies (for \coldenovi$\,\ga 14.3$). \citet{Werk2016} using the COS-halos sample show that very broad and strong \ovi\ absorbers are defining characteristics of the CGM of star-forming $L^{\star}$ galaxies at $z\sim$\,0.2. \citet{Berg2023} show HM gas occurs solely around star-forming, relatively massive galaxies. Given the behavior of $R<\rvir$ absorption, that the strong \ovi\ absorbers in our sample are found exclusively in the HM population, and the corresponding large velocity breadths of much of our sample, there seems to be a robust link between strong \ovi\ absorption and enriched gas within halos of star-forming galaxies. 

As noted by \citet{Lehner2014} (see also \citealt{Lehner2017}), the breadth ($\Delta v_{90} ({\ovi}) \ga 100$ \kms)  and strength (\coldenovi$\,\ga 14.3$) of the \ovi\ absorption in strong \hi\ absorbers at high redshifts ($z\sim 2.4$--3.5) are strikingly similar to those observed in starburst galaxies at low redshift \citep[e.g.,][]{Grimes2009,Muzahid2015,Tripp2008,Rosenwasser2018}. Strong and broad \ovi\ absorbers could, therefore, be a signature of large-scale feedback (outflowing or recycling) in these high redshift galaxies. This could also be the case for the strong low-redshift \ovi\ absorbers in view of their connection to the massive star-forming galaxies, especially for the 50\% of the strong \ovi\ absorbers that have a large breadth with $\Delta v_{90} \ga 100$ \kms\ (see Fig.~\ref{fig:logNOVI_Z_cbar}b). Recently, \citet{Qu2024} show that the \ovi\ absorption is stronger and broader in star-forming galaxies than passive galaxies within \rvir\ at $z\sim 0.4$--0.7. These could arise in large-scale radiatively cooling gas behind a supersonic wind produced by galaxy starburst-driven winds \citep[e.g.,][]{Heckman2002,Werk2016}. Alternatively, some of these \ovi\ absorbers could also be signatures of the hot halo corona themselves \citep[e.g.,][]{Werk2016,Sameer2024CMBM}, radiative cooling of galactic fountain gas \citep[e.g.,][]{Sembach2001,Wakker2005,Lehner2022MNRAS.513.3228L,Marasco2022MNRAS.515.4176M}, or possibly active galactic nucleus (AGN) feedback \citep[e.g.,][]{Oppenheimer2016,Suresh2016}. Future observations of the galaxies in some of these fields should help better discern the origins of the strong \ovi\ associated with these \hi-selected absorbers. 
\section{Summary}\label{s-summary}

We studied the kinematics and absorption properties of {\ovi} gas and their relationship with lower ionization gas in a {\hi}-selected survey by analyzing the {\ovi} absorption in a sample of 126 low-$z$ absorption systems ($0.14 < z < 0.73$) observed in {\hst}/COS G130M and/or G160M spectra with $\sn \ga 8$ per resolution element and complete to \coldenovi$\,\ga 13.6$ at the $2\sigma$ level. The main results are as follows:
  
\begin{enumerate}[wide, labelwidth=!, labelindent=0pt]

\item The sample comprises 100 SLFSs and 26 pLLSs with  \ovi\ detection rates at a sensitivity level of \coldenovi$\,\ga 13.6$ of about 50\%  and 60\%, respectively. This difference is largely driven by the low-metal (LM; $\xh \leq -1.4$) SLFSs since their \ovi\ detection rate is 20\% compared to 60\% for the high-metal (HM; $\xh > -1.4$) SLFSs and HM or LM pLLSs.

\item An empirical characterization of the profile kinematics shows that $\langle|{\vcii} - {\vhi}|\rangle=9 \pm 7$ {\kms} and $\langle|{\vciii} - {\vhi}|\rangle = 7 \pm 8$ {\kms} compared to $\langle|{\vovi} - {\vhi}|\rangle = 14 \pm 11$ {\kms} (the error represents the $1\sigma$ dispersion). On average, larger velocity offsets between \ovi\ and lower ions indicates that \ovi\ does not necessarily trace the same phases of gas as the lower ions. On the other hand, SLFSs and pLLSs with no \ovi\ are consistent with the gas being in a single phase. 

\item We measure the velocity velocity widths, $\Delta v_{90}$, for {\hi}, {\cii}, {\ciii}, and {\ovi}. The mean observed velocity width and dispersions are  $\langle\Delta v_{90}\rangle = 82 \pm 22$ {\kms} for {\hi}, $74 \pm 27$ {\kms} for {\cii}, $67 \pm 21$ {\kms} for {\ciii}, and $89 \pm 32$ {\kms} for {\ovi}  (the error represents the $1\sigma$ dispersion). HM {\ovi} absorbers show larger mean velocity widths, $\Delta v_{90}$ =  92 $\pm$ 31 {\kms}, compared with those of LM {\ovi} absorbers, $\Delta v_{90}$ =  75 $\pm$ 16 {\kms}. The higher line width of {\ovi} in HM lower-ionization gas compared with LM lower-ionization gas suggests that HM SLFSs and pLLSs typically exhibit more pronounced multiphase characteristics. 
    
\item  We observe a strong correlation between the {\ovi} column density and metallicity, but not between {\ovi} column density and {\hi} column density. In particular, LM absorbers show weaker {\ovi} absorption than HM absorbers, with mean (and uncertainty on the mean) values of  $\meancoldenovi = 13.46 \pm 0.02$ and $14.11 \pm 0.01$, respectively, hinting at possible differences in the physical processes and origins of the {\ovi}-bearing gas between these two populations. 

\end{enumerate}

With these findings, we group the SLFSs and pLLSs in three categories: (1) absorbers with non-detections of \ovi\ with \coldenovi$\, < 13.6$; (2) absorbers with \ovi\ detections with \coldenovi$\,\la 14$; and (3) absorbers with \ovi\ detections with \coldenovi$\,\ga 14$. Comparing the distributions of the \ovi\ and metallicity of the cool gas in surveys of \ovi\ absorbers and galaxies (e.g., \citetalias{Tchernyshyov2022}; \citet{Berg2023}; \citetalias{Savage2014}; \citealt{Johnson2015}), we infer that: 

\begin{enumerate}[wide, labelwidth=!, labelindent=0pt]

\item SLFSs with no \ovi\ are predominantly found in metal-poor ($\xh \le -1.4$) gas and likely trace dense regions of the IGM. 

\item  SLFSs and pLLSs with \ovi\ absorbers with {\coldenovi}$\,\la 14$ are found at any metallicities and most likely probe dense IGM regions, extended CGM beyond $\rvir$. 

\item SLFSs and pLLSs with \ovi\ absorbers with {\coldenovi}$\,\ga 14$ are mostly found in metal enriched gas ($\xh > -1.4$) and  most likely probe the CGM  within {\rvir} of star-forming galaxies. About 50\% of these absorbers are also broad with  $\Delta v_{90}($\ovi$) \ga 100$ \kms, which could be possibly signatures of active galaxy feedback.

\end{enumerate}

Based on the above, we conclude that the combination of metallicity of the cool gas {\it and}\ \ovi\ in surveys like CCC or KODIAQ provides a strong diagnostic of the origins of {\hi}-selected absorbers and their association with galaxies. Further characterization of the galaxies in the fields of these absorbers would be extremely beneficial for a deeper understanding of the galaxies and their environments and use these absorbers as signatures of specific phenomena and environments.

Further knowledge on these absorbers can also be gained using cosmological simulations, as our entire dataset is well suited to be modeled by a high-resolution cosmological simulation volume. For example, IllustrisTNG50~\citep{Nelson2018,Pillepich2019}, FIREbox~\citep{Feldmann2023}, and E-MOSAICS~\citep{Pfeffer2018} provide ideal volumes to simulate both the IGM and CGM absorbers (see, e.g., the recent study by \citealt{Weng2024} for \hi\ absorbers at $z\sim0.5$). By forward modeling COS sightlines using a mock sightline generator (e.g., Trident, \citealt{Hummels2017}), one can create full absorption systems, develop an algorithm to select \hi-selected samples that mimic our pLLS/SLFS sample, and generate measurements of \ovi\ column density, metallicity, velocity width, and velocity offsets between different ion species. This exercise would test whether the $N($\ovi$)$--\xh\ correlation naturally arises in these simulations helping elucidate its origin. Moreover, it would provide additional insight into whether strong \ovi\ in SLFSs/pLLSs arise from outflows and result in the observed velocity widths and offsets, and how that may depend on the impact parameter (e.g., \citealt{Shen2013}). \citet{Mallik2023,Mallik2024,Maitra2024} using Sherwood simulations~\citep{Bolton2017} find that the statistical properties such as distribution functions of {\ovi} column density, velocity spread, fraction of {\lya} absorbers with detectable metal lines are influenced by feedback model implemented and the EUVB used. They find that changing the feedback model (for example, from stellar feedback only to a combination of stellar and active galactic nucleus feedback) can produce changes in the distribution functions that are comparable to those seen when varying the EUVB for a given feedback prescription.

\section*{Acknowledgements}

We thank the referee for their detailed and helpful comments, which have strengthened and improved the manuscript. Support for this research was provided by NASA through grant HST-AR-15634 from the Space Telescope Science Institute (STScI), which is operated by the Association of Universities for Research in Astronomy, Incorporated, under NASA contract NAS5-26555. Data presented in this work were obtained from CCC, which was funded through NASA grant HST-AR-12854 from STScI. 
This research was supported by the Notre Dame Center for Research Computing through the Grid Engine software and the Notre Dame Cooperative Computing Lab through the HTCondor software. This research made use of \software{Astropy \citep{astropy:2022}, Matplotlib \citep{Hunter2007}, PyIGM \citep{pyigm}}, NASA's Astrophysics Data System (\url{https://ui.adsabs.harvard.edu/}), adstex (\url{https://github.com/yymao/adstex}).

\facilities{HST(COS)}

\bibliography{ms}

\clearpage
\appendix

\begin{figure*}[tbp]
\centering
\includegraphics[scale=0.65]{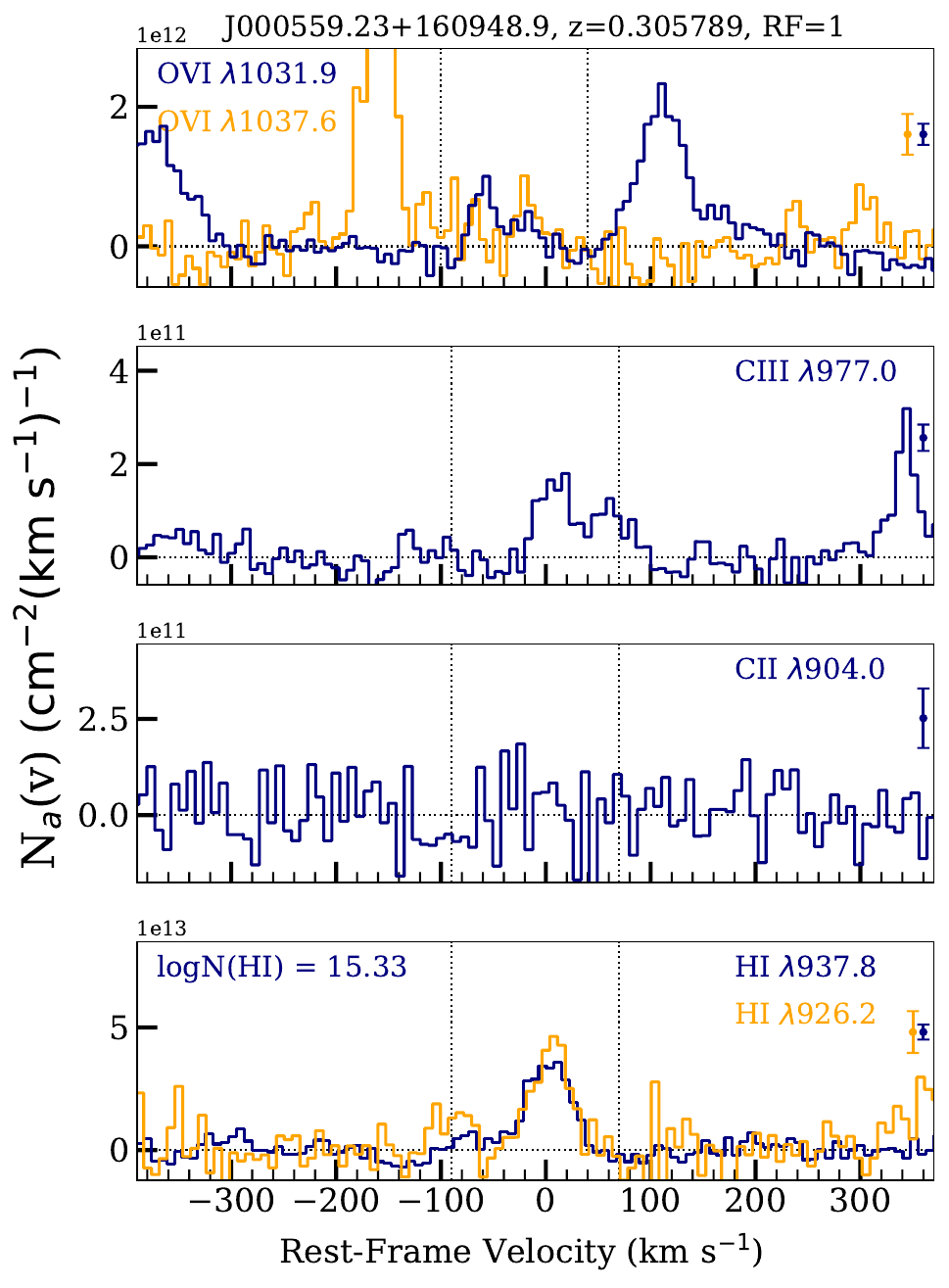}
\caption{Apparent optical column density profiles for {\ovi}, {\cii}, {\ciii}, and {\hi}, for the  $z= 0.305789$ absorber toward the quasar J000559.23$+$160948.9. The integration velocity ranges are indicated by two vertical lines in each panel. The average symmetric uncertainty on the data is indicated to the right in each of the panels. The profiles are binned by 3 pixels for display purposes. The complete figure set (170 images) is available as supplementary material.} \label{fig:example_NaV}
\end{figure*}

\begin{figure*}[tbp]
\centering
\includegraphics[scale=0.65]{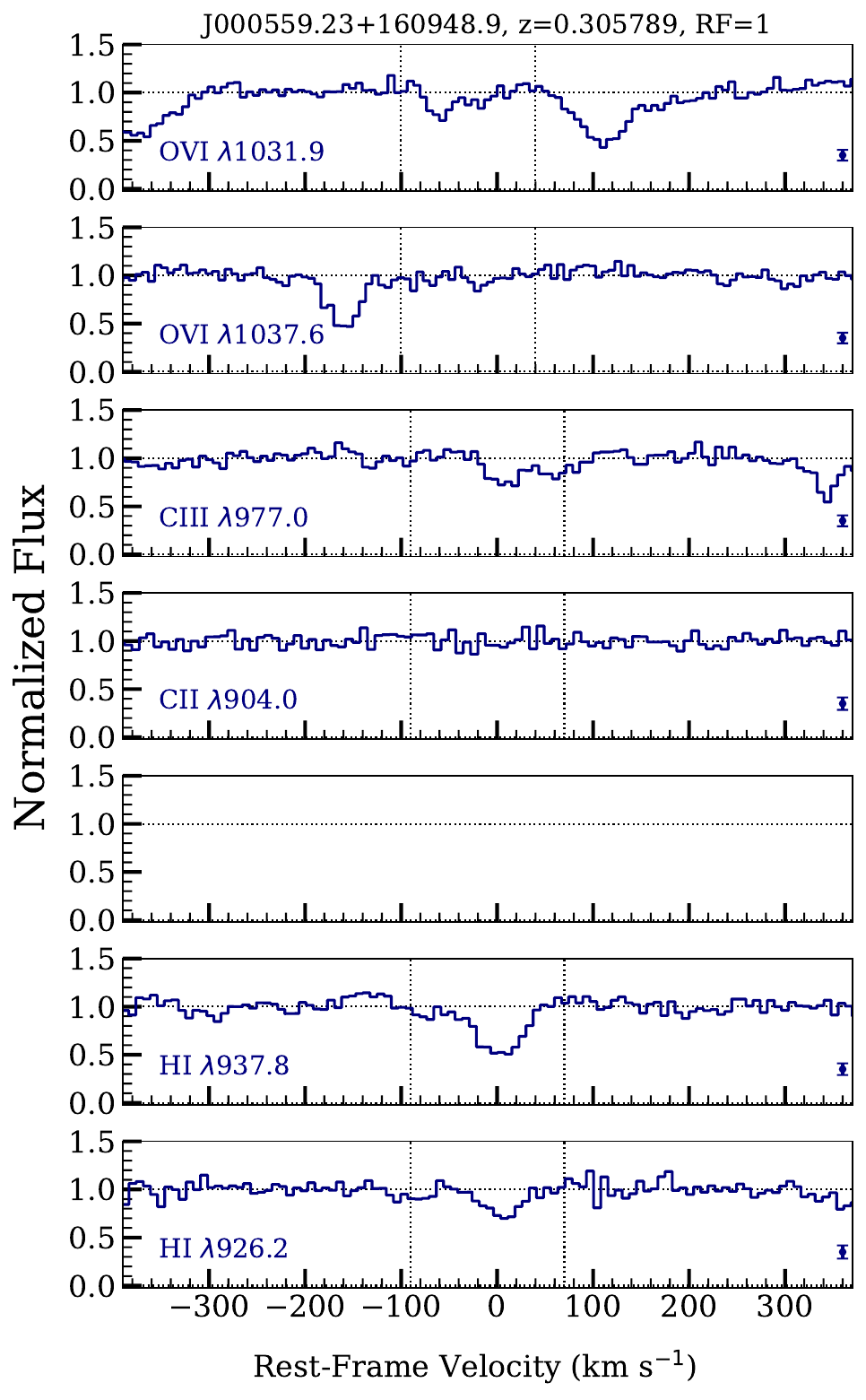}
\caption{Normalized absorption lines of {\ovi}, {\cii}, {\ciii}, and {\hi}, as a function of rest-frame velocity centered on the absorber at $z = 0.305789$ toward J000559.23$+$160948.9. These transitions are from COS G130M and G160M. The vertical dashed lines shows the velocity range of the absorption over which the velocity profile was integrated to derive the column densities and kinematics. The average symmetric uncertainty on the data is indicated to the right in each of the panels. The profiles are binned by 3 pixels for display purposes. The complete figure set (170 images) is available as supplementary material.} \label{fig:example_NF}
\end{figure*}

We provide apparent column density, {\nav}, profiles of {\ovi}, {\cii}, {\ciii}, and {\hi}, as illustrated in Figure~\ref{fig:example_NaV}, for each absorber in this study (including those with lower S/N) as supplementary online material. These {\nav} plots include the adopted {\hi} column densities from \citetalias{Lehner2018} in the panel showing absorption in one of the weak {\hi} transitions. We also provide normalized flux profiles of {\ovi}, {\cii}, {\ciii}, and {\hi} for each absorber. Additional ions or transitions are available in \citetalias{Lehner2018}. Figure~\ref{fig:example_NF} shows an example of these profiles.  Vertical dashed lines in these figures indicate the velocity range encompassing the absorption, which was integrated to derive column densities and kinematics, including velocity and line-width. The complete set (170 images) of {\nav} and normalized flux profiles for the entire sample of {\ovi} absorbers is provided as online supplementary material.

\makeatletter
\renewcommand{\thetable}{A\@arabic\c@table}
\setcounter{table}{0}
\thispagestyle{plain}
\phantom{This is needed for the table to fully resolve.}

\end{document}